\documentclass[12pt]{elsart}

\usepackage{graphicx}
\usepackage{subfigure}
\usepackage{amsmath}
\usepackage{cite}
\usepackage{tikz}
\usepackage{setspace}

\DeclareMathOperator*{\argmin}{arg\,min}

\journal{Physica E}

\begin{document}

\begin{frontmatter}

\title{Thermodynamic and magnetocaloric properties of geometrically frustrated Ising nanoclusters}
\author{M. \v{Z}ukovi\v{c}\corauthref{cor}}
\ead{milan.zukovic@upjs.sk}
\address{Department of Theoretical Physics and Astrophysics, Faculty of Science,\\ 
P. J. \v{S}af\'arik University, Park Angelinum 9, 041 54 Ko\v{s}ice, Slovakia}
\corauth[cor]{Corresponding author.}

\begin{abstract}
Thermodynamic and magnetocaloric properties of geometrically frustrated Ising spin clusters of selected shapes and sizes are studied by exact enumeration. In the ground state the magnetization and the entropy show step-wise variations with an applied magnetic field. The number of steps, their widths and heights depend on the cluster shape and size. While the character of the magnetization plateau heights is always increasing, the entropy is not necessarily decreasing function of the field, as one would expect. For selected clusters showing some interesting ground-state properties, the calculations are extended to finite temperatures by exact enumeration of densities of states in the energy-magnetization space. In zero field the focus is laid on a peculiar behavior of some thermodynamic quantities, such as the entropy, the specific heat and the magnetic susceptibility. In finite fields various thermodynamic functions are studied in the temperature-field parameter plane and particular attention is paid to the cases showing an enhanced magnetocaloric effect. The exact results on the finite clusters are compared with the thermodynamic limit behavior obtained from Monte Carlo simulations.
   
\end{abstract}


\end{frontmatter}

\section{Introduction}
A two-dimensional triangular lattice Ising antiferromagnet is a typical geometrically frustrated spin system with a long history of investigation~\cite{wann,wann2,hout,step,metc1,schick,netz}. The ground state (GS) of this fully frustrated system is highly degenerate with non-vanishing entropy density of $0.32306$~\cite{wann,wann2}, which results in lack of long-range ordering at any finite temperature~\cite{wann,wann2,step}. Nevertheless, the presence of an external magnetic field can partially lift this degeneracy and the system can show a ferrimagnetic long-range order manifested by a magnetization plateau at one third of the saturation value~\cite{metc1,schick,netz,zuko}. At the saturation field there is another substantial increase in the entropy due to the situation in which a large number of non-interacting spins are free to reorient at no energy cost\cite{metc2}. The value of this saturation field entropy density $0.3332427$ has been found exactly~\cite{milo,varm} through the solution of the hard-hexagon model~\cite{baxt}. Such a large entropy change can result in an enhanced magnetocaloric effect and offers the opportunity to exploit the system as a magnetic refrigerant. Several recent studies have demonstrated that in the vicinity of the saturation field, huge cooling rates can be achieved for certain frustrated topologies due to the large (sometimes even macroscopic) degeneracy of states~\cite{zhit1,zhit2,derz,hone,schn0}.\\
\hspace*{5mm} In case of systems comprising a finite number of spins located in domains of different shapes, the zero-field GS degeneracy strongly depends on domain sizes and boundary conditions~\cite{griff,aizen}. By increasing the domain size to infinity the ground-state entropy in the thermodynamic limit is recovered only if the boundary conditions giving maximum degeneracy are considered~\cite{aizen,simon}. If it is not the case, such as for domains of a shape of a rhombus, the entropy density vanishes in the thermodynamic limit~\cite{milla1}. In zero field the domain size- and shape-dependences of the entropy density have been investigated for domains of various shapes and sizes~\cite{milla1,zuko3}. In the presence of a magnetic field, the effect of frustration in finite antiferromagnetic spin clusters has been shown to lead to unusual and interesting magnetization processes at low temperatures, quite different from their nonfrustrated counterparts~\cite{schm,schn1,schn2,kons}. For example, jumps to the saturation magnetization or metamagnetic phase transitions at zero temperature have been observed in some antiferromagnetic finite-size spin systems with icosahedral symmetry. Furthermore, they displayed an enhanced cooling rate in comparison with non-frustrated (bipartite) spin rings either at the saturation field or elsewhere, depending on the cluster geometry~\cite{schn0}. Using of molecular magnets consisting of such spin clusters as magnetic refrigerants has some advantages, such as the possibility to synthesize them in a great variety of structures. In some cases the clusters do not interact with each other due to large distances between the magnetic centers of different molecules. Then, magnetic properties of a single cluster, which due to the small size can be often calculated exactly, can represent the macroscopic sample.\\
\hspace*{5mm} In the present paper we consider geometrically frustrated Ising spin clusters of various shapes and sizes and by exact enumeration study GS magnetization processes and degeneracies as functions of an external magnetic field. For several cases that show some interesting GS properties, we extend our calculations to finite temperatures by exact evaluation of densities of states in the energy-magnetization space. In zero field we focus on a peculiar behavior of the entropy, the specific heat and the magnetic susceptibility, and in finite fields we study various thermodynamic functions in the temperature-field parameter plane and pay particular attention to the cases showing an enhanced magnetocaloric effect. The exact results on the finite clusters are compared with the thermodynamic limit behavior obtained from Monte Carlo simulations.  

\section{Model and methods} 
\begin{figure}[t]
\centering
    \subfigure[R]{\includegraphics[scale=0.5,clip]{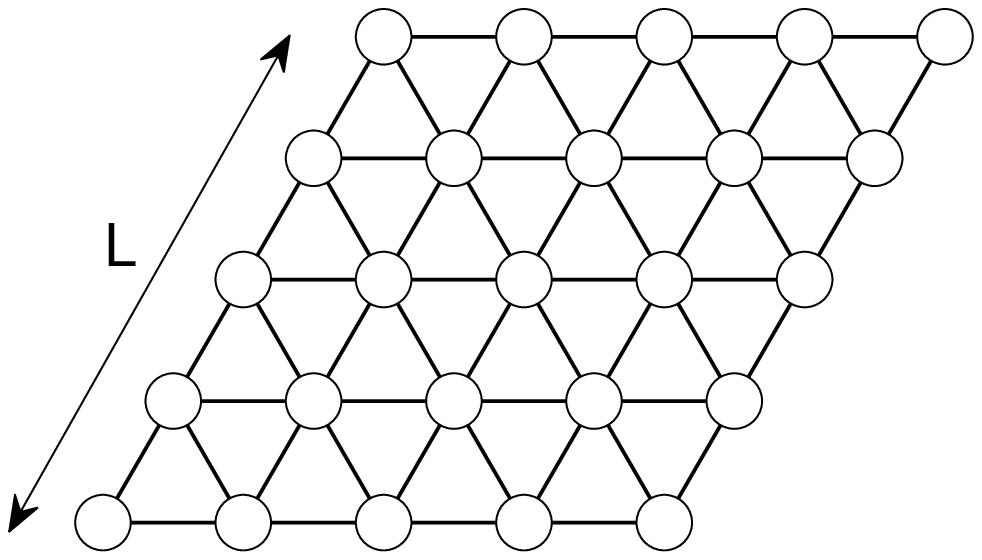}\label{fig:rhomb}}
    \subfigure[R1]{\includegraphics[scale=0.5,clip]{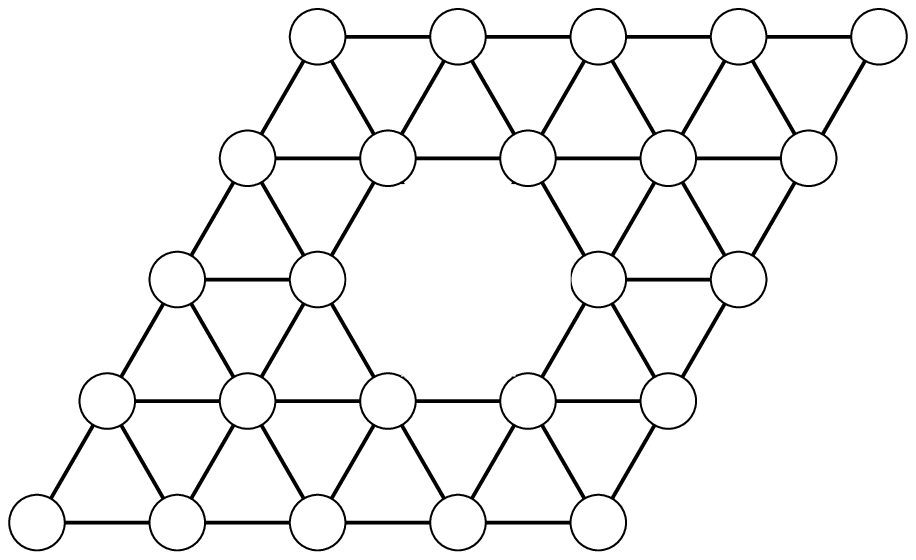}\label{fig:rhomb1}} \\
    \subfigure[H]{\includegraphics[scale=0.5,clip]{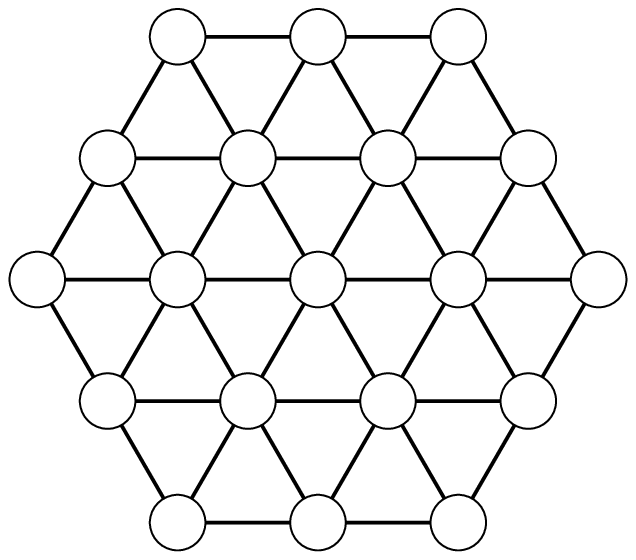}\label{fig:hexa}} 
    \subfigure[H1]{\includegraphics[scale=0.5,clip]{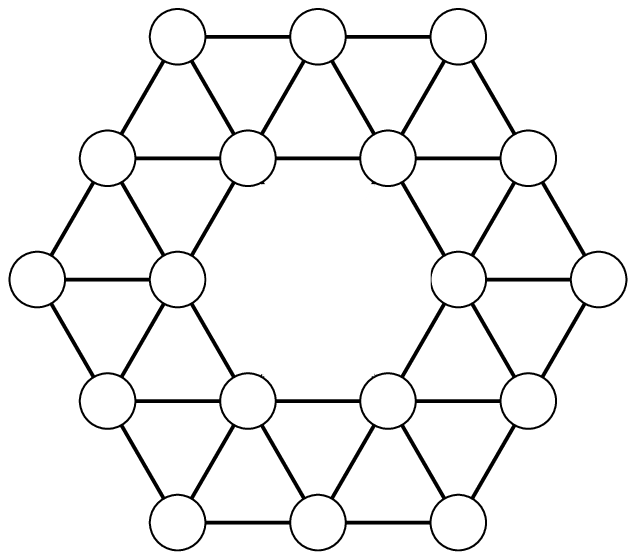}\label{fig:hexa1}} \\
    \subfigure[T]{\includegraphics[scale=0.5,clip]{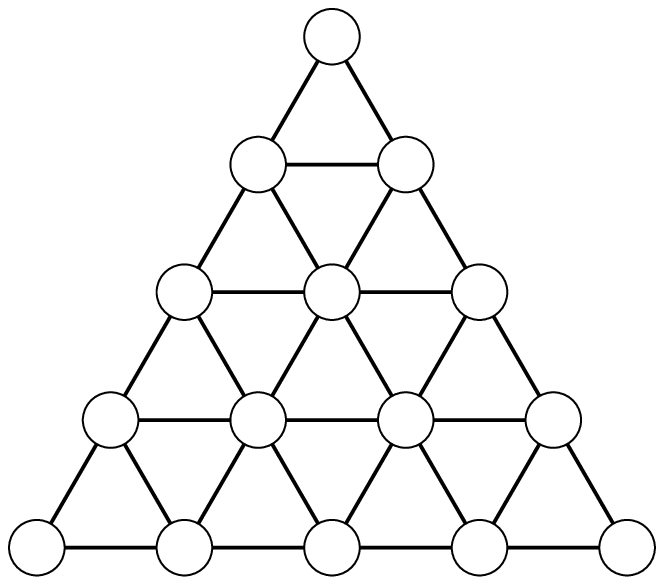}\label{fig:tria0}}
    \subfigure[T1]{\includegraphics[scale=0.5,clip]{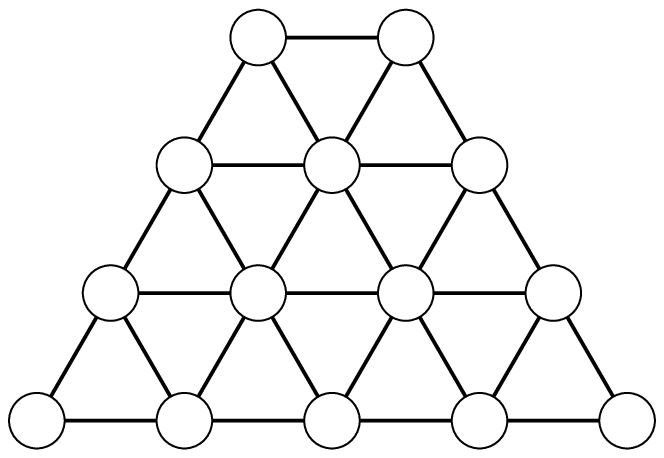}\label{fig:tria1}}
    \subfigure[T2]{\includegraphics[scale=0.5,clip]{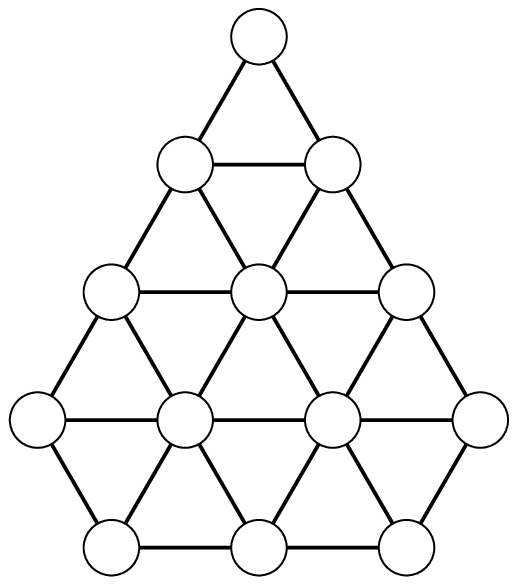}\label{fig:tria2}}
    \subfigure[T3]{\includegraphics[scale=0.5,clip]{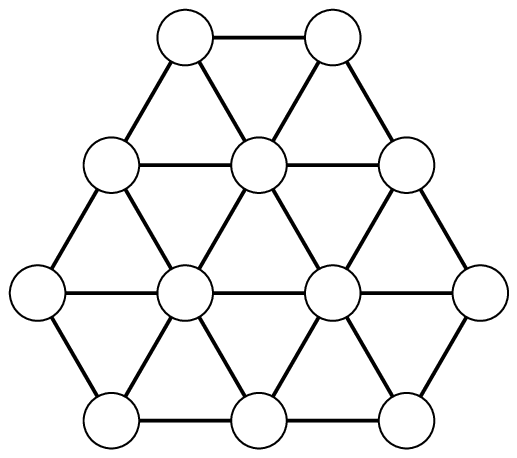}\label{fig:tria3}}\\
\caption{Various clusters derived from the rhombus with the side length $L=5$: (a) rhombus (R), (b) rhombus without the central spin (R1), (c) hexagon (H), (d) hexagon without the central spin (H1), (e) triangle (T), triangles with (f) one (T1), (g) two (T2) and (h) three (T3) vertices removed.}\label{fig:clusters}
\end{figure}

The Ising spin cluster on a lattice with triangular geometry in an external magnetic field can be modeled by the Hamiltonian 
\begin{equation}
\label{Hamiltonian}
\mathcal H=-J\sum_{\langle i,j \rangle}s_{i}s_{j} - h\sum_{i}s_{i},
\end{equation}
where the spin on the $i$th cluster site $s_{i}=\pm 1/2$, the summation $\langle i,j \rangle$ runs over nearest-neighbor sites, $J<0$ is an antiferromagnetic exchange interaction parameter (for simplicity, hereafter, we put $J=-1$) and $h$ is the external magnetic field. The cluster consists of a finite number of spins arranged in domains of various shapes, as shown in Fig.~\ref{fig:clusters}. For such relatively small clusters it is possible to fully explore the state space, exactly determine the density of states $g(M,E)$ and calculate any thermodynamic quantities of interest that are functions of the magnetization $M$ and/or the internal energy $E$. In order to understand the effects of different shapes and sizes of the finite clusters it is interesting to compare the obtained results with those for the thermodynamic limit, i.e., for an infinite lattice, which can be estimated from Monte Carlo (MC) simulation. 

\subsection{Exact enumeration}
GS spin configurations ${\bf s_{\mathrm{GS}}}=\{s_1,\ldots,s_{N}\}$, where $N$ is a number of spins in the cluster, can be found as configurations\footnote{There are in total $2^{N}$ configurations to be considered.} that minimize the energy functional~(\ref{Hamiltonian}), i.e.:
\begin{eqnarray}
\label{GS} {\bf s_{\mathrm{GS}}} = \argmin_{{\bf s}} \, \mathcal H({\bf s}).
\end{eqnarray}
For different field values, we record the number of GS configurations $W$ (degeneracy) and their magnetizations $M_j=\sum_{i=1}^{N}{\bf s_{\mathrm{GS}}}$, $j=1,\ldots,W$, and evaluate the magnetization per spin $m=\sum_{j=1}^{W}{M_j}/(WN)$ and the entropy per spin $S/N=\ln W/N$ (here and hereafter, we put the Boltzmann constant $k_B=1$).\\
\hspace*{5mm} At finite temperatures, having obtained the exact density of states $g(M,E)$, as a function of the total magnetization $M=\sum_{i=1}^{N}{s_i}$ and the total exchange energy $E=\sum_{\langle k,l \rangle}s_{k}s_{l}$, where the summation $\langle k,l \rangle$ runs over the nearest neighbors $s_{k}$ and $s_{l}$ on the considered cluster, one can calculate a mean value of a thermodynamic quantity $A$ in the temperature-field parameter plane as
\begin{equation}
\label{mean}
A(T,h)=\frac{\sum_{M,E}{Ag(M,E)e^{-\frac{(E-hM)}{T}}}}{Z(T,h)},
\end{equation}
where 
\begin{equation}
\label{PF}
Z(T,h)=\sum_{M,E}{g(M,E)e^{-\frac{(E-hM)}{T}}} 
\end{equation}
is the partition function. Using the above equations, we can calculate the internal energy $U(T,h)=\langle E(T,h)-hM(T,h) \rangle$, and magnetization per spin $m=\langle M(T,h) \rangle/N$. The entropy per spin can be then obtained as
\begin{equation}
\label{entropy}
\frac{S(T,h)}{N}=\frac{U(T,h)-F(T,h)}{NT},
\end{equation}
where $F(T,h)=-T\ln{Z(T,h)}$ is the free energy. Finally, the specific heat $c$ and the susceptibility $\chi$ per spin are calculated as
\begin{equation}
\label{spec_heat}
c(T,h)=\frac{\langle \left[E(T,h)-hM(T,h)\right]^2 \rangle-\langle E(T,h)-hM(T,h) \rangle^2}{NT^2}
\end{equation}
and
\begin{equation}
\label{susc}
\chi(T,h)=\frac{\langle M(T,h)^2 \rangle - \langle M(T,h) \rangle^2}{NT}.
\end{equation}
The adiabatic temperature change due to the change of the magnetic field from $0$ to $h$ can be obtained from thermodynamic relations as
\begin{equation}
\label{temp_change}
\Delta T_{ad}(T,h)=-\int_{0}^{h}\frac{T}{c}\Big(\frac{\partial m}{\partial T}\Big)_hdh.
\end{equation}

\subsection{Monte Carlo simulation} 
MC approach with Metropolis dynamics is employed to simulate larger lattice sizes with periodic boundary conditions to approximate the thermodynamic limit behavior. It turns out that magnetization processes show little dependence on the system size and, therefore, for this purpose we use a relatively small linear lattice size of $L = 24$. Consequently, the equilibration is relatively fast and $20,000$ MC sweeps were confirmed to be sufficient to bring the system to the equilibrium. For thermal averaging we use another $100,000$ MC sweeps.
To simulate the magnetization process in the increasing field at a fixed temperature, the simulations start from zero field using a random initial state and then the field is gradually increased with the step $\Delta h = 0.01$ and the simulations start from the final configuration obtained at the previous field value. We calculate the total magnetization per spin $m$ and the specific heat per spin $c$ according to the above definitions, however, the entropy cannot be evaluated in MC simulations directly. Therefore, the isothermal magnetic entropy change per spin $\Delta S(T,h)/N$, which occurs at changing the field from 0 to $h$, is estimated from thermodynamic Maxwell equation by numerical integration as
\begin{equation}
\label{entr_change}
\frac{\Delta S(T,h)}{N}=\int_{0}^{h}\Big(\frac{\partial m}{\partial T}\Big)_hdh.
\end{equation}

\section{Results}
\subsection{Ground state}
\begin{figure}[]
\centering
\subfigure[T]{\includegraphics[scale=0.37,clip]{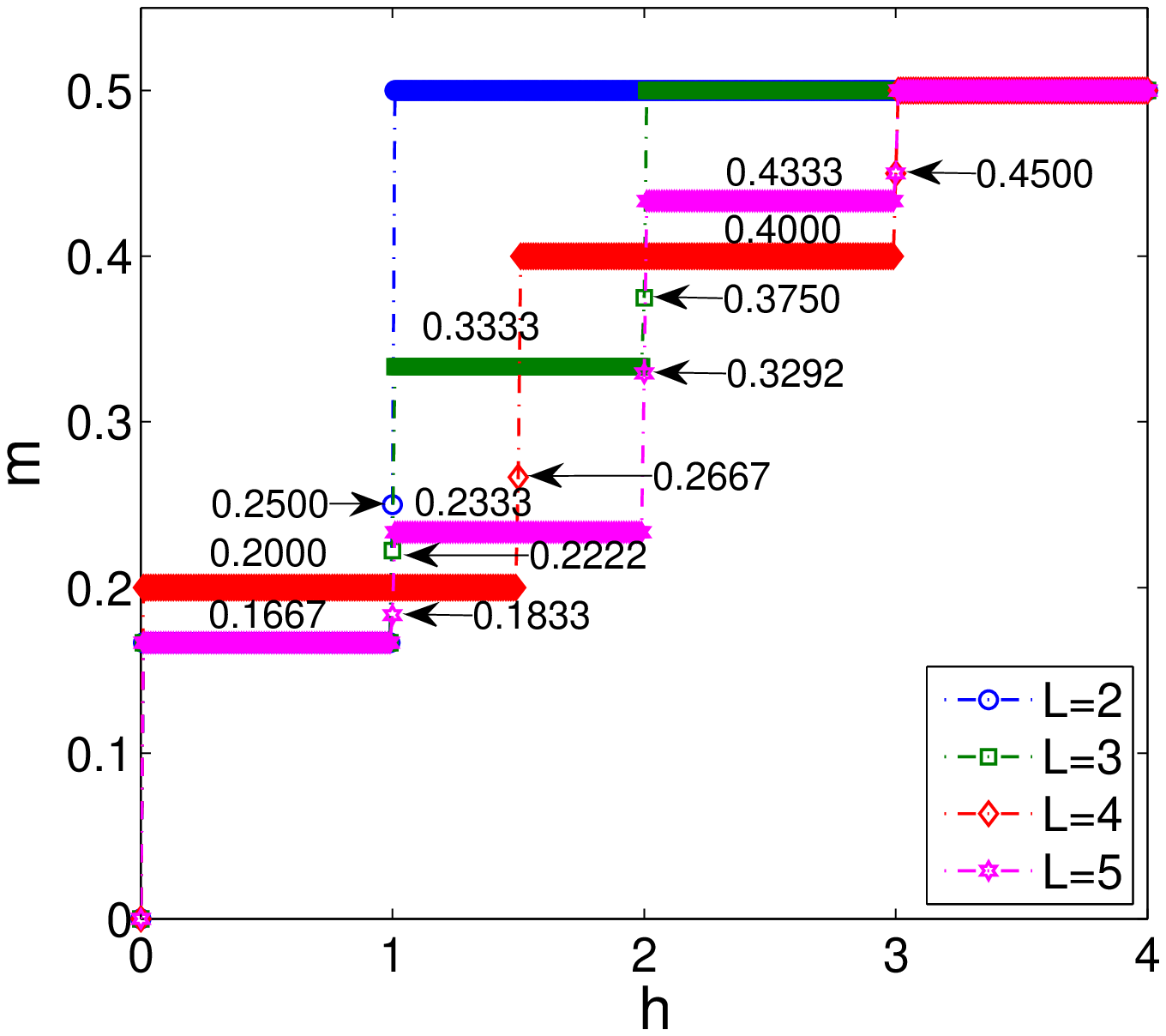}\label{fig:gs_m_tria0}}
\subfigure[T]{\includegraphics[scale=0.37,clip]{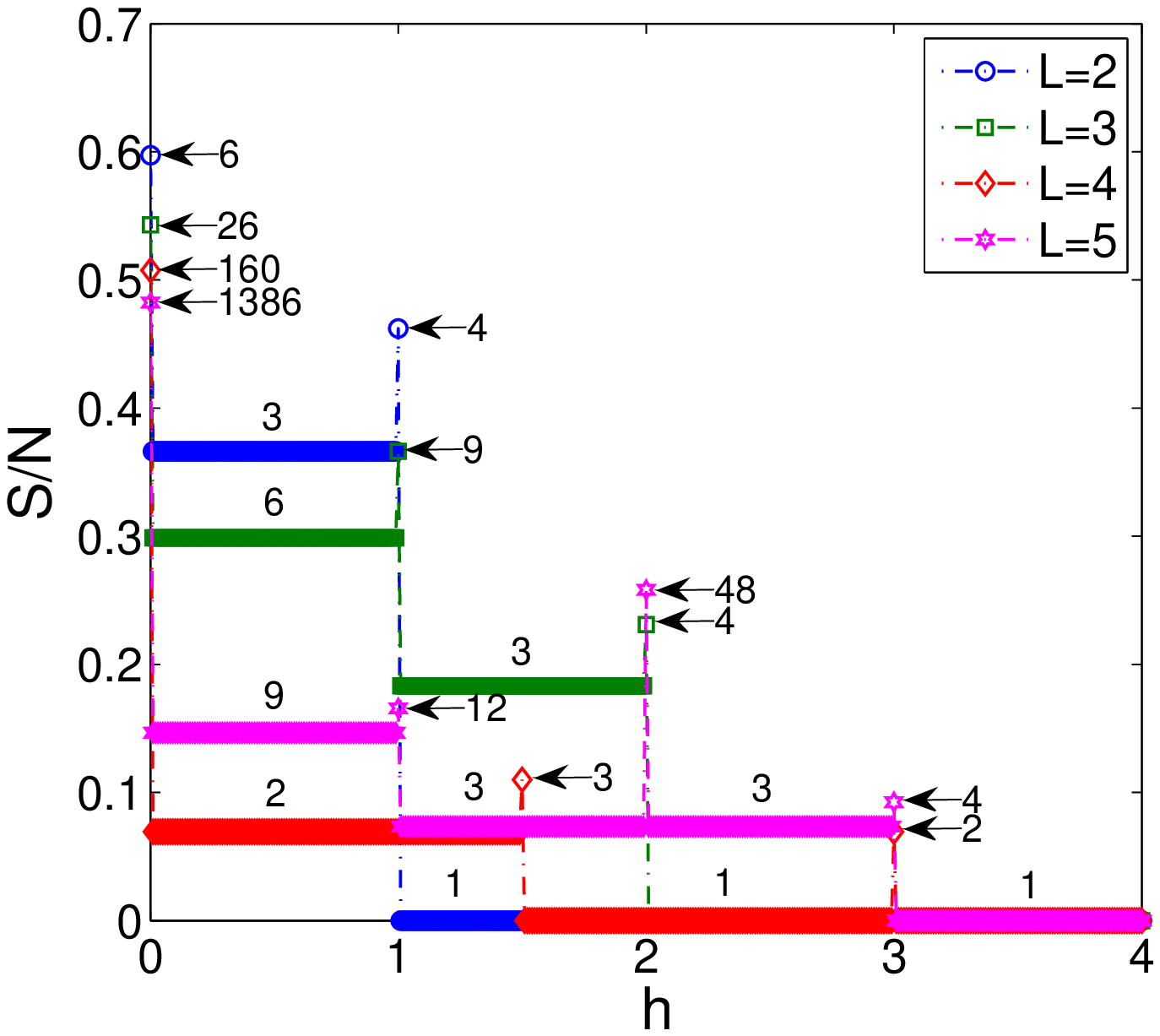}\label{fig:gs_S_tria0}}
\subfigure[T1]{\includegraphics[scale=0.37,clip]{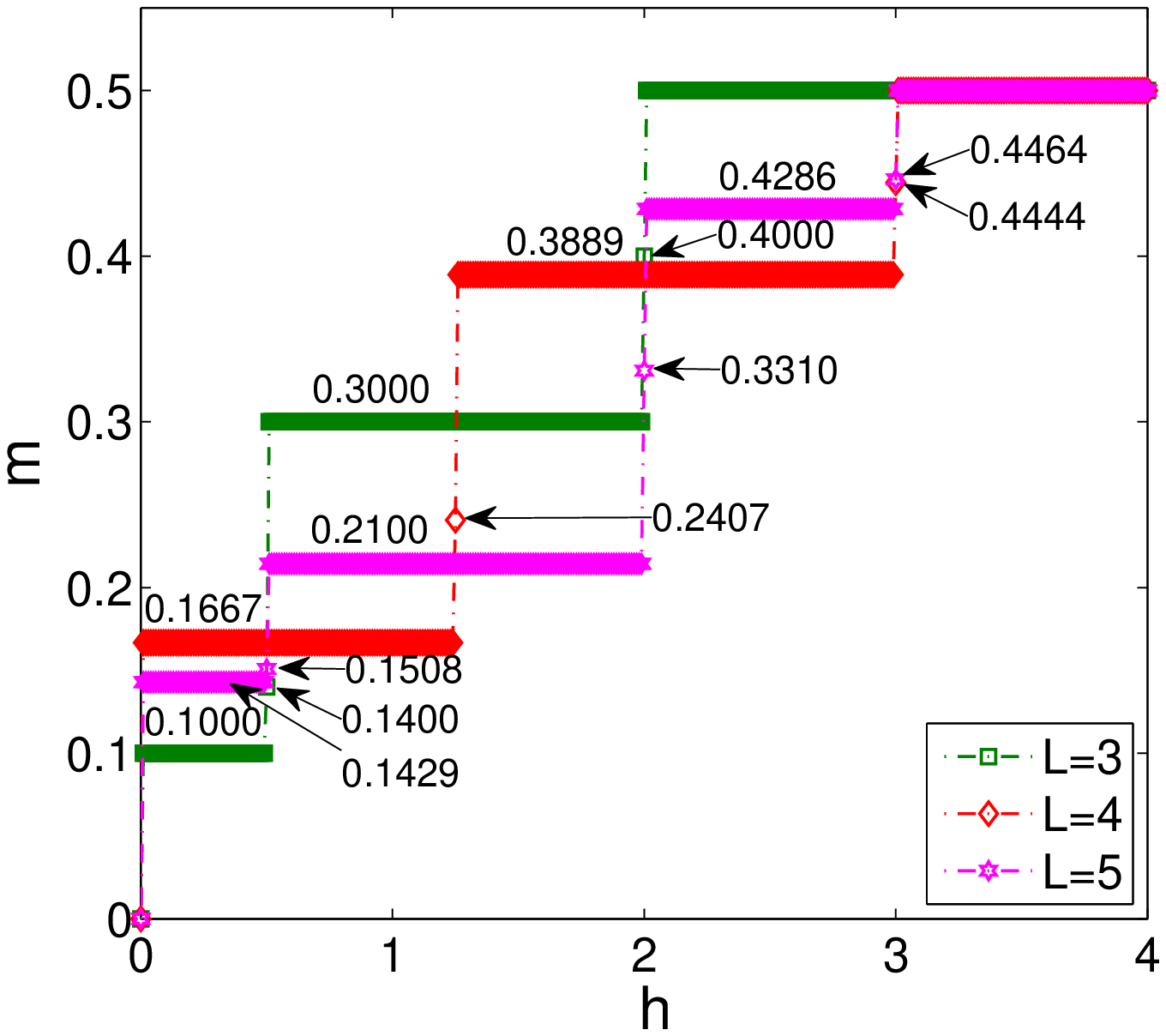}\label{fig:gs_m_tria1}}
\subfigure[T1]{\includegraphics[scale=0.37,clip]{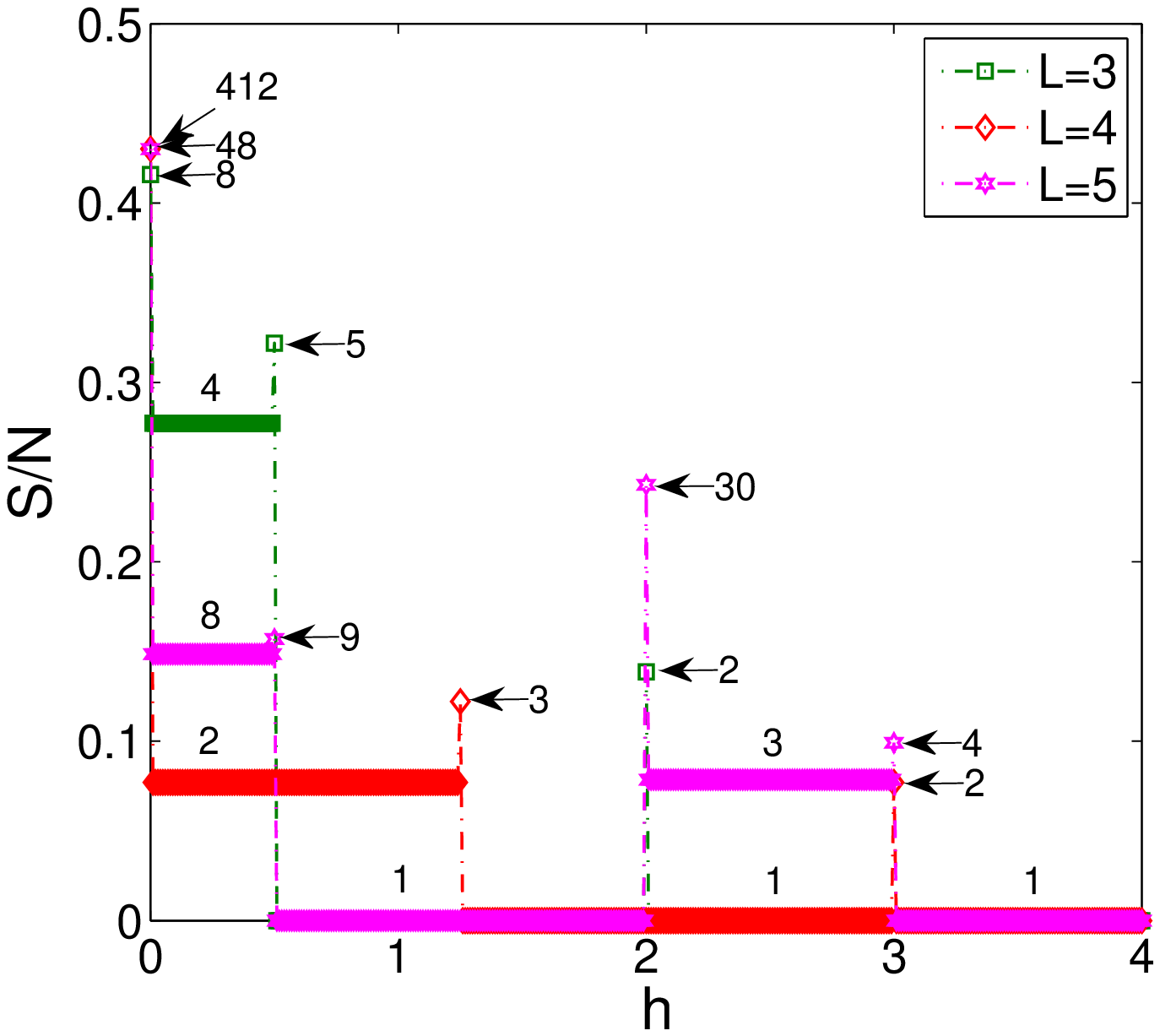}\label{fig:gs_S_tria01}}
\subfigure[T2]{\includegraphics[scale=0.37,clip]{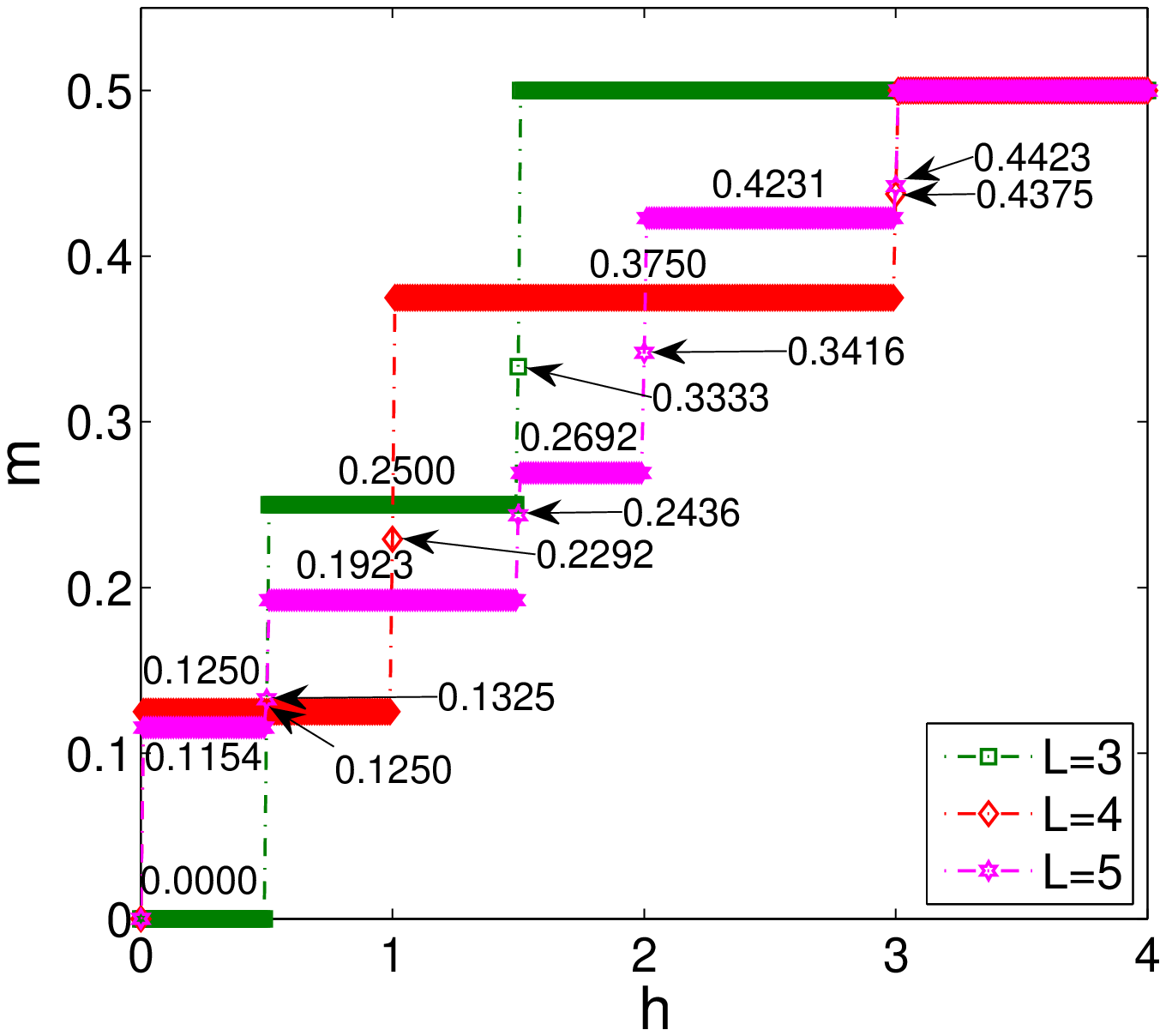}\label{fig:gs_m_tria2}}
\subfigure[T2]{\includegraphics[scale=0.37,clip]{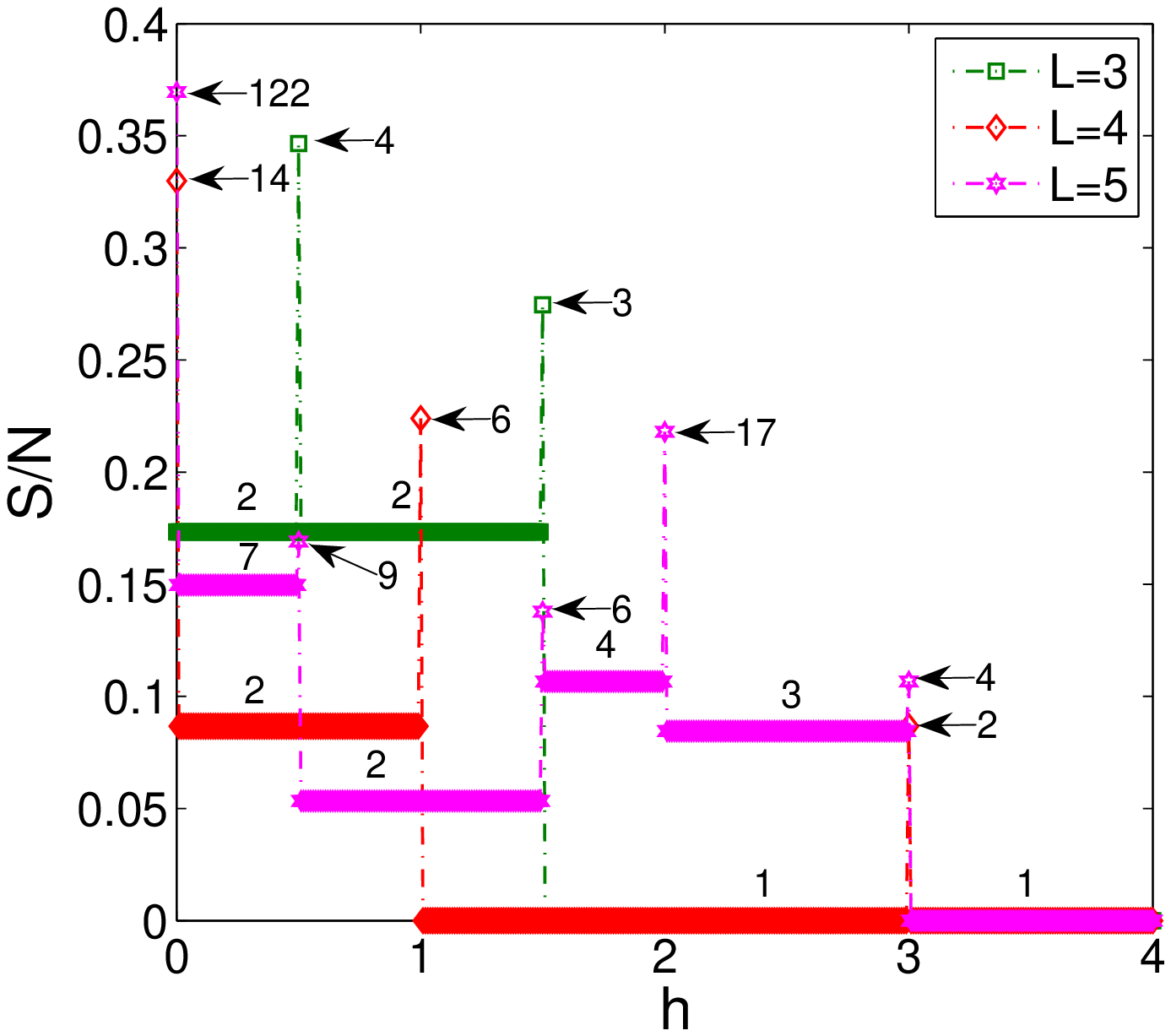}\label{fig:gs_S_tria2}}
\subfigure[T3]{\includegraphics[scale=0.37,clip]{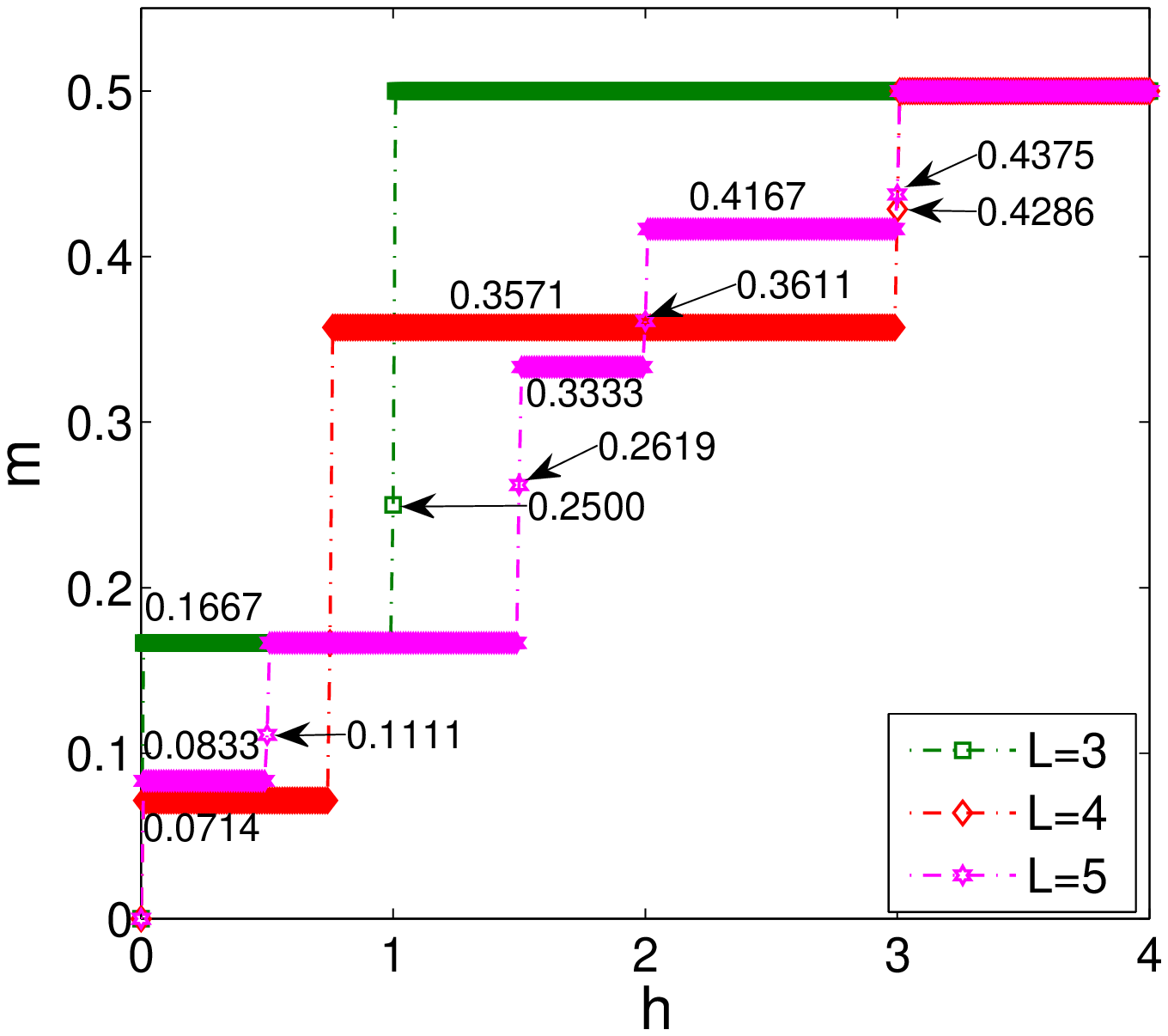}\label{fig:gs_m_tria3}}
\subfigure[T3]{\includegraphics[scale=0.37,clip]{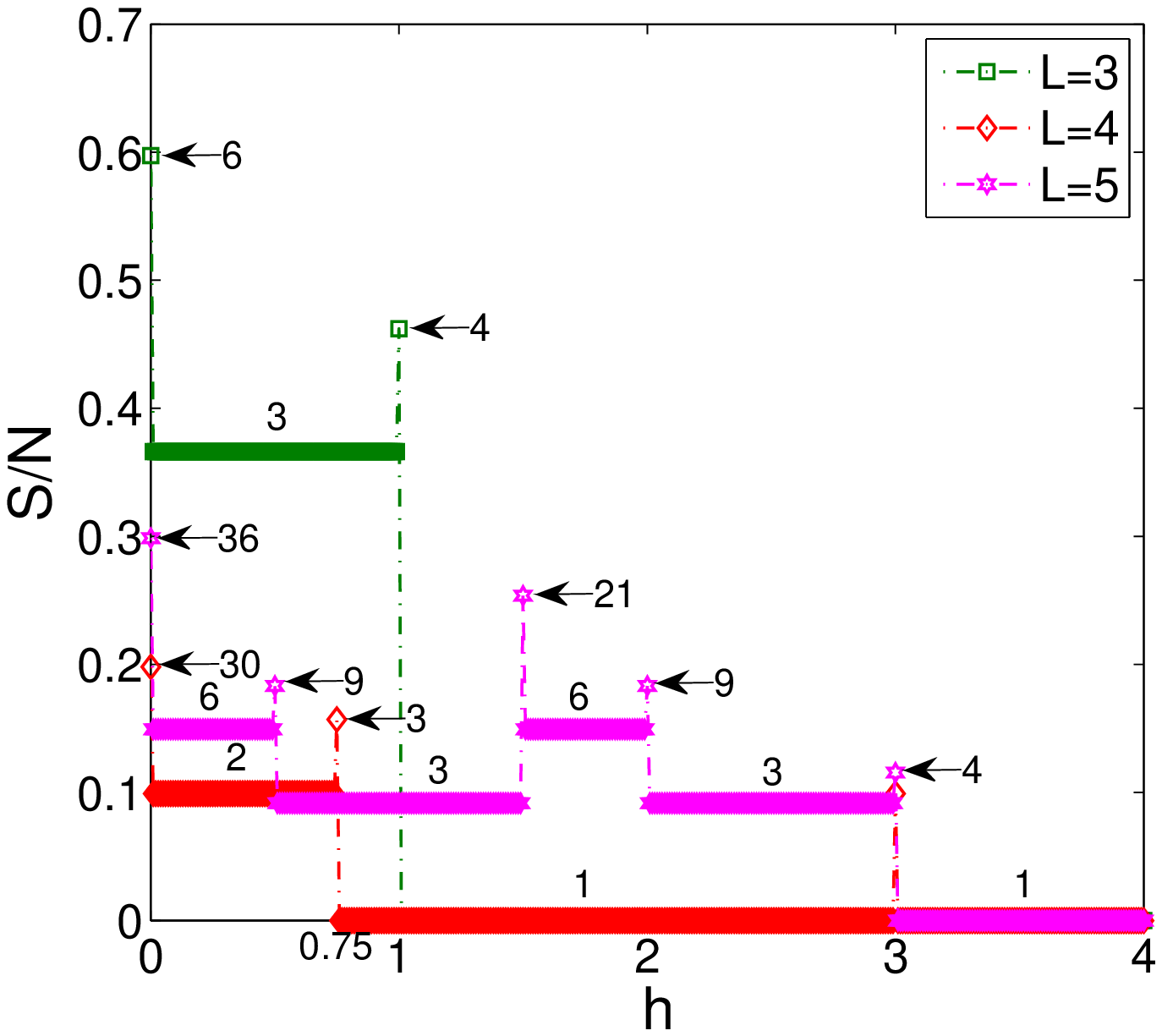}\label{fig:gs_S_tria3}}
\caption{Magnetization (left column) and entropy density (right column) for the triangular T-T3 domains with different sizes $L$.}\label{fig:gs_tria}
\end{figure}

\begin{figure}[]
\centering
\subfigure[R]{\includegraphics[scale=0.37,clip]{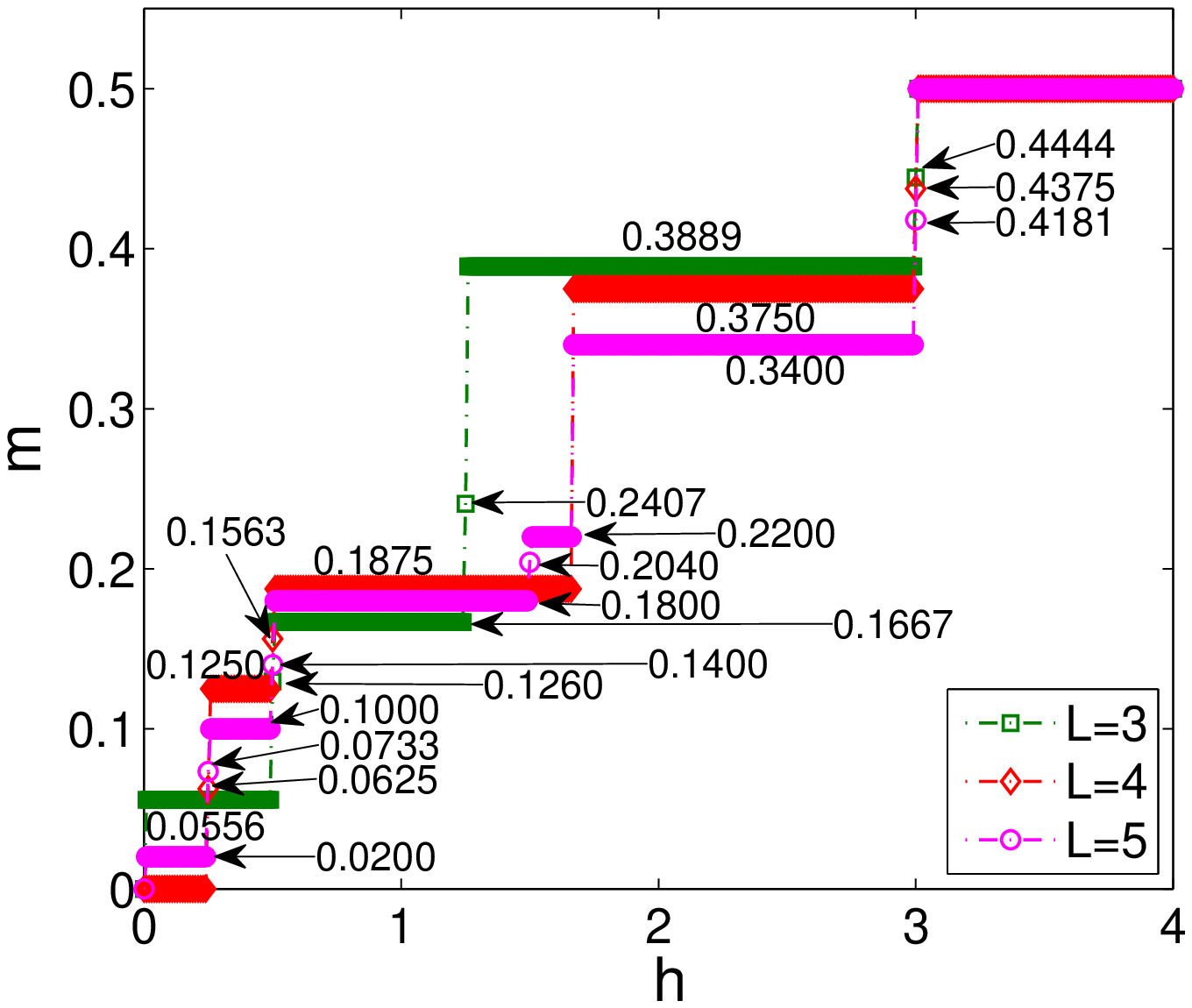}\label{fig:gs_m_rhomb}}
\subfigure[R]{\includegraphics[scale=0.37,clip]{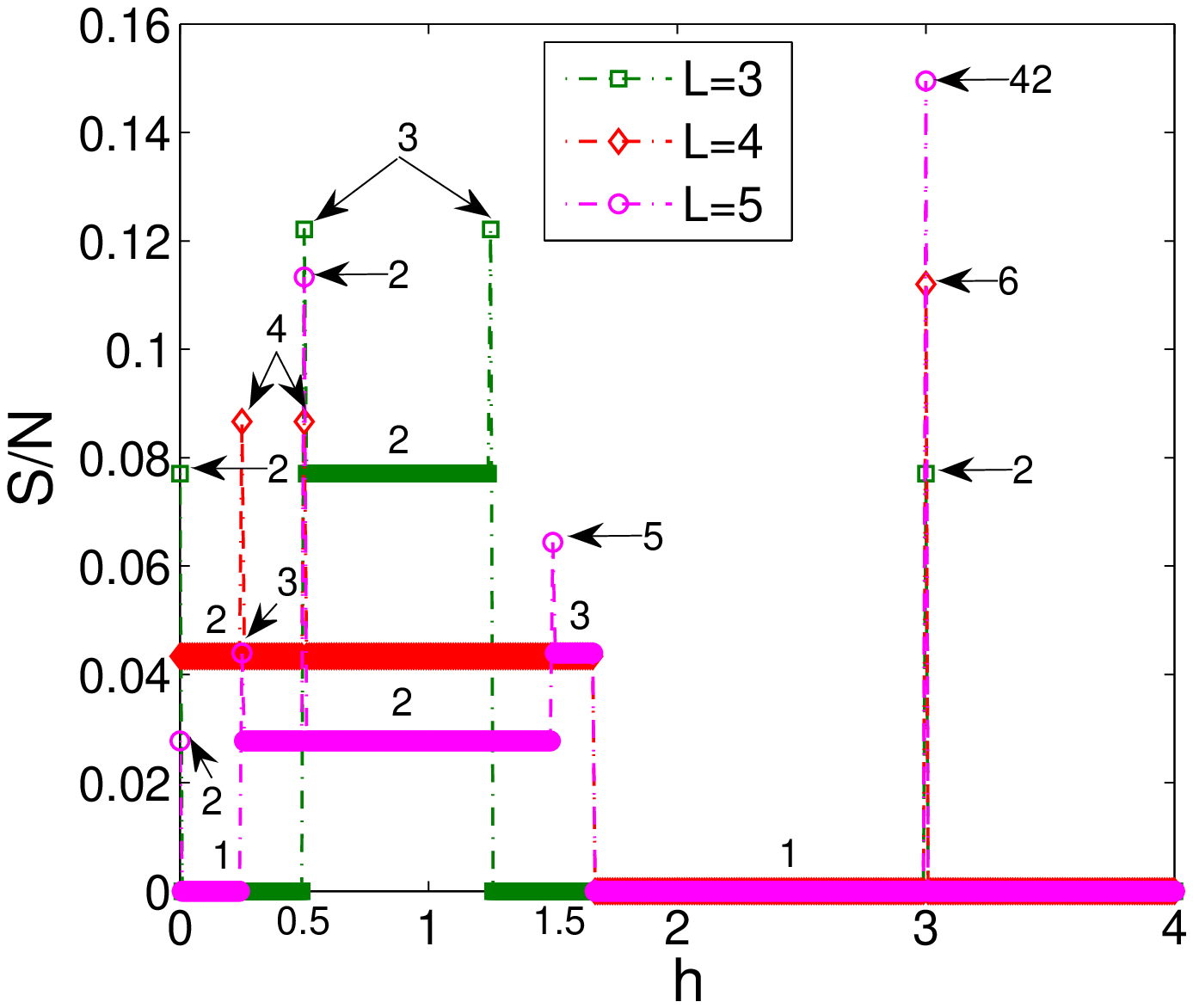}}
\subfigure[R1]{\includegraphics[scale=0.37,clip]{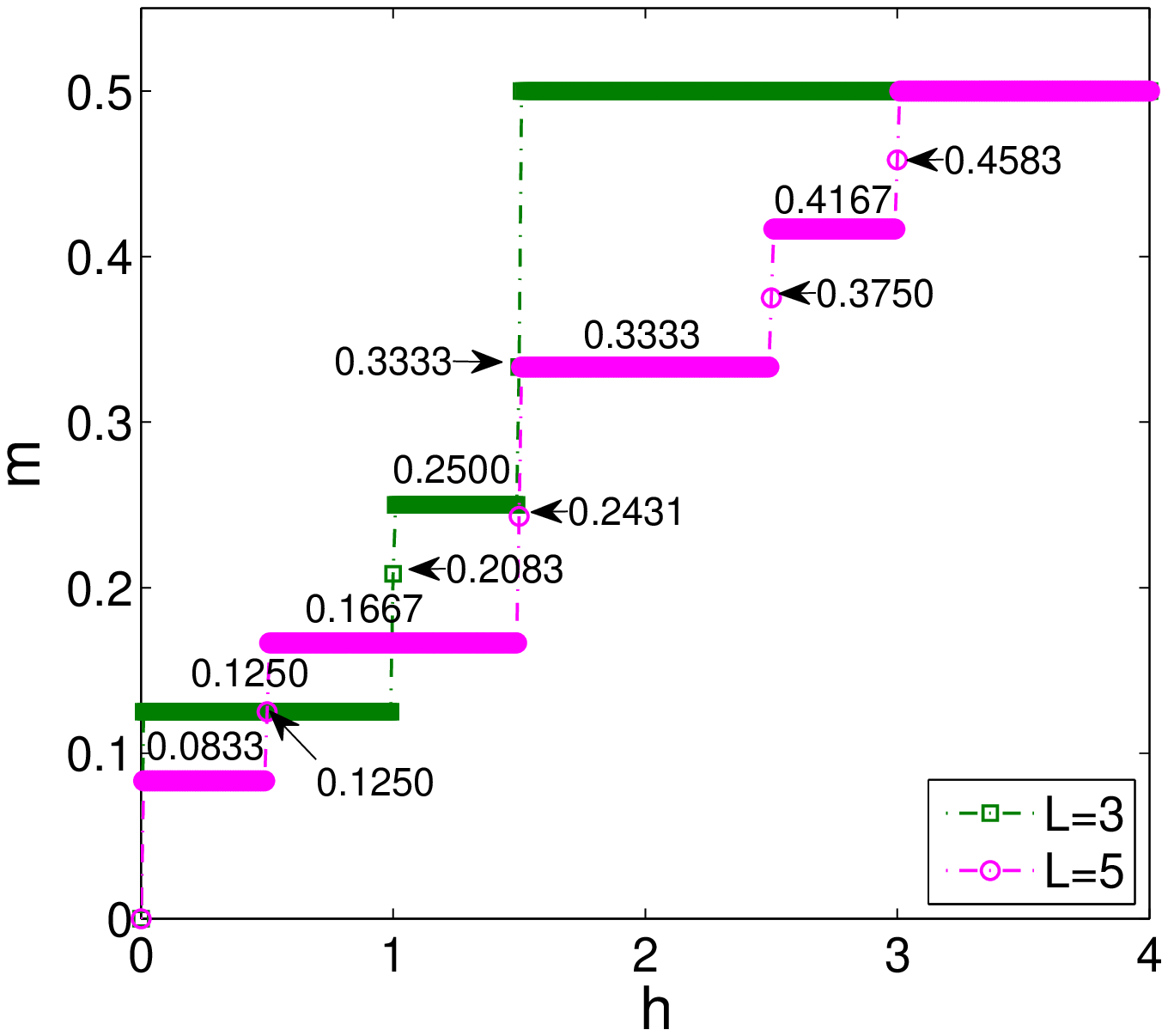}}
\subfigure[R1]{\includegraphics[scale=0.37,clip]{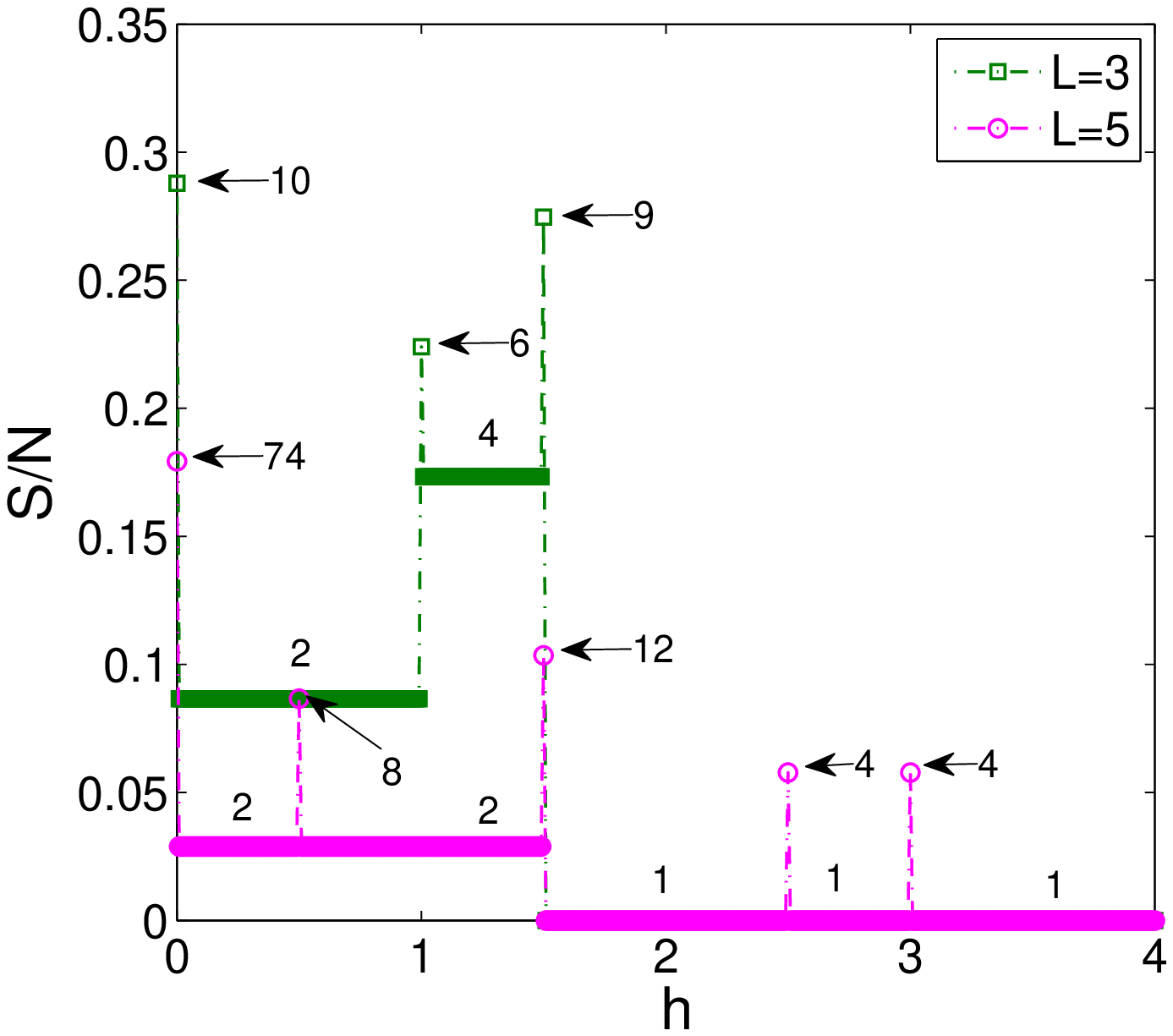}}
\subfigure[H]{\includegraphics[scale=0.37,clip]{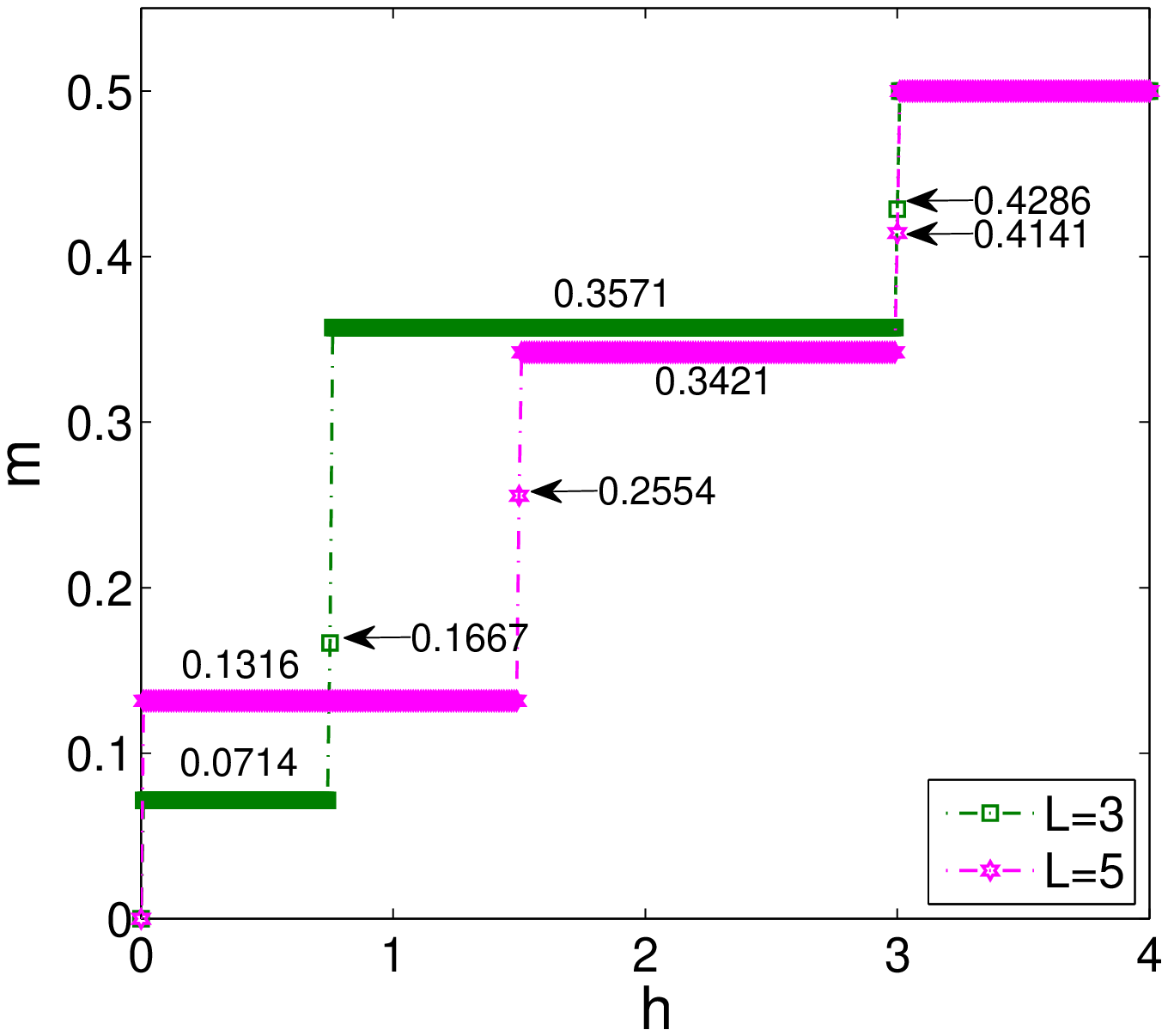}}
\subfigure[H]{\includegraphics[scale=0.37,clip]{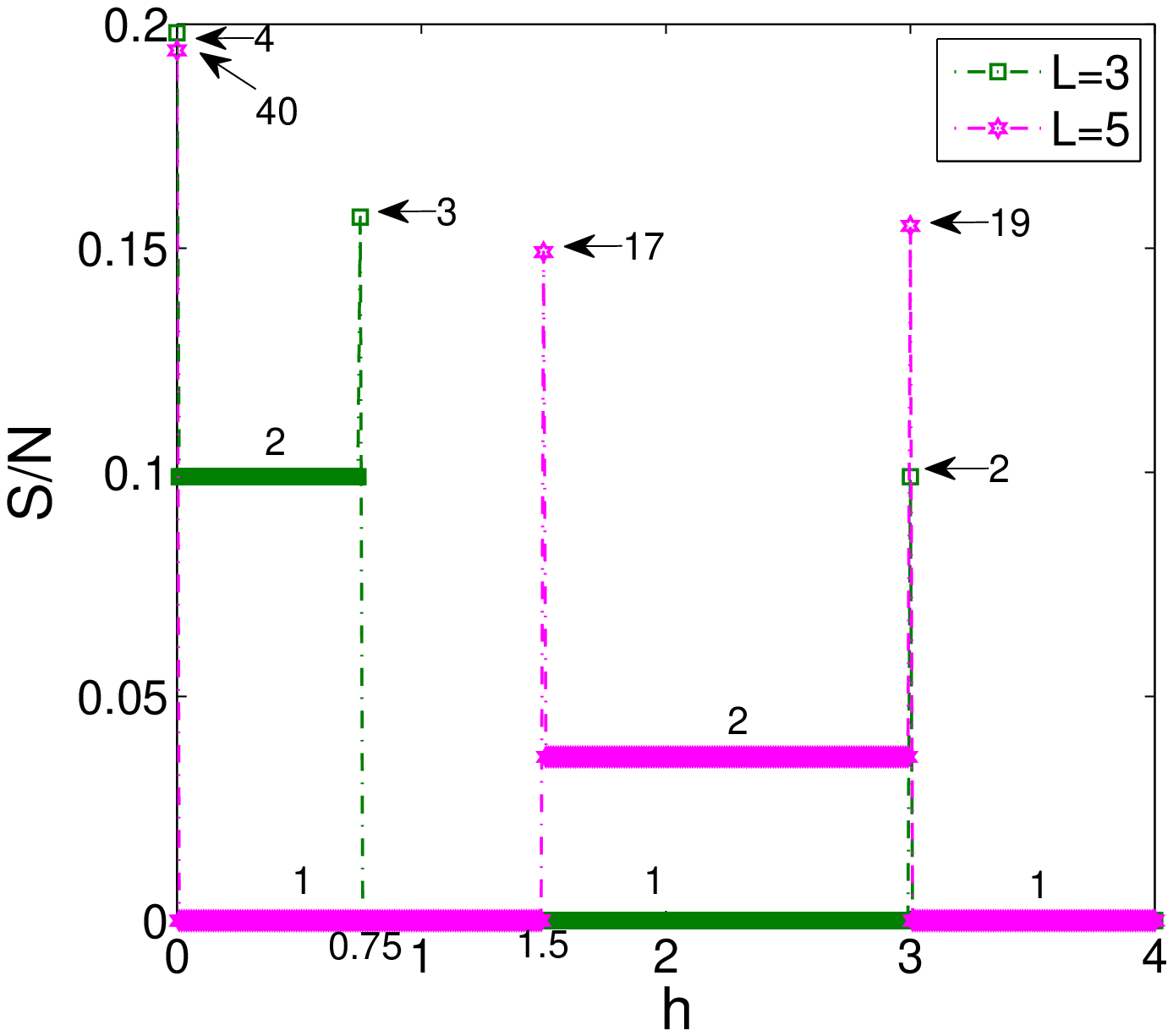}}
\subfigure[H1]{\includegraphics[scale=0.37,clip]{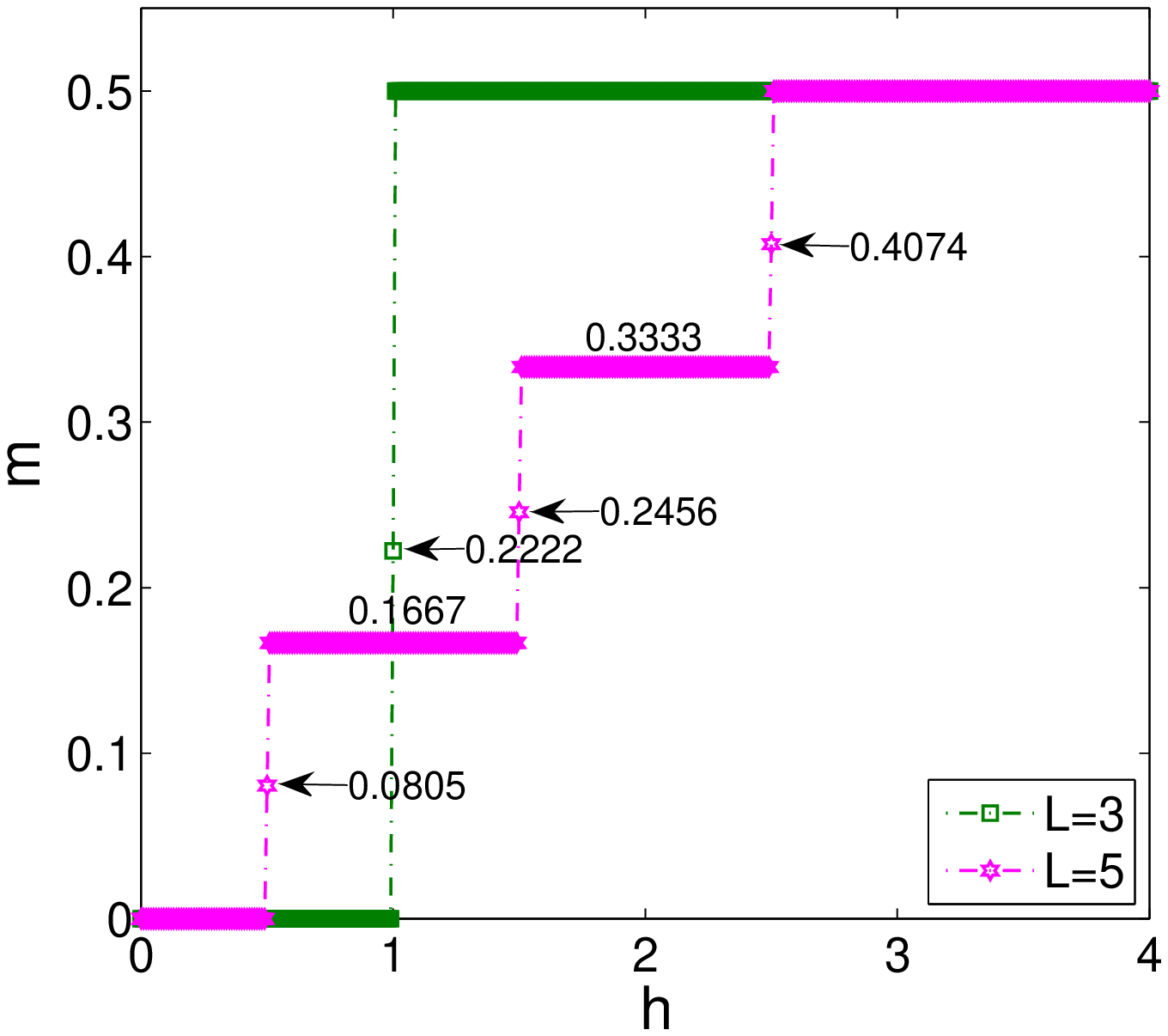}\label{fig:gs_m_hexa1}}
\subfigure[H1]{\includegraphics[scale=0.37,clip]{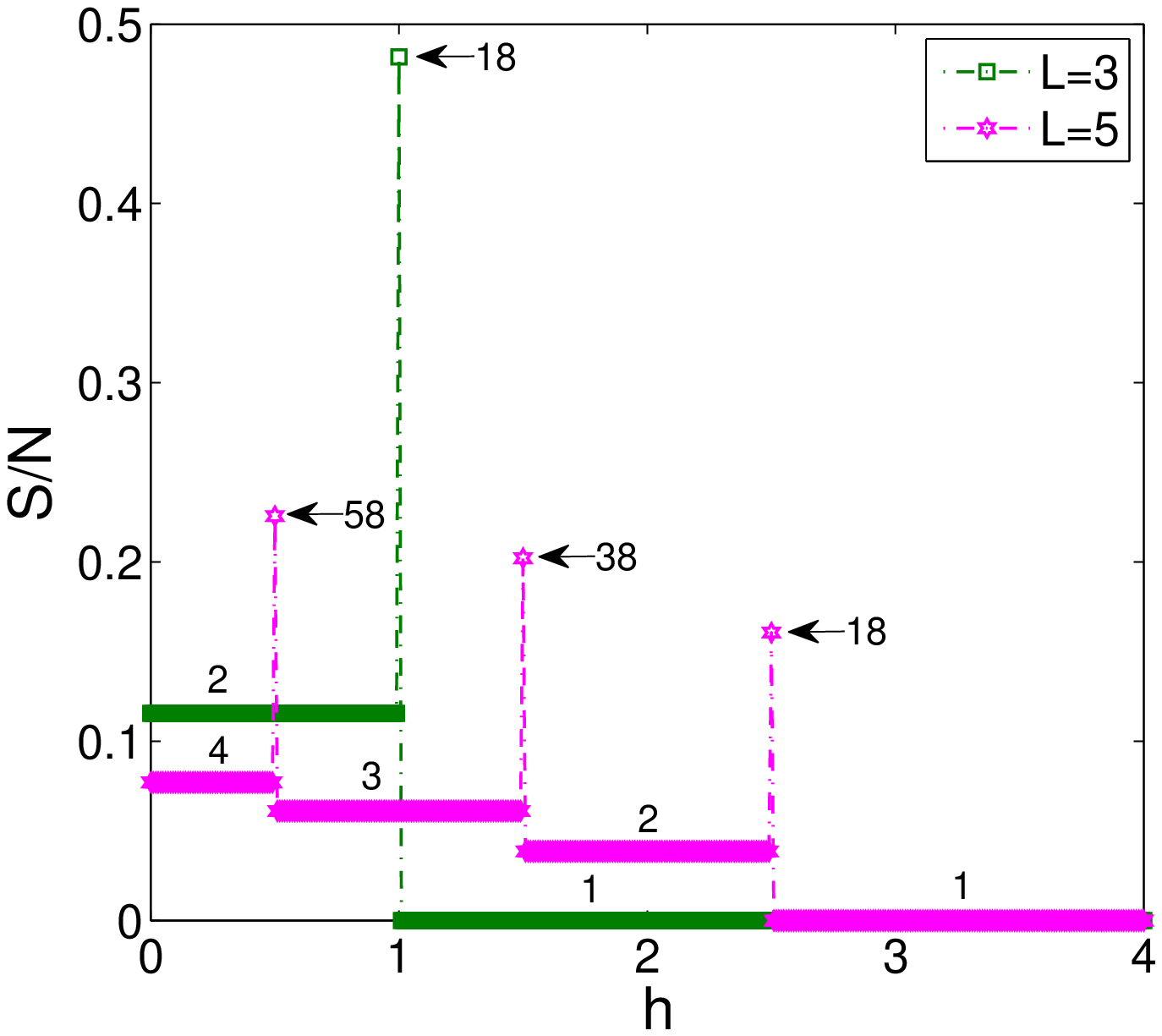}}
\caption{Magnetization (left column) and entropy density (right column) for the rhombic R, R1 and hexagonal H, H1 domains with different sizes $L$.}\label{fig:gs_rhomb_hexa}
\end{figure}

In zero field, the ground-state configurations of finite clusters, their energies and degeneracies have been shown to depend on the domain shapes and sizes~\cite{milla1,milla2,milla3}. An applied magnetic field is expected to gradually align the spins with the field direction and remove the respective degeneracies. In the infinite lattice this happens in two steps. Upon application of a small field intensity spins on two sublattices align with the field, resulting in a 1/3 magnetization plateau, and the large GS degeneracy, corresponding to the residual entropy density $S_{0}/N = 0.32306$~\cite{wann,wann2}, is reduced to the three-fold generate ferrimagnetic state~\cite{metc1,schick,netz}. The ferrimagnetic state persists up to the saturation field $h_s$ at which all spins get fully aligned with the field direction and the degeneracy is completely lifted. However, right at the saturation field the spin reversal is accompanied with an enormous entropy increase of $S_{s}/N = 0.3332427$~\cite{milo,varm}.\\
\begin{figure}[t]
\centering
\includegraphics[scale=0.5,clip]{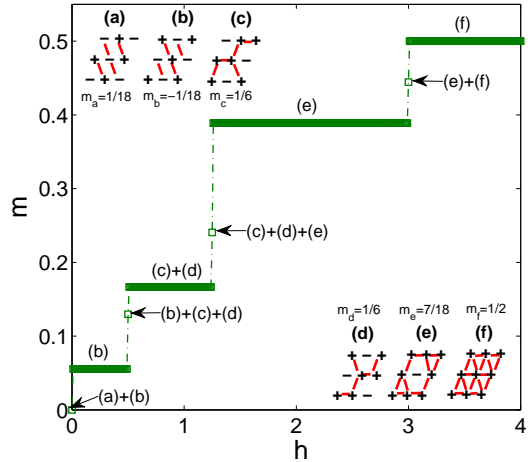}
\caption{Magnetization processes for the rhombic domain with $L=3$. The involved states (a)-(f) are shown in the inset with the red lines representing unsatisfied bonds. The jumps occur at the field values $h_1=0$, $h_2=0.5$, $h_3=1.25$ and $h_4 \equiv h_s=3$.}\label{fig:gs_rhombL3}
\end{figure}
\hspace*{5mm} In finite clusters, magnetization and entropy processes in a field can be expected to consist of multiple steps. Depending on the domain shape and size, one can anticipate a variety of states with non-trivial magnetizations and degeneracies. The magnetization process occurring in the rhombic R cluster with the side length $L=3$ is demonstrated in Fig.~\ref{fig:gs_rhombL3}. The involved states (a)-(f) are shown in the inset with the red color marking unsatisfied bonds. The system evolves from a two-fold degenerate phase at zero field (states (a) and (b)) through a sequence of several phases up to the saturated phase (f). The magnetization curve shows four jumps at the critical field values $h_1=0$, $h_2=0.5$, $h_3=1.25$ and $h_4 \equiv h_s=3$, which separate four magnetization plateaux. The jumps occur when the Zeeman contribution in the Hamiltonian~(\ref{Hamiltonian}), corresponding to the portion of the spins with the weakest coupling (typically, but not necessarily, at the domain boundary), overcomes their exchange energy. At the critical fields $h_i$, the system's degeneracy is enhanced owing to the contributions from the neighboring phases. The plateaux heights represent mean magnetization values obtained from the respective degenerate states. It is interesting to notice how the field lifts the degeneracy in this particular case. The states at $h_i$ put aside, the initial zero-field two-fold degeneracy is lifted by a small field and the state (b) is non-degenerate. However, further increase in the field intensity within $h_2<h<h_3$ counterintuitively {\it increases} the degeneracy and thus the entropy of the system. \\
\hspace*{5mm} Magnetization and entropic processes in the respective domains shown in Fig.~\ref{fig:clusters} for several sizes are presented in Figs.~\ref{fig:gs_tria} and \ref{fig:gs_rhomb_hexa}. At the entropy density curves also exact numbers of the degenerate states are provided. Depending on the domain shape and size we can observe a variety of states with different numbers of magnetization plateaux of different heights and widths, as well as degeneracies. Generally, the number of plateuaux increases with the system size, from two for a simple triangle, i.e. T with $L=2$ or a non-frustrated hexagon, i.e. H1 with $L=3$, up to six for the rhombic shape R with $L=5$. The simple triangle also shows the largest entropy density among the considered shapes and sizes. As already mentioned above, in some cases the entropy is not a monotonically decreasing function of the applied field, even if we do not consider the enhanced entropy points at the critical fields $h_i$. Besides the presented case of the rhombic R cluster for $L=3$, similar phenomenon was observed in R1 for $L=3$ and in T1, T2, T3, H and R for $L=5$. For R with $L=5$, with the increasing field the entropy was even repeatedly increased across two consecutive phases.\\
\hspace*{5mm} In the infinite lattice, the saturation-field entropy $S_s/N=0.3332427$~\cite{milo,varm} is somewhat larger than the zero-field entropy $S_0/N=0.32306$~\cite{wann,wann2}. For most of the present finite clusters it is vice versa, presumably due to the fact that the spin reversal does not occur instantly at the saturation field but it is split into more stages occurring at the critical fields $h_i$, $i=1,2,\ldots,s$. On the other hand, there are some cases for which $S_0/N<S_i/N, \forall i=1,2,\ldots,s$, because the zero-field system is either non- or trivially degenerate, (T2 for $L=3$, R for $L=3,4,5$ and H1 for $L=3$) or the degeneracy is very low (H1 for $L=5$).

\subsection{Finite temperatures}
\subsubsection{Zero-field behavior}
\begin{figure}[t]
\centering
    \subfigure[]{\includegraphics[scale=0.5,clip]{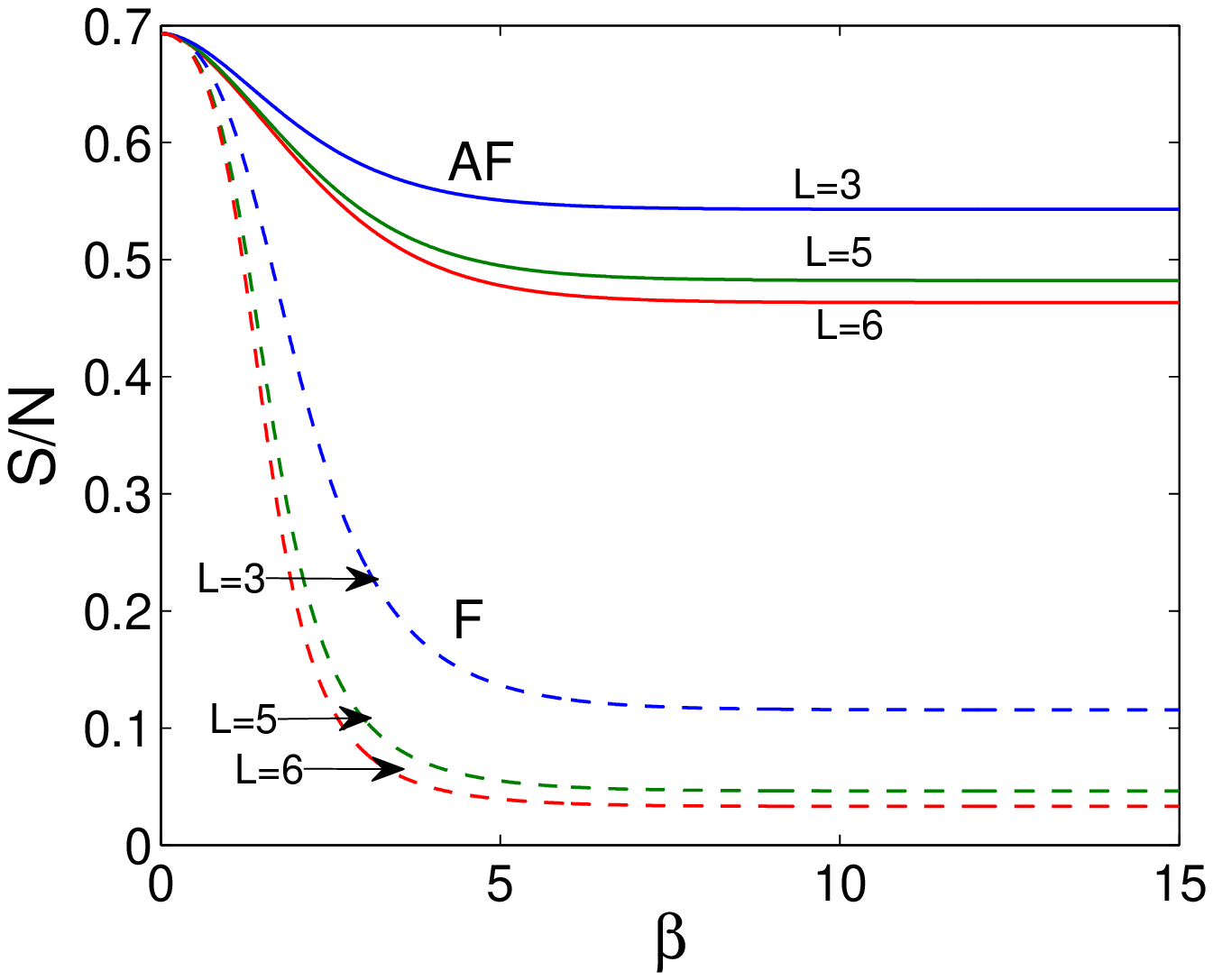}\label{fig:ent-fin_T_T0}}
    \subfigure[]{\includegraphics[scale=0.5,clip]{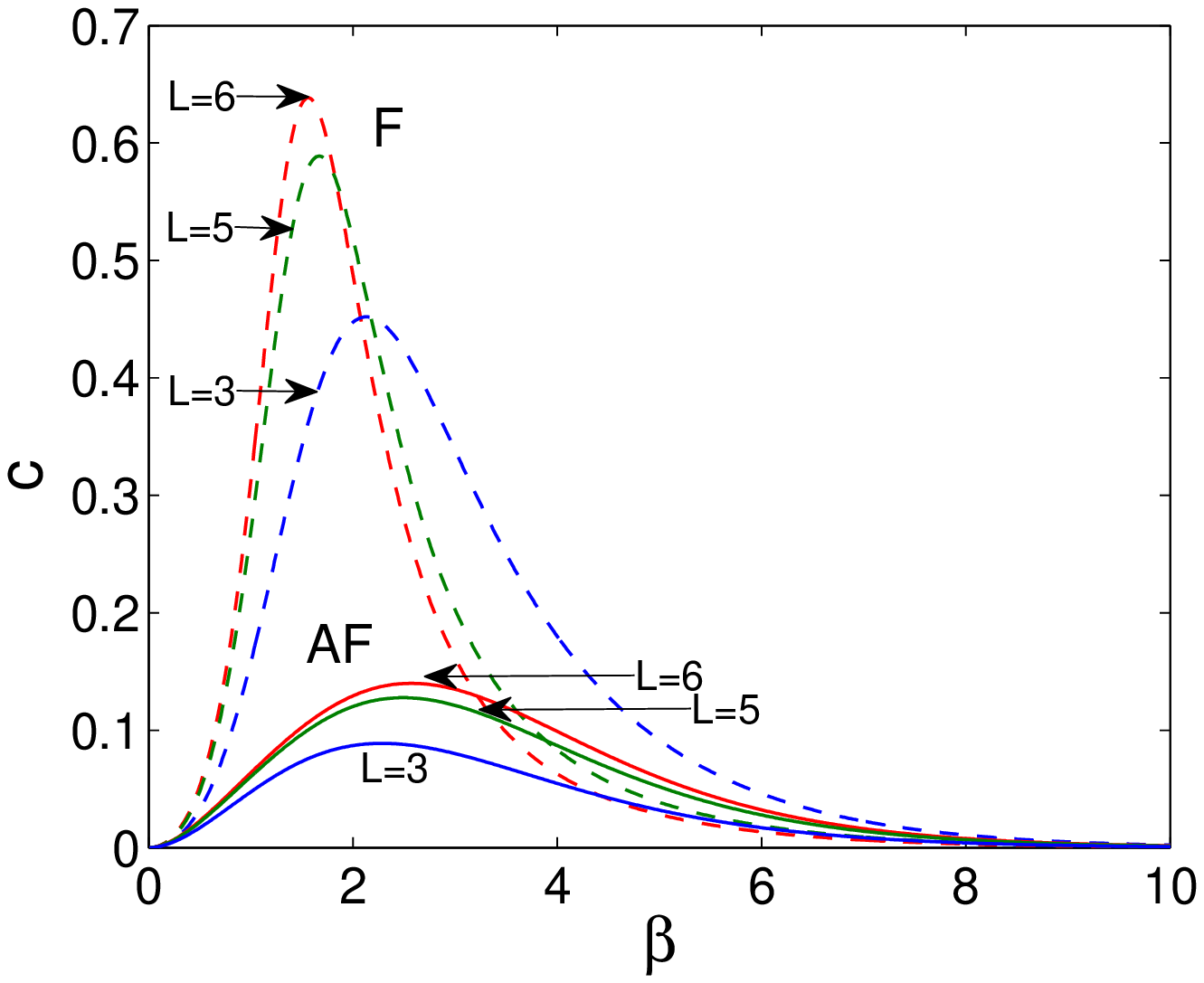}\label{fig:c-fin_T_T0}}
\caption{(a) Entropy density and (b) specific heat per spin as functions of the inverse temperature $\beta$ for T ferromagnetic (F) and antiferromagnetic (AF) clusters with different sizes $L$.}\label{fig:ent-c_fin_T_T0}
\end{figure}

Exhaustive scanning of the entire state space allows an exact calculation of different thermodynamic functions also at finite temperatures. In Fig.~\ref{fig:ent-fin_T_T0} we show how the entropies of the T clusters approach their residual values for various cluster sizes. To focus on the behavior in the most interesting low-temperature region we plot the values as functions of the inverse temperature $\beta=1/T$ (solid curves) and to see the frustration effect we also include the results for the corresponding non-frustrated ferromagnetic (F) clusters (broken curves). The decay for the antiferromagnetic (AF) clusters is slower than for the ferromagnetic ones, due to larger residual values of the former, but otherwise they look similar. Namely, a sharp monotonic decrease occurs at moderate temperatures ($\beta \approx 2$) and the residual value is achieved already well above the ground state ($\beta \approx 10$). Nevertheless, some qualitative difference in the behavior of the anomalies related to the sharp entropy decrease between AF and F clusters can be seen looking at the corresponding specific heat curves, shown in Fig.~\ref{fig:c-fin_T_T0}. While with the increasing cluster size the peaks of the F systems move towards higher temperatures (the critical value of $\beta_c=\ln(3)/4$ in the thermodynamic limit~\cite{hout}), for the frustrated AF clusters the peak moves toward lower temperatures ($\beta = 3.2067$ with the height $C_{max}/N=0.2161$ in the thermodynamic limit~\cite{domb}).
\begin{figure}[t]
\centering
    \subfigure[]{\includegraphics[scale=0.5,clip]{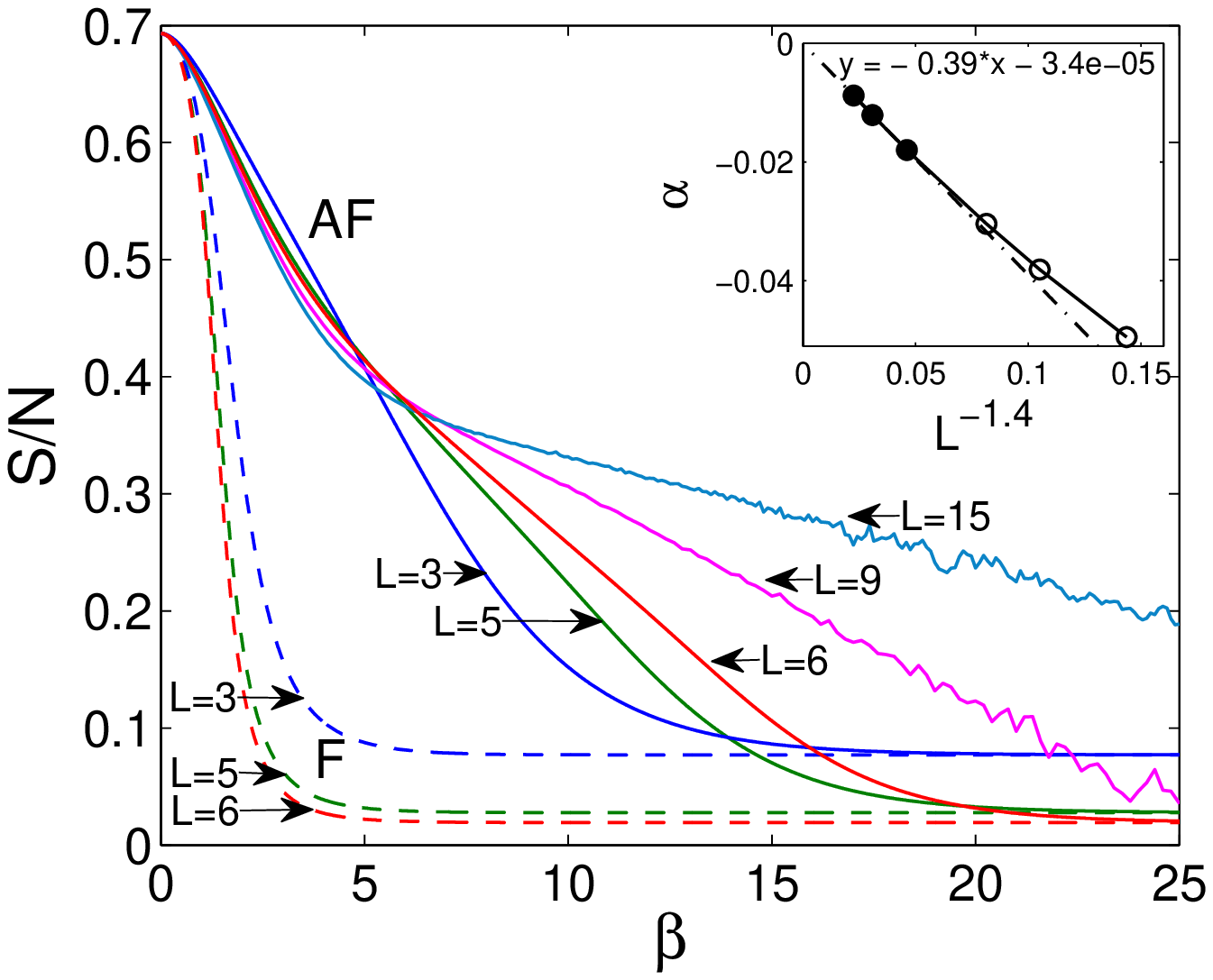}\label{fig:entr_rhomb}}
    \subfigure[]{\includegraphics[scale=0.5,clip]{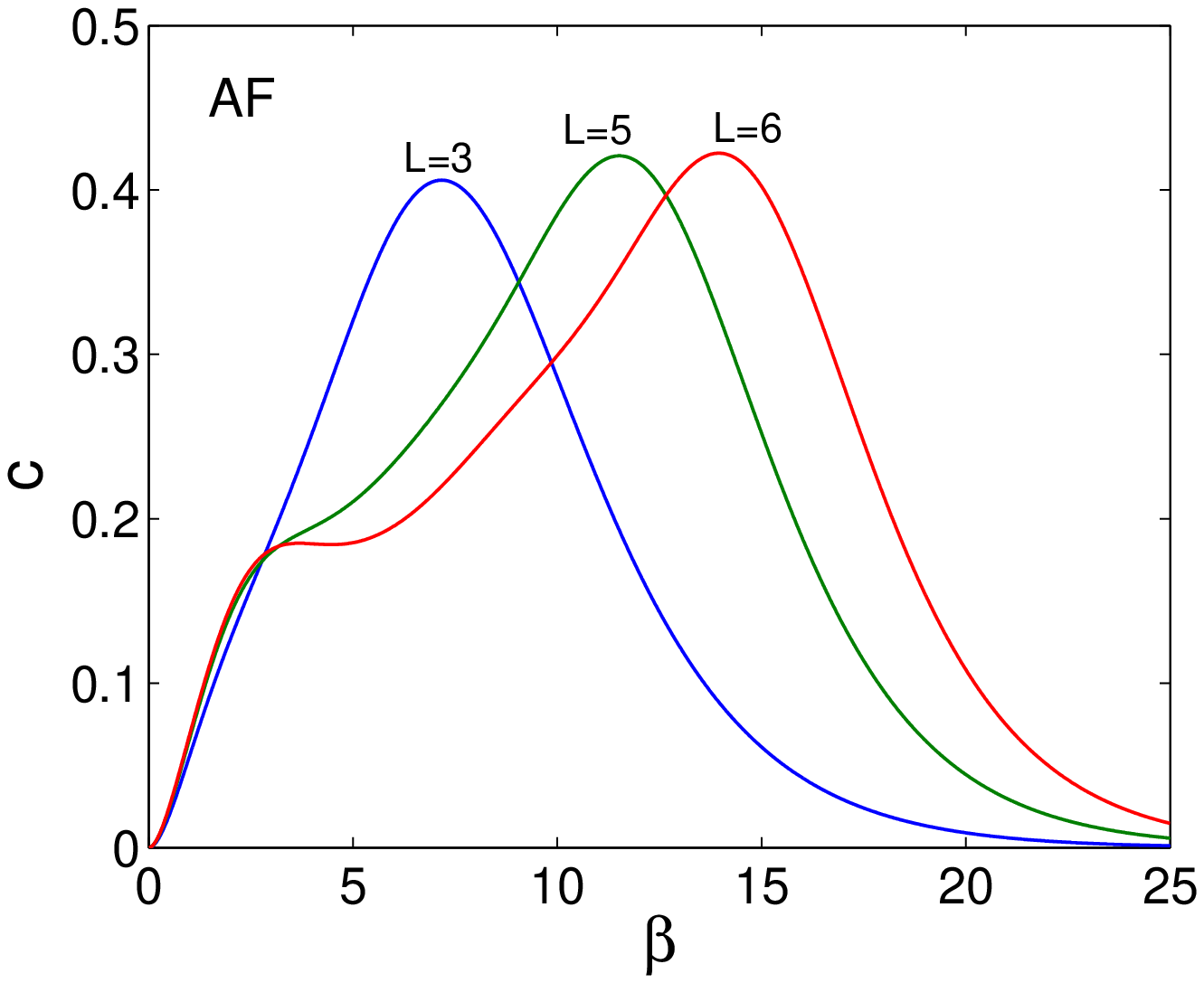}\label{fig:c_rhomb}}
\caption{(a) Entropy density as a function of the inverse temperature $\beta$ for R ferromagnetic and antiferromagnetic clusters with different sizes $L$. The cases of $L=9$ and $12$ were obtained from Monte Carlo simulation. The inset shows the slopes of the linear parts of the curves vs. $L^{-1.4}$ and the dashed line represents the best linear fit for $L \geq 9$. (b) Specific heat per spin corresponding to the antiferromagnetic clusters with $L=3,5$ and 6.}\label{fig:ent-c_rhomb}
\end{figure}\\
\hspace*{5mm} Similar behavior can also be observed in the remaining cluster shapes, except for the rhombus, the entropies of which are presented in Fig.~\ref{fig:entr_rhomb}. As expected from the ground-state discussion, at sufficiently low temperatures the curves of F and AF clusters of a given size merge and tend to zero as the cluster size increases. Nevertheless, the entropy values of AF R clusters do not approach the zero-temperature values in the same manner as those, for example, for AF T clusters. More specifically, after the initial sharp decrease the $\beta$-dependence changes to linear for a range of temperatures before the residual value is reached. The slopes of the linear parts of the curves become more gentle with the increasing cluster size and the dependence appears to be power law. In the inset of Fig.~\ref{fig:entr_rhomb} the slopes  $\alpha$ determined from the exact calculations for $L=3,5,6$ (open circles) and from MC simulations by applying the thermodynamic integration method~\cite{roma,zuko2} for $L=9,12,15$ (filled circles) are plotted against $L^{-1.4}$. Apparently the linear regime is reached only for sufficiently large cluster sizes. The dotted line represents the best linear fit based on the sizes $L \geq 9$. The thermodynamic limit extrapolation suggests that the zero entropy is reached only in the ground state, where AF and F curves merge. This would indicate much higher sensitivity of this non-degenerate ground state to thermal fluctuations, compared with both F and AF infinite-lattice systems for which the ground-state manifolds seem little affected for some rage of temperatures above zero~\cite{wann}, such as shown in Fig.~\ref{fig:ent-fin_T_T0}. As a result, the specific heat curves for R clusters (see Fig.~\ref{fig:c_rhomb}) show, besides the higher-temperature anomaly, another low-temperature peak which moves to $\beta \to \infty$ for $L \to \infty$.\\
\begin{figure}[t]
\centering
\subfigure[T]{\includegraphics[scale=0.48,clip]{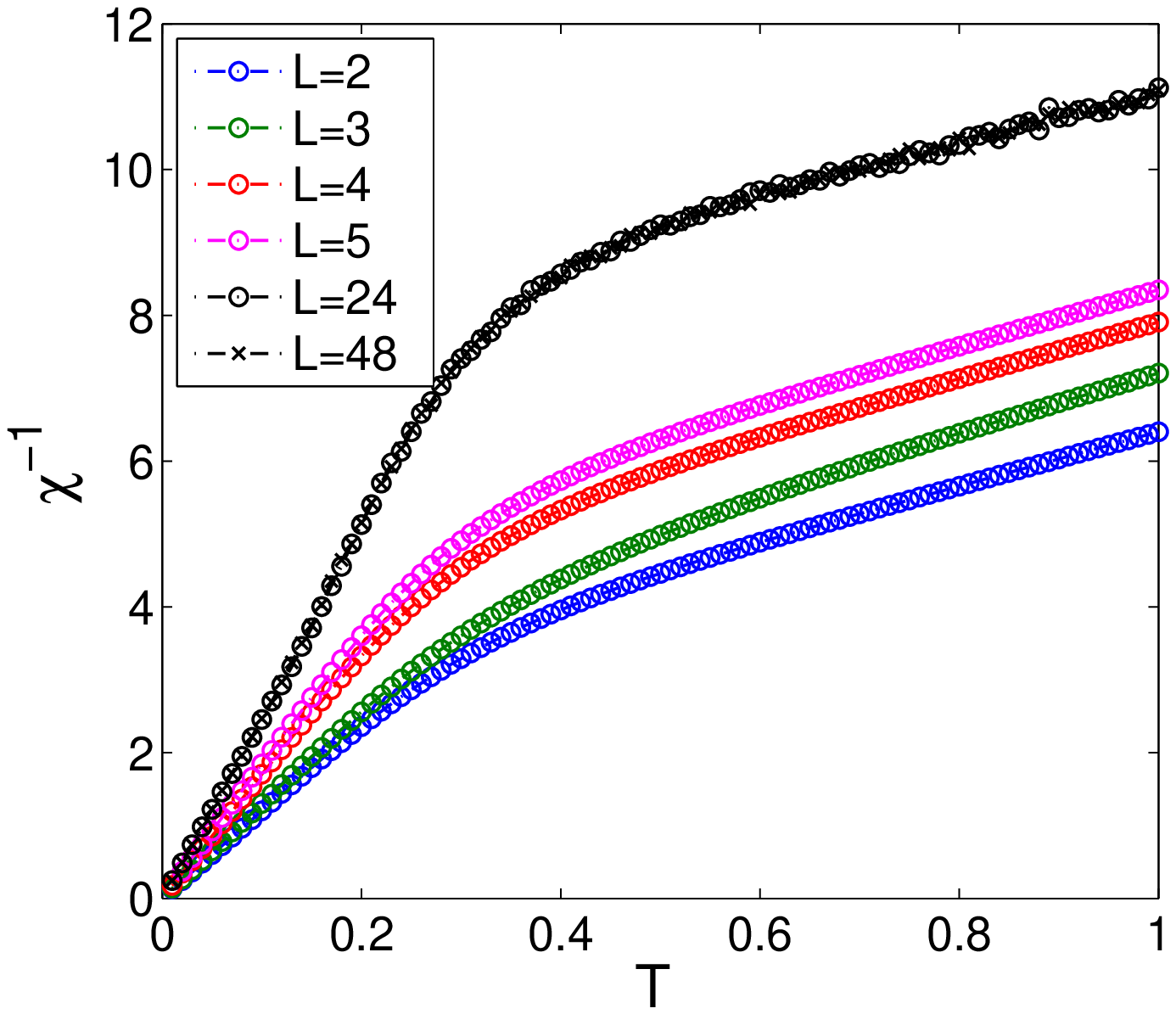}\label{fig:xi_tria0_s1_2}}
\subfigure[H]{\includegraphics[scale=0.48,clip]{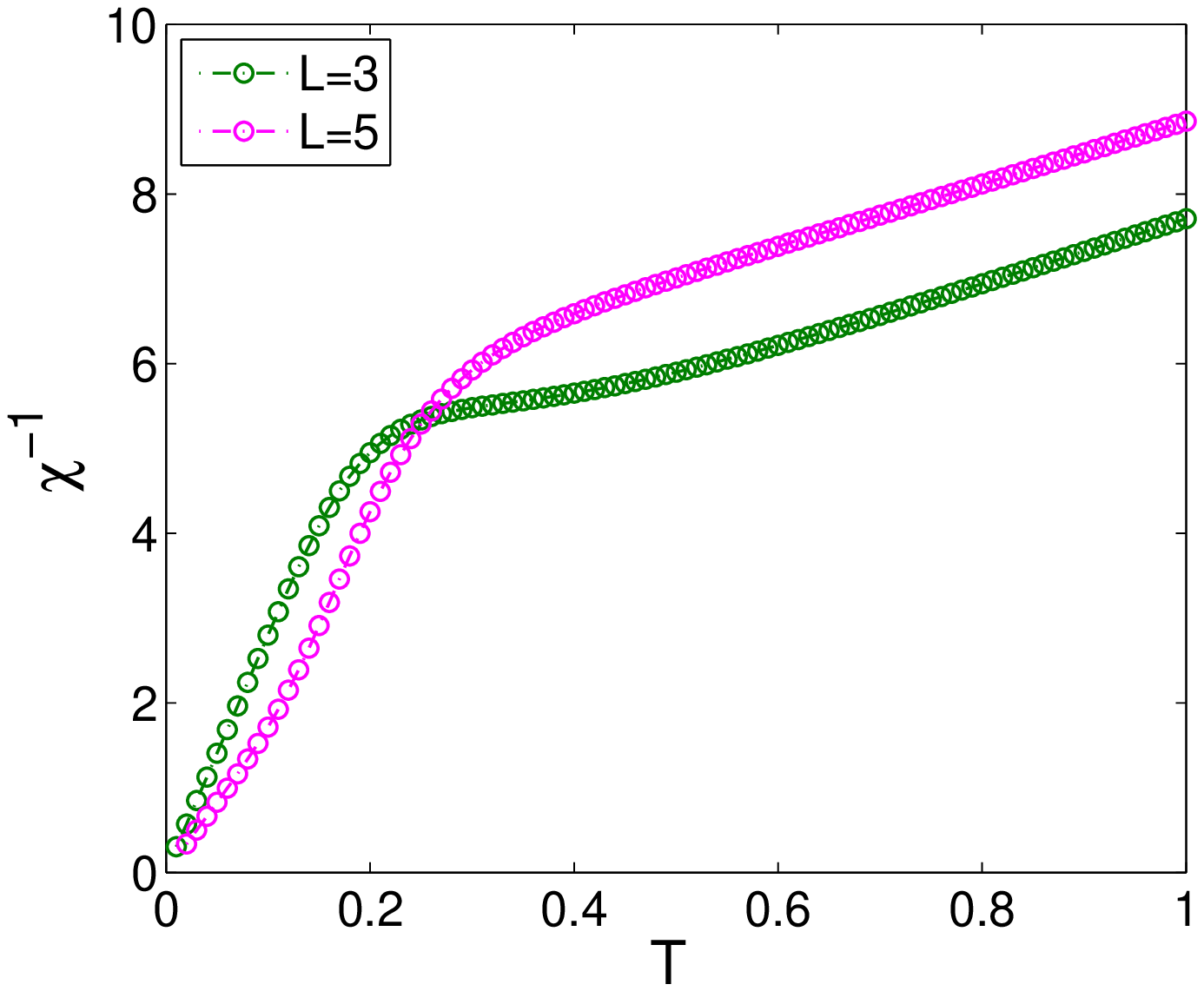}\label{fig:xi_hexa_s1_2}}
\subfigure[H1]{\includegraphics[scale=0.48,clip]{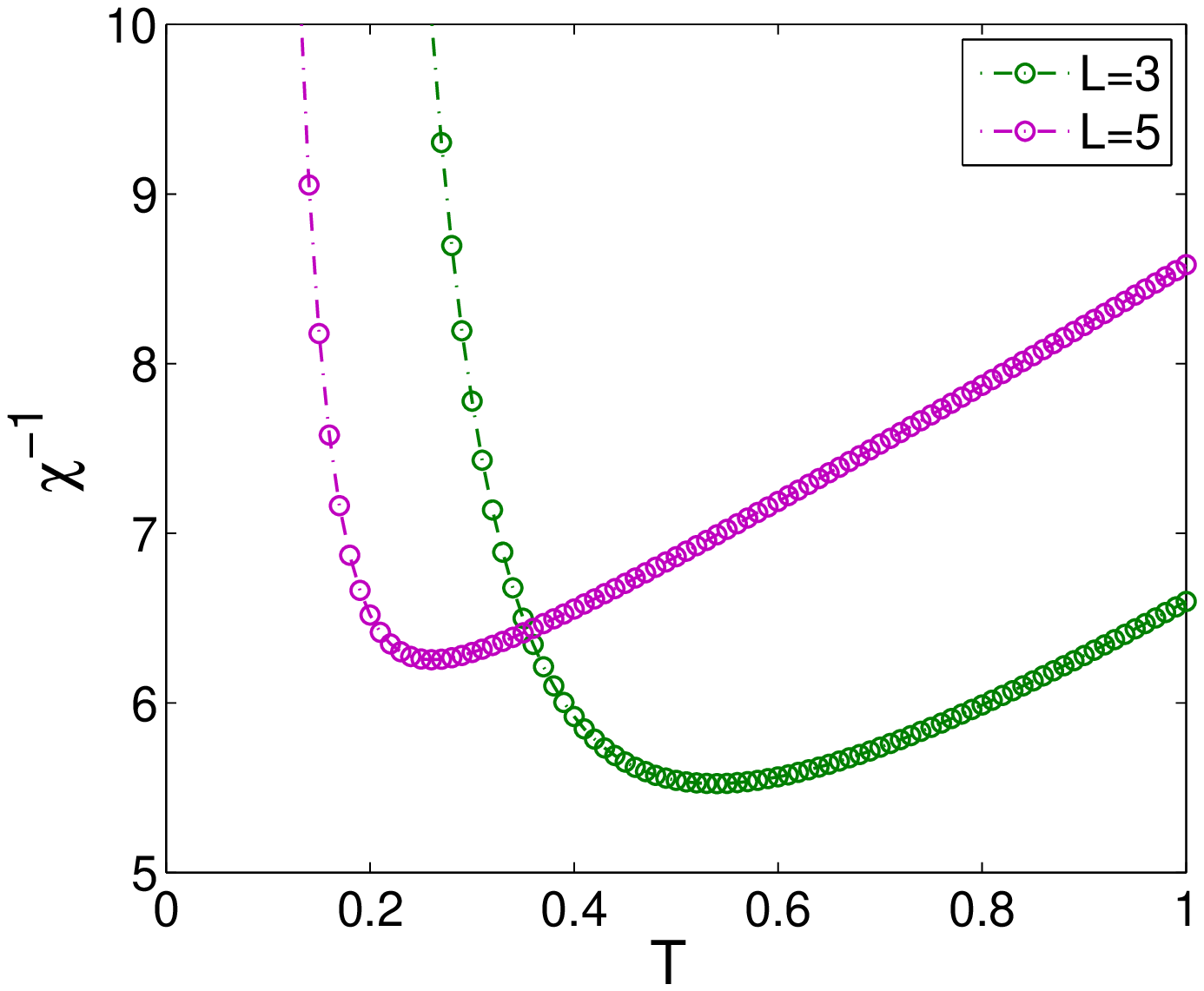}\label{fig:xi_hexa1_s1_2}}
\subfigure[R]{\includegraphics[scale=0.48,clip]{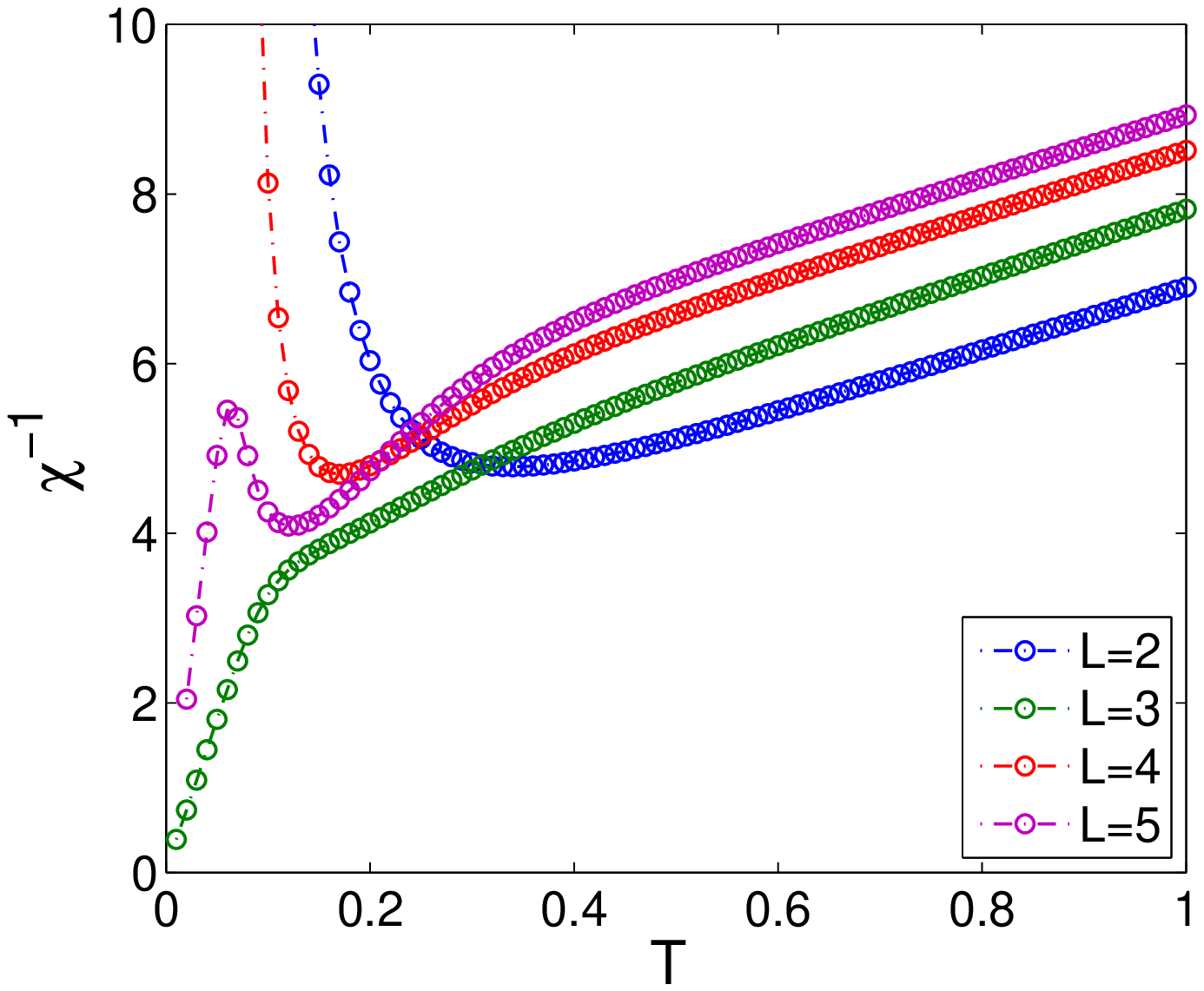}\label{fig:xi_rhomb_s1_2}}
\caption{Temperature dependencies of the inverse susceptibility for (a) T (b) H (c) H1 and (d) R domains with different sizes $L$. In (a), the values of $L=24$ and $48$ represent the thermodynamic limit approximations obtained from the MC simulations.}\label{fig:xi_s1_2}
\end{figure}
\hspace*{5mm} The magnetic susceptibility of the infinite two-dimensional triangular lattice Ising antiferromagnet has been shown in the low-temperature region to diverge as $1/T$ for $T \to 0$~\cite{sano}. However, at higher temperatures there is a bending to the Curie-Weiss like dependence. Interestingly, similar dependence of the susceptibility was also found in a simple triangle Ising spin cluster~\cite{isoda}. Considering also the similarity in the magnetization plateaux, displayed by both the infinite lattice and the cluster, the author speculates that they might originate in the local low energy states. Motivated by these findings, we were interested whether this picture could be qualitatively changed by changing the cluster shape and/or size.\\
\hspace*{5mm} In Fig.~\ref{fig:xi_s1_2} we present temperature dependencies of the inverse magnetic susceptibility $\chi^{-1}$ for different cluster shapes and sizes. We selected four shapes which show qualitatively different behaviors. In Fig.~\ref{fig:xi_tria0_s1_2}, the triangular T clusters show the variation with the downward bending below a certain temperature, as described above. Such a behavior is typical also for the infinite-lattice limit, which in the figure is represented by two curves obtained from the MC simulations for the values of $L=24$ and $48$, with the periodic boundary conditions applied to suppress finite-size effects\footnote{The two curves, which nicely collapse, are both shown to demonstrate virtual independence of the present results on the lattice size and thus to justify the presentation of the finite-size Monte Carlo calculations as a good approximation of the thermodynamic limit results.}. The respective curves differ in the slopes of the low-temperature linear dependences, which tend to increase with the increasing cluster size from $k\approx 11.8$ for $L=2$ up to $k \approx 25.7$ for $L=48$. Similar behavior is observed also in T1, T2 and T3 clusters. On the other hand, in the H clusters there seem to exist three distinct regimes of approximately linear dependence, but there are apparent differences even between the sizes $L=3$ and $L=5$, as can be seen in Fig.~\ref{fig:xi_hexa_s1_2}. The removal of the central nod in the H $\to$ H1 shape transition also removes the frustration either completely (for $L=3$) or at least partially (for $L=5$) and the susceptibility behavior changes drastically. Namely, instead of diverging it tends to zero for $T \to 0$, as depicted in Fig.~\ref{fig:xi_hexa1_s1_2}. In fact, this could be expected from the corresponding GS magnetization process (Fig.~\ref{fig:gs_m_hexa1}), which, in contrast with the triangular T, T1, T2 and T3 shapes, demonstrates the magnetization insensitivity to mild field intensities. Following the same line of thought, from Fig.~\ref{fig:gs_m_rhomb} for the rhombic R shape magnetization process, one can expect either diverging or vanishing susceptibility, depending on the cluster size. Indeed, the susceptibility vanishes for $L=2,4$ and diverges for $L=3,5$. Moreover, in the latter case it shows anomalous behavior in the intermediate temperature range, as evidenced in Fig.~\ref{fig:xi_rhomb_s1_2}. Actually, the zero-temperature susceptibility divergence (vanishing) can be expected for any finite rhombic cluster of an odd (even) valued linear size $L$. The zero-field state is trivially two-fold degenerate, characterized by two unique configurations with all the bonds along the domain satisfied~\cite{milla1}. For even $L$, the numbers of spins up and spins down are equal and small fields do not break this symmetry, do not remove the degeneracy and the magnetization remains zero. On the other hand, configurations with odd valued $L$ have prevalence of either spins up or down and zero magnetization results from the presence of two degenerate configurations with non-zero magnetization of opposite signs. Then, an infinitesimal field lifts the degeneracy and the preferred state will have non-zero magnetization.        

\subsubsection{Magnetocaloric properties}
\begin{figure}[]
\centering
\subfigure[$m$]{\includegraphics[scale=0.43,clip]{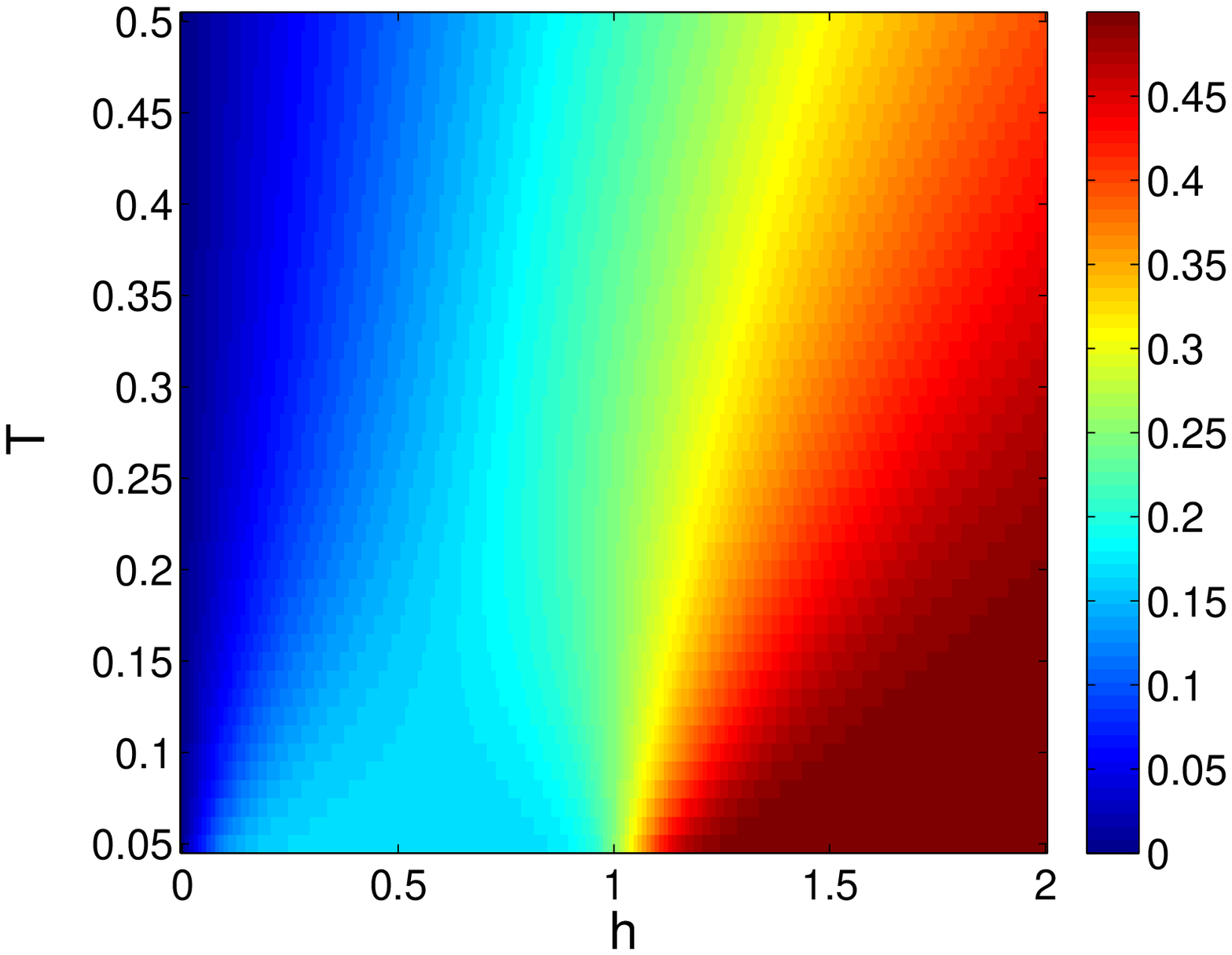}\label{fig:T0L2_m}}
\subfigure[$(\partial m/\partial T)|_h$]{\includegraphics[scale=0.43,clip]{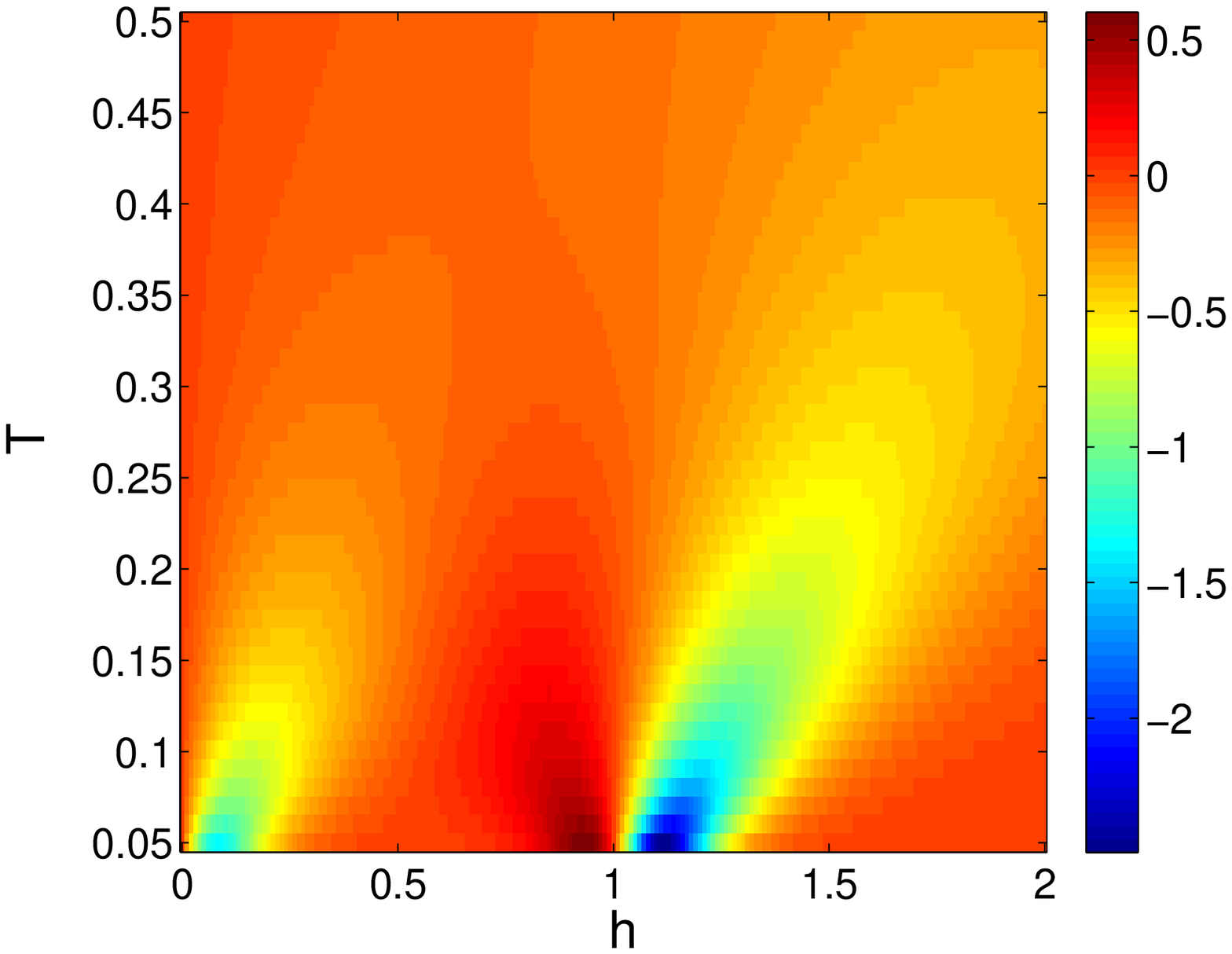}\label{fig:T0L2_dmdt}}
\subfigure[$\Delta S/N$]{\includegraphics[scale=0.43,clip]{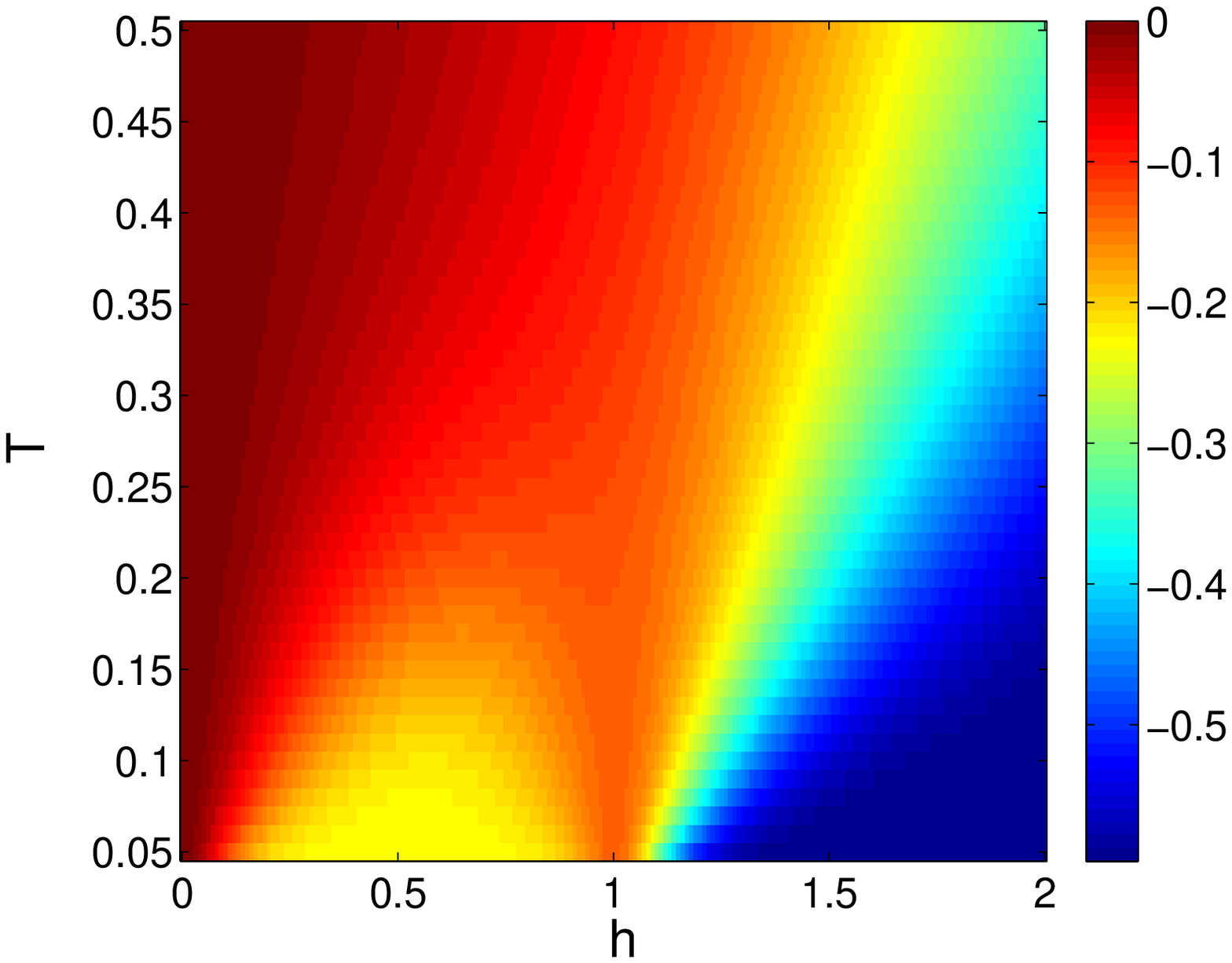}\label{fig:T0L2_ds}}
\subfigure[$c$]{\includegraphics[scale=0.43,clip]{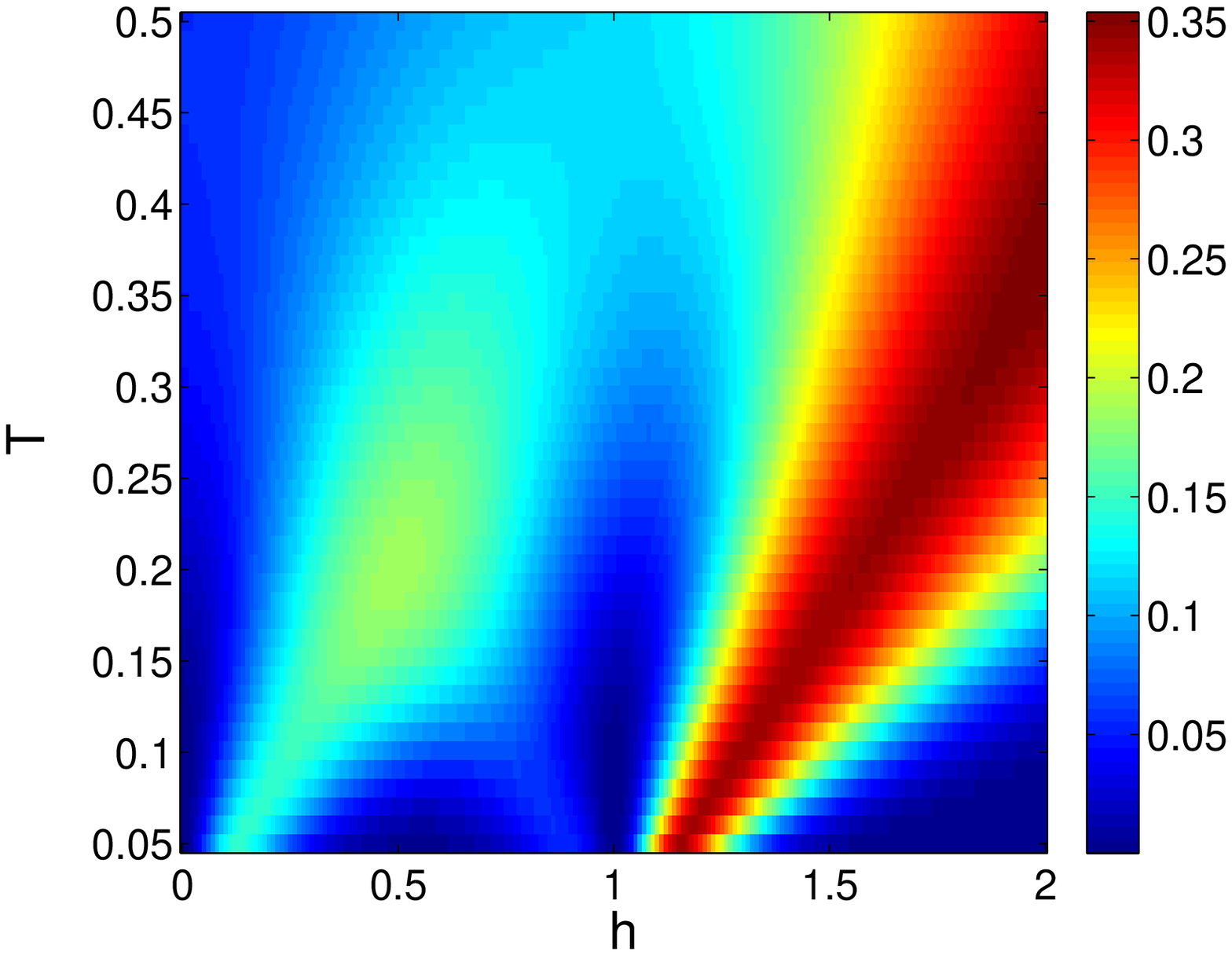}\label{fig:T0L2_c}}
\caption{Temperature-field dependencies of (a) the magnetization $m$, (b) the magnetization derivative with respect to temperature at a constant field $(\partial m/\partial T)|_h$, (c) the entropy density change $\Delta S/N$, and (d) the specific heat per spin $c$, for the triangular domain T with $L=2$.}\label{fig:T0L2}
\end{figure}
\begin{figure}[t]
\centering
\subfigure[$m$]{\includegraphics[scale=0.43,clip]{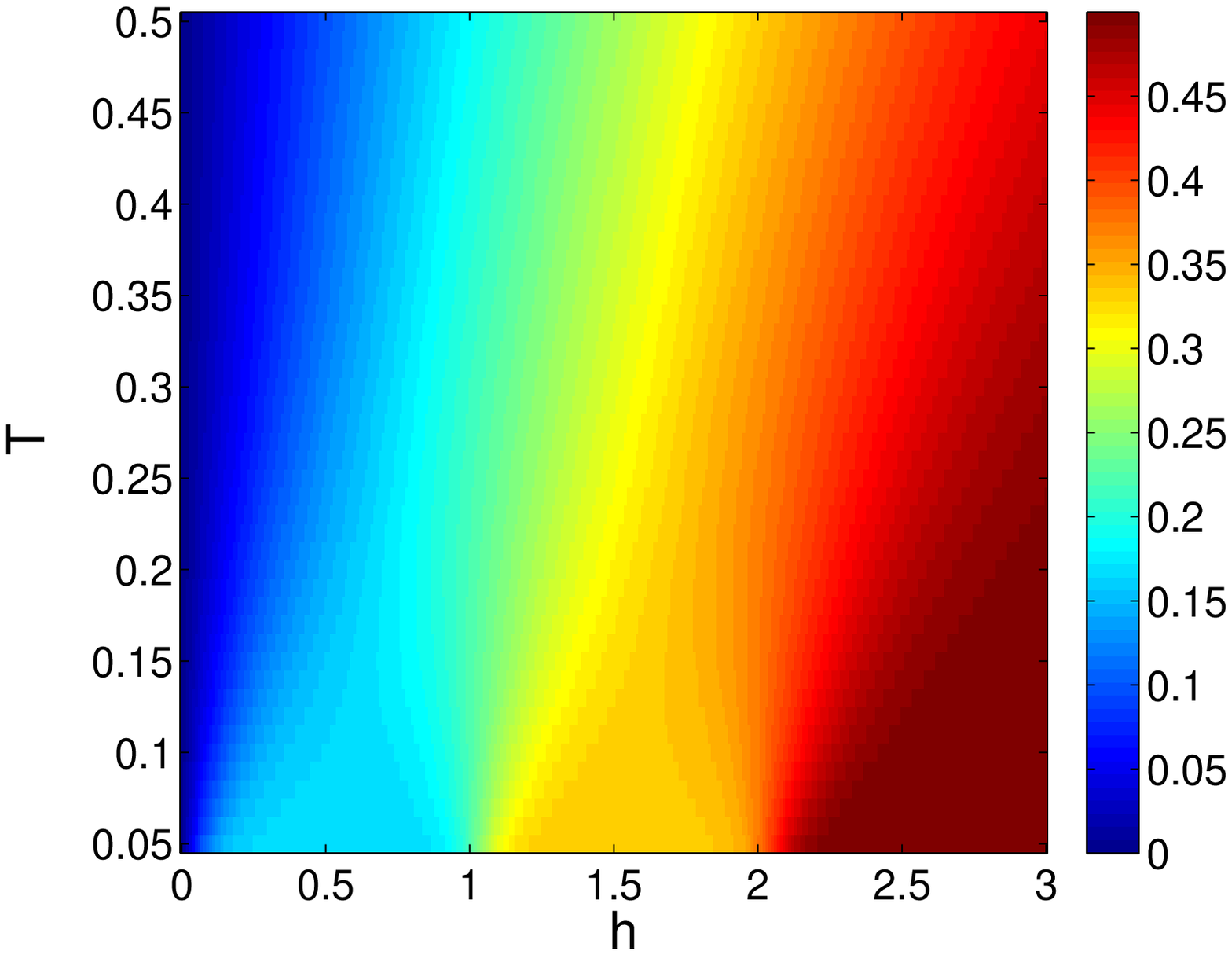}\label{fig:T0L3_m}}
\subfigure[$(\partial m/\partial T)|_h$]{\includegraphics[scale=0.43,clip]{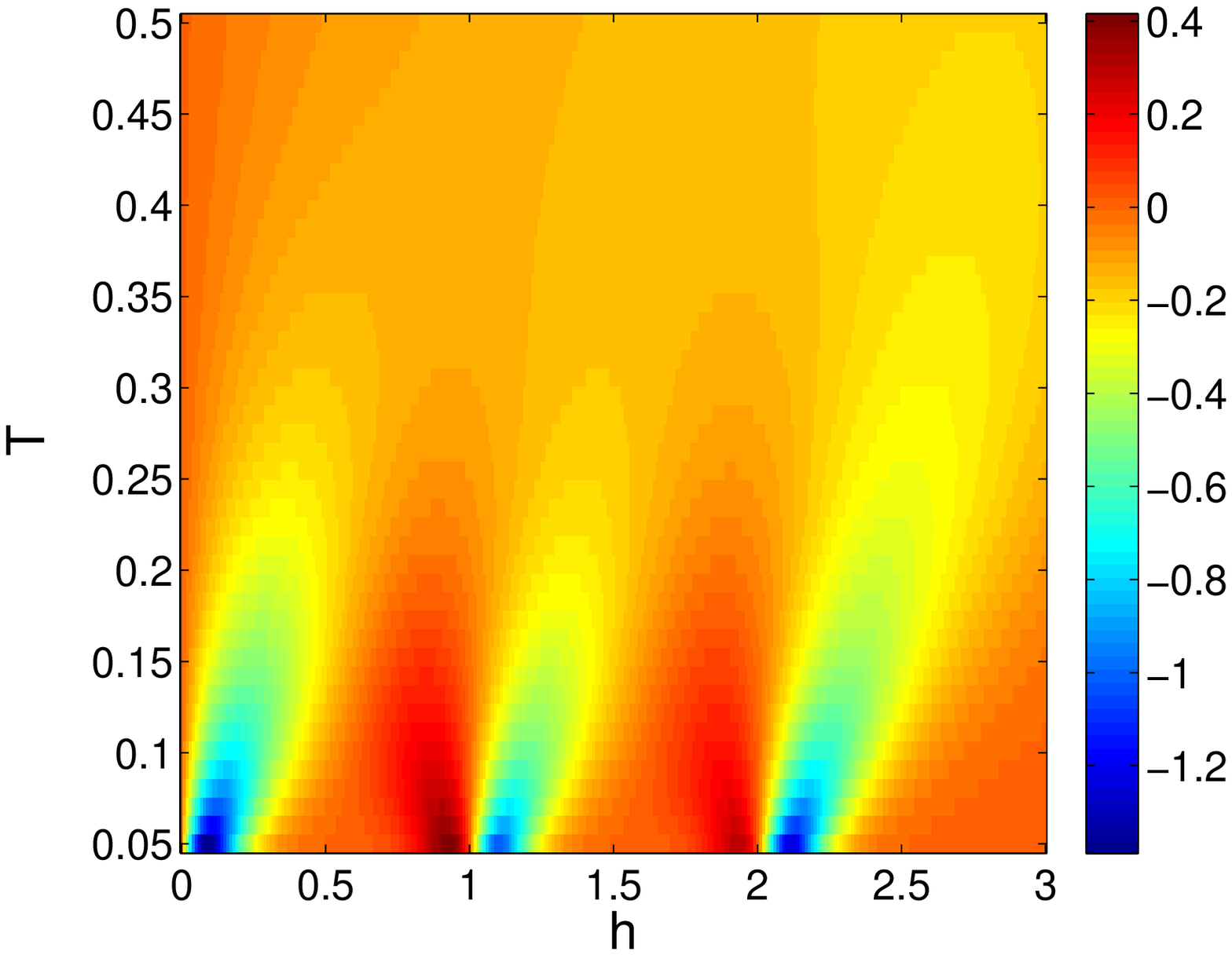}\label{fig:T0L3_dmdt}}
\subfigure[$\Delta S/N$]{\includegraphics[scale=0.43,clip]{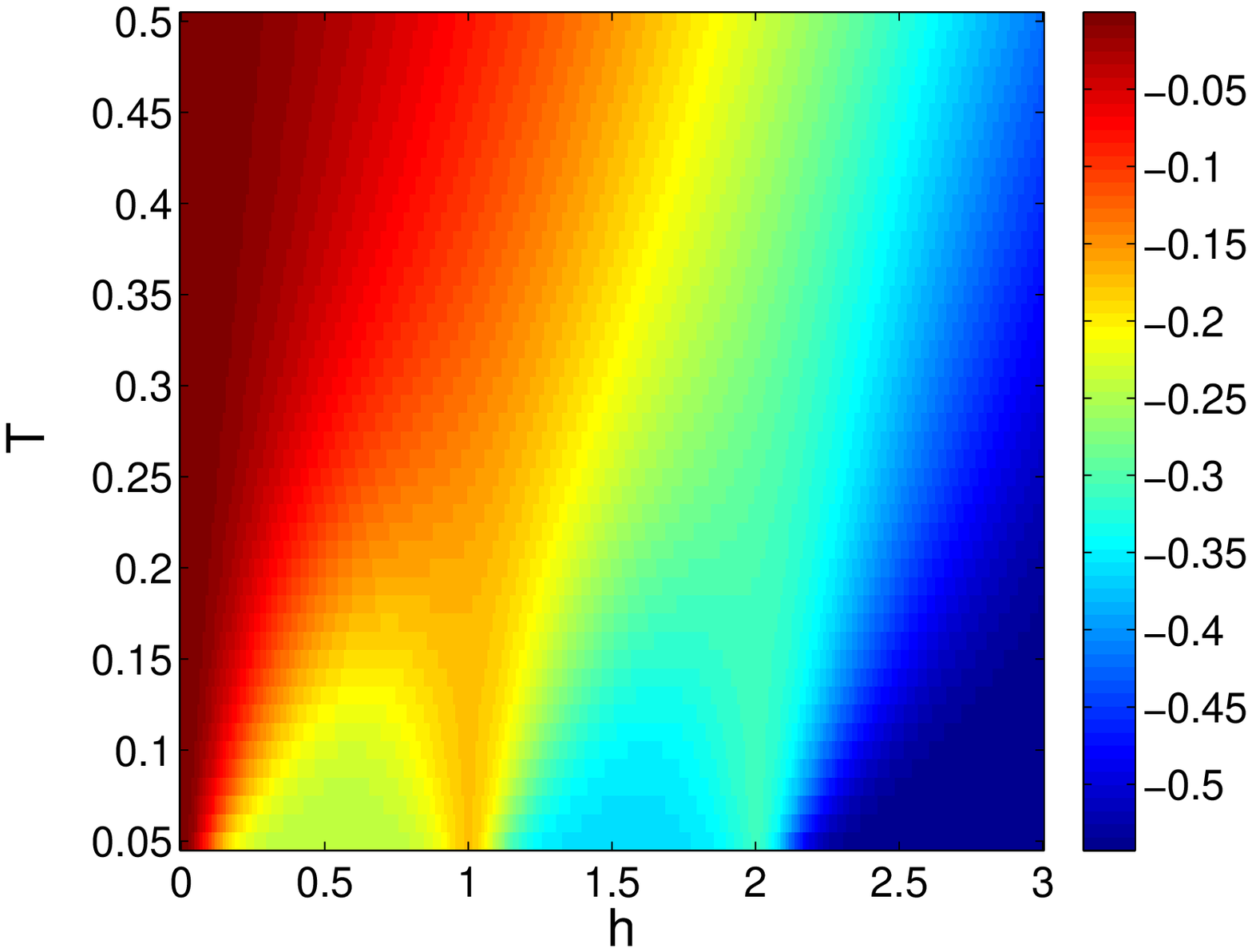}\label{fig:T0L3_ds}}
\subfigure[$c$]{\includegraphics[scale=0.43,clip]{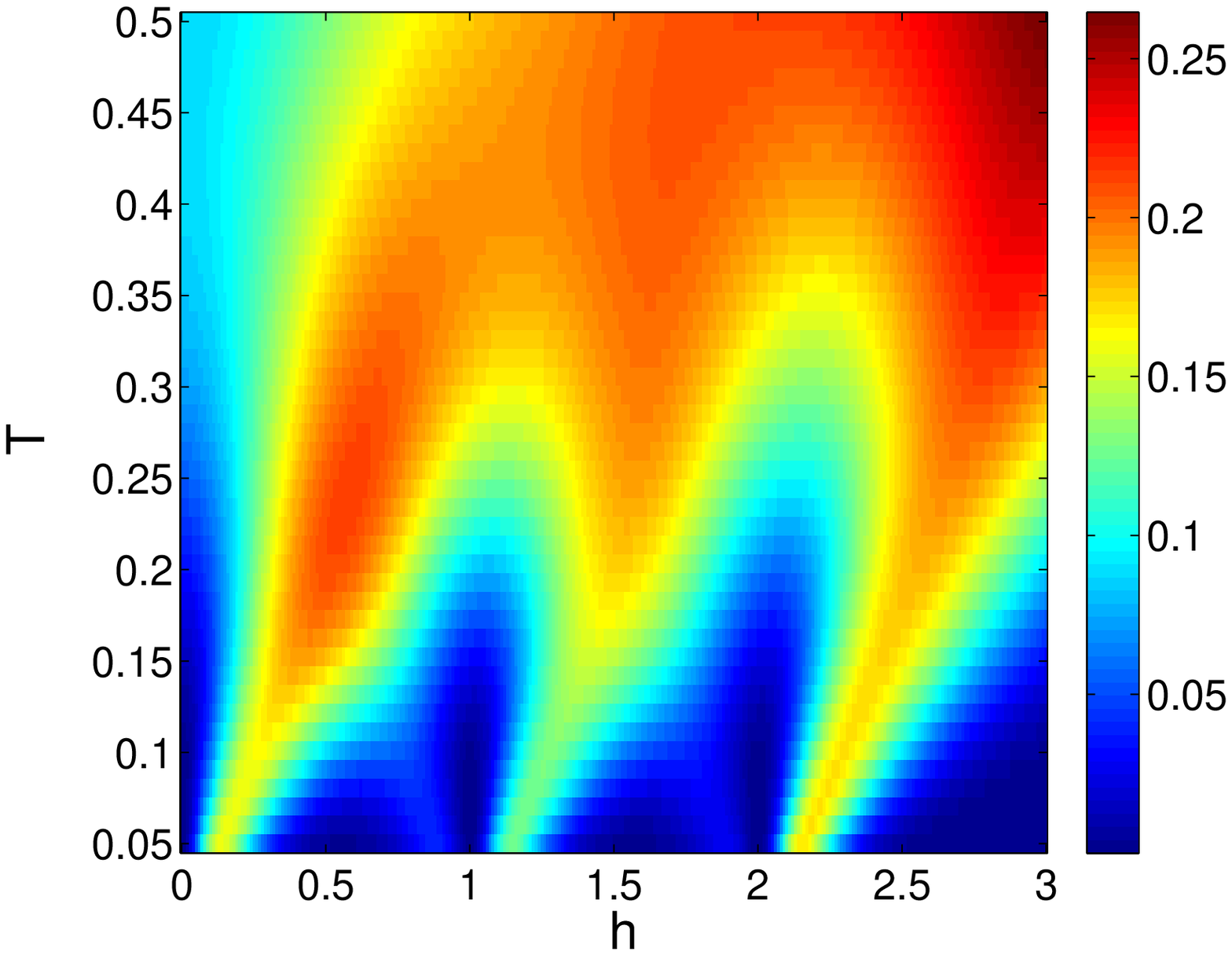}\label{fig:T0L3_c}}
\caption{The same quantities as in Fig.~\ref{fig:T0L2} for $L=3$.}\label{fig:T0L3}
\end{figure}
\begin{figure}[t]
\centering
\subfigure[$m$]{\includegraphics[scale=0.43,clip]{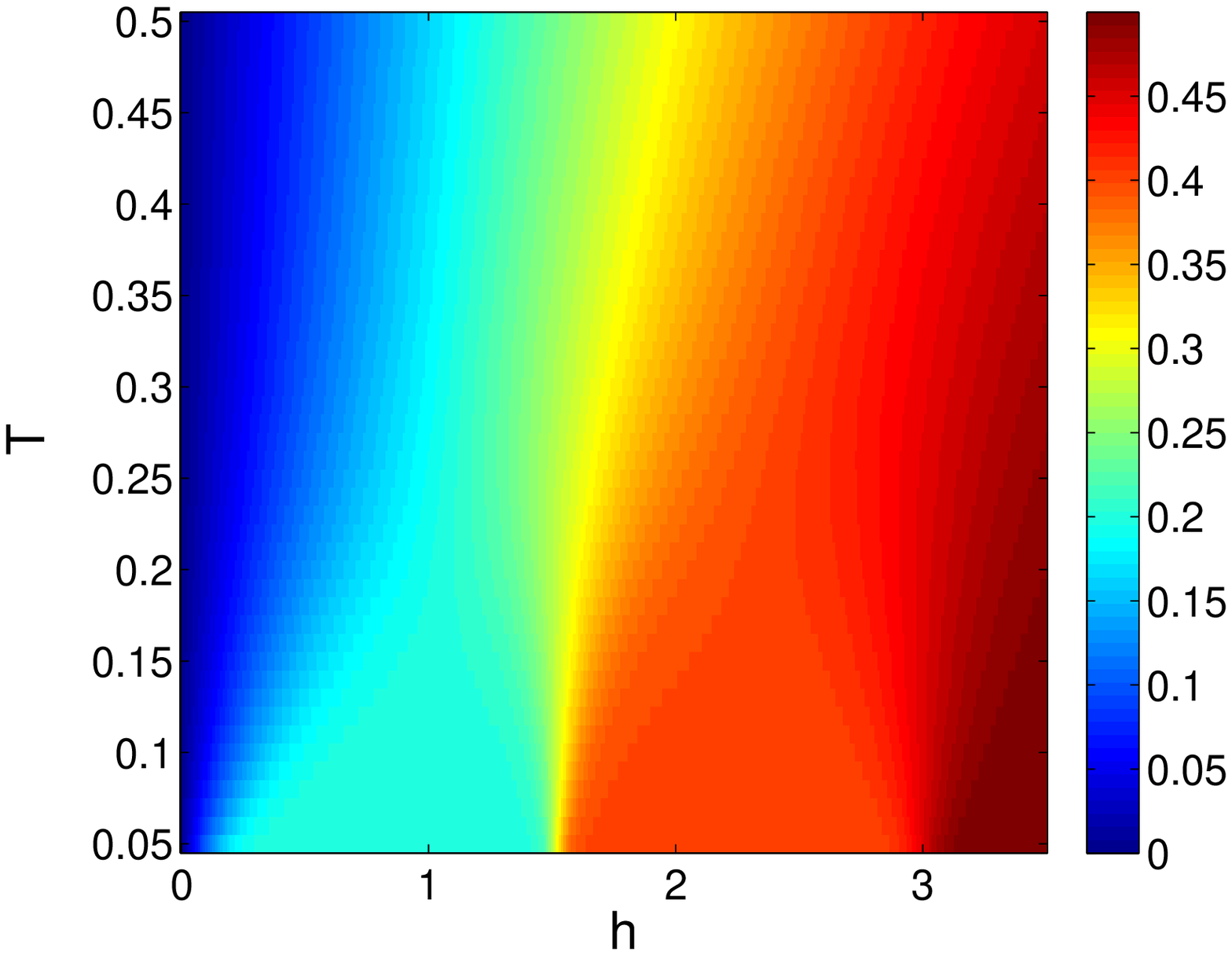}\label{fig:T0L4_m}}
\subfigure[$(\partial m/\partial T)|_h$]{\includegraphics[scale=0.43,clip]{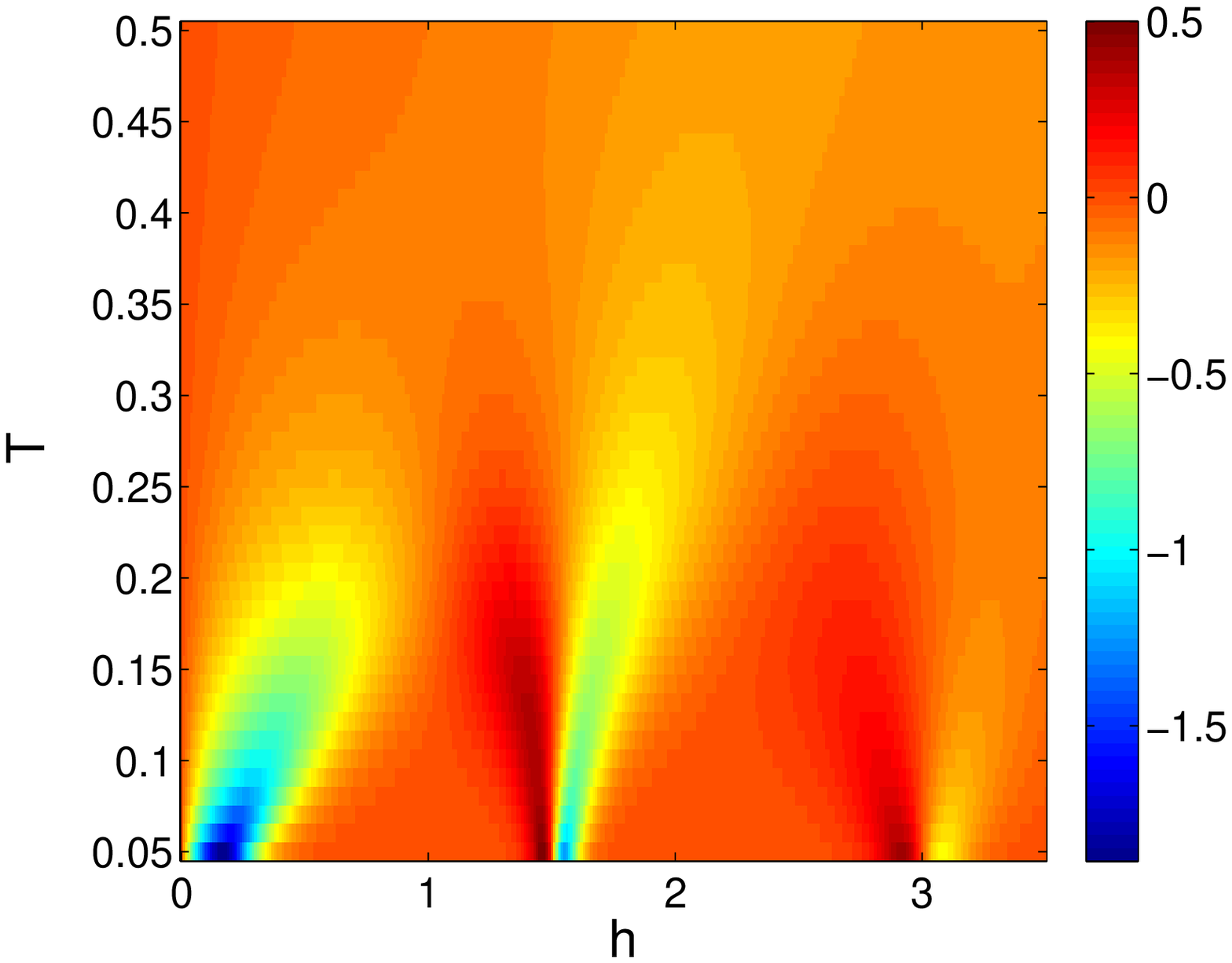}\label{fig:T0L4_dmdt}}
\subfigure[$\Delta S/N$]{\includegraphics[scale=0.43,clip]{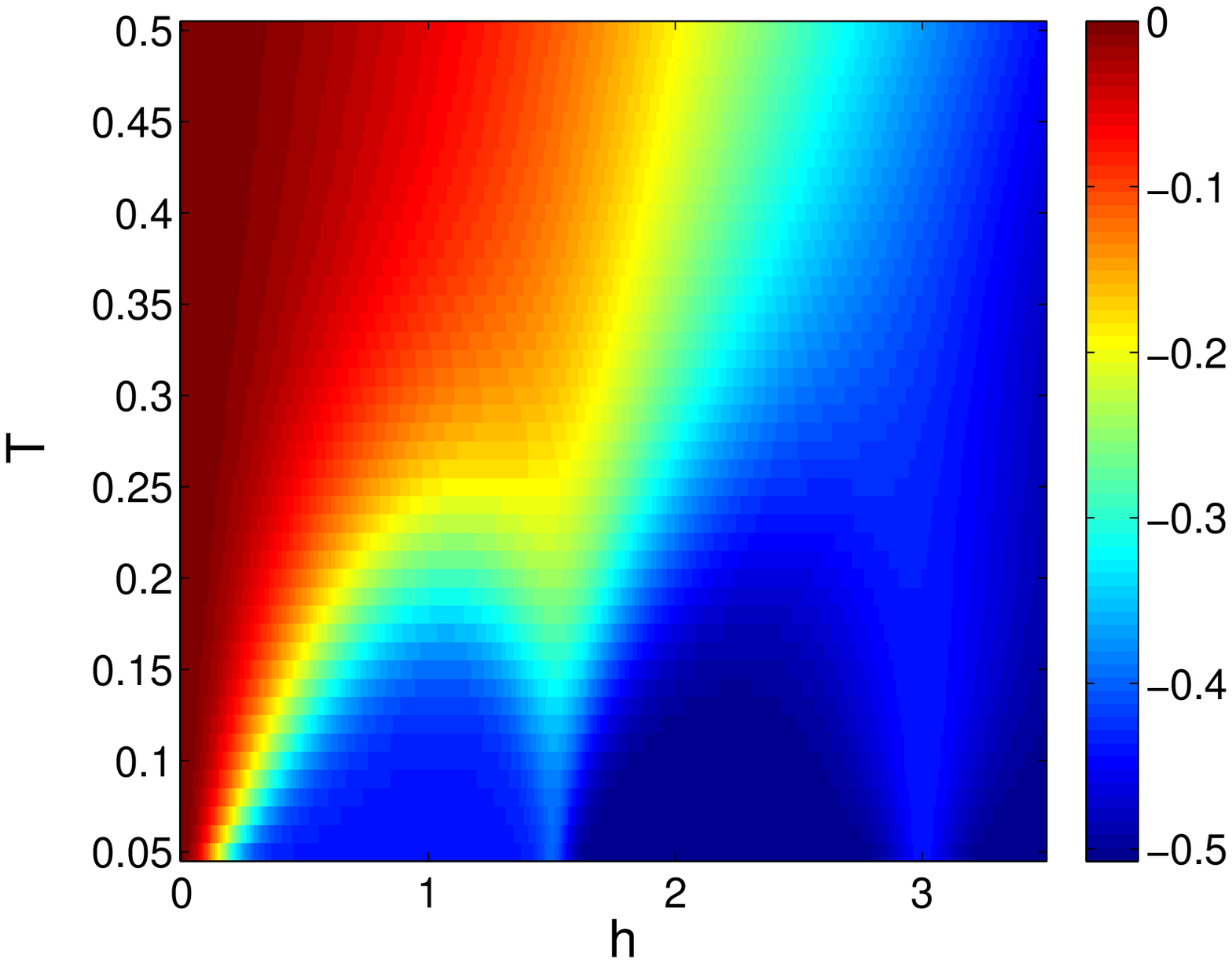}\label{fig:T0L4_ds}}
\subfigure[$c$]{\includegraphics[scale=0.43,clip]{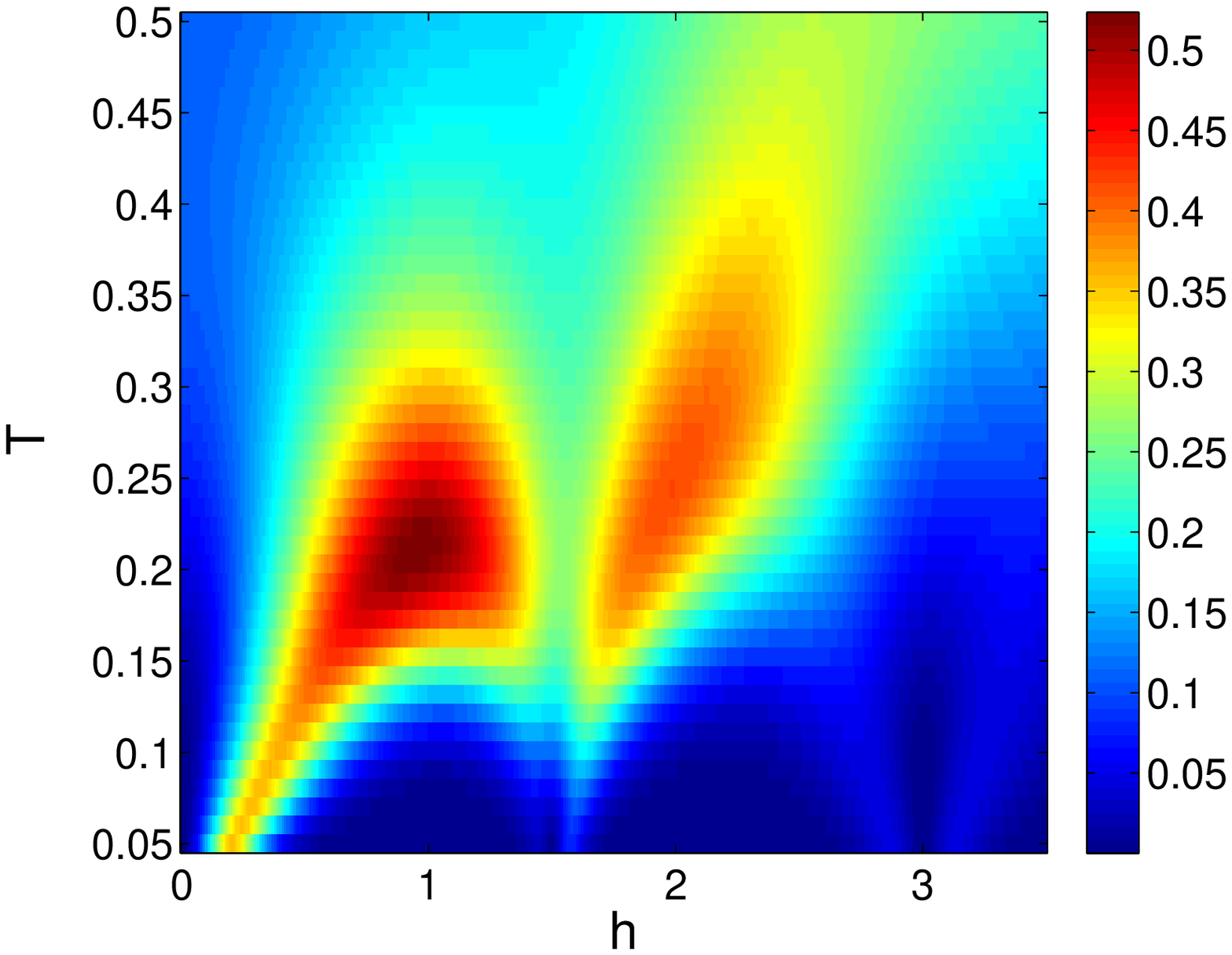}\label{fig:T0L4_c}}
\caption{The same quantities as in Fig.~\ref{fig:T0L2} for $L=4$.}\label{fig:T0L4}
\end{figure}
\begin{figure}[t]
\centering
\subfigure[$m$]{\includegraphics[scale=0.43,clip]{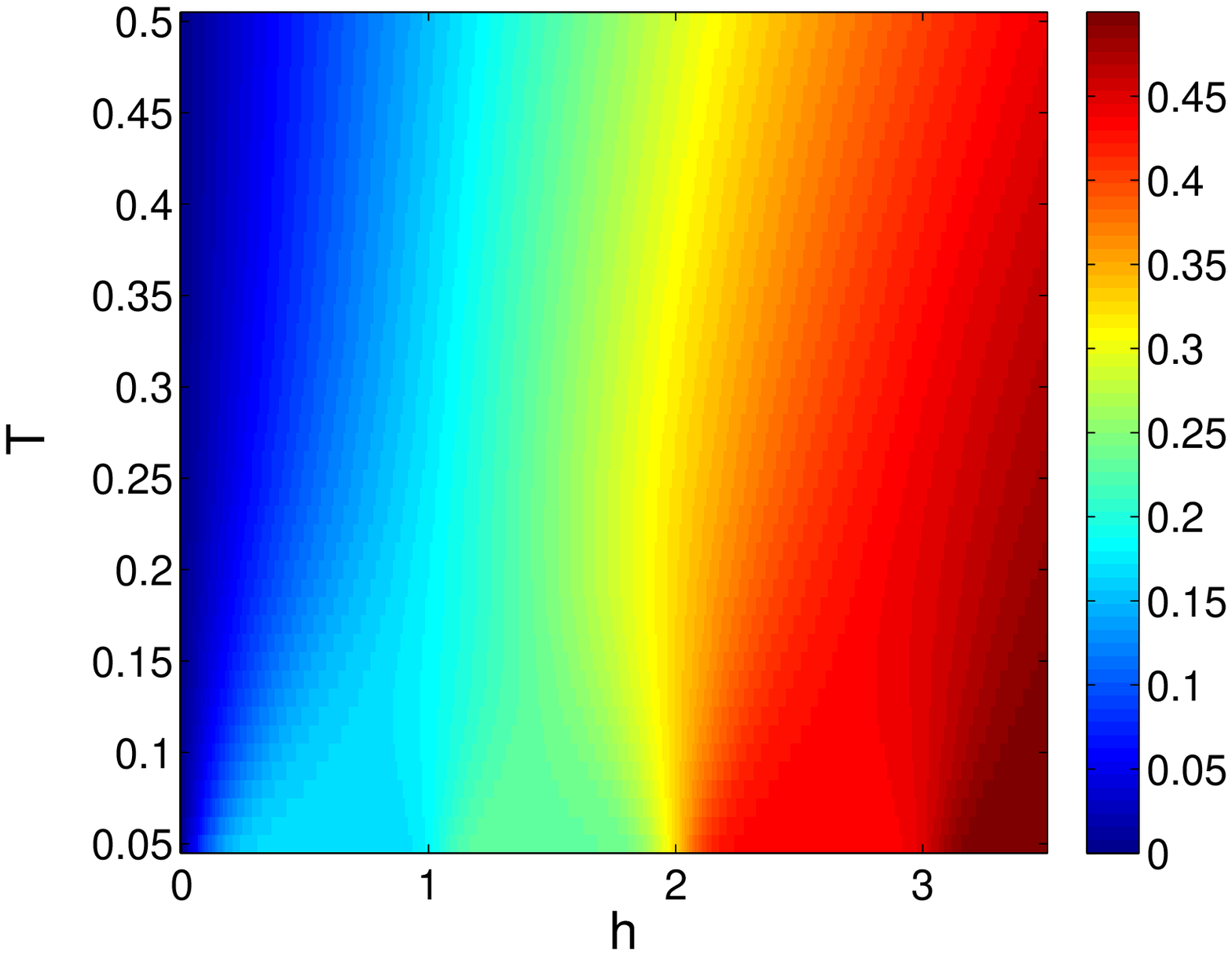}\label{fig:T0L5_m}}
\subfigure[$(\partial m/\partial T)|_h$]{\includegraphics[scale=0.43,clip]{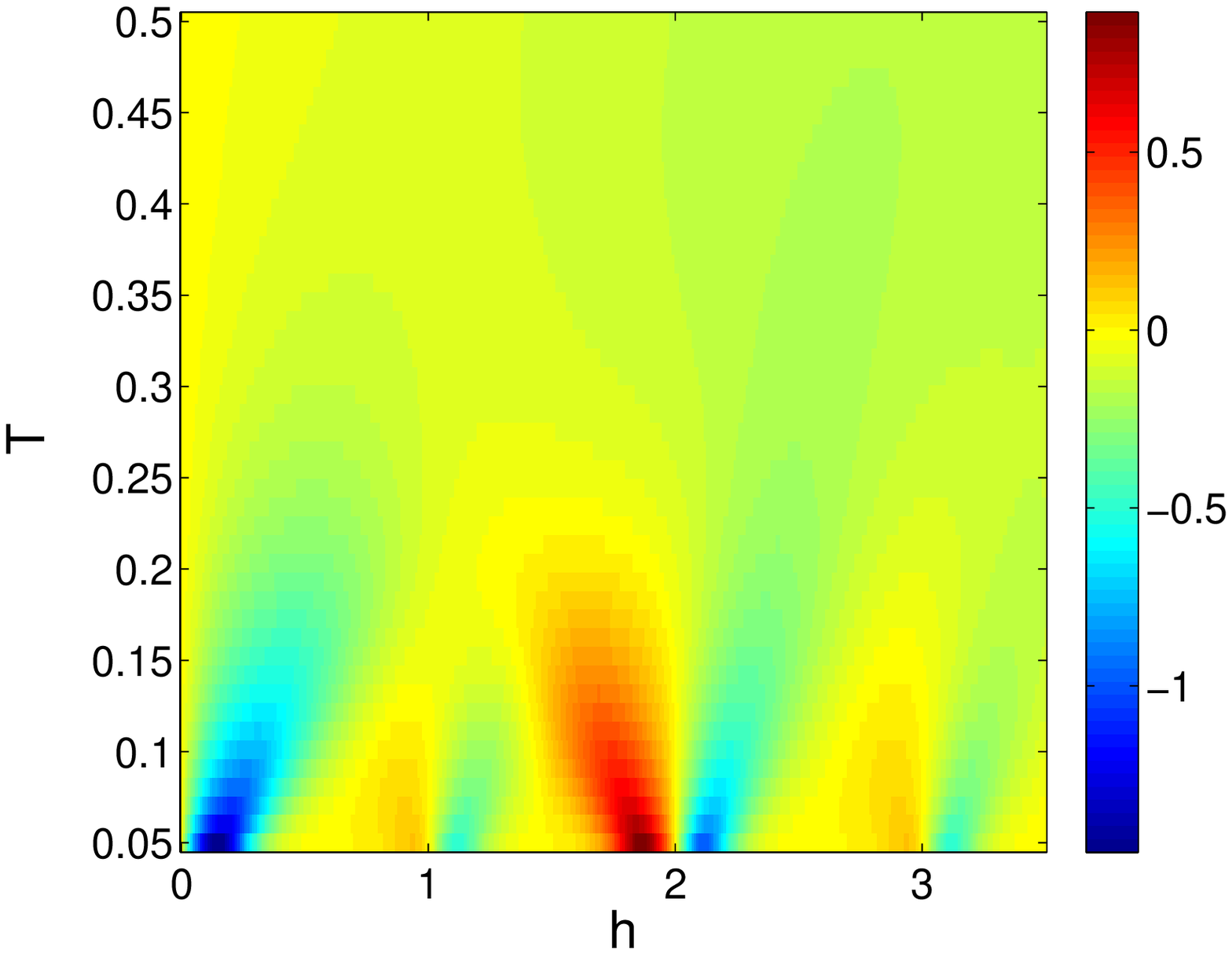}\label{fig:T0L5_dmdt}}
\subfigure[$\Delta S/N$]{\includegraphics[scale=0.43,clip]{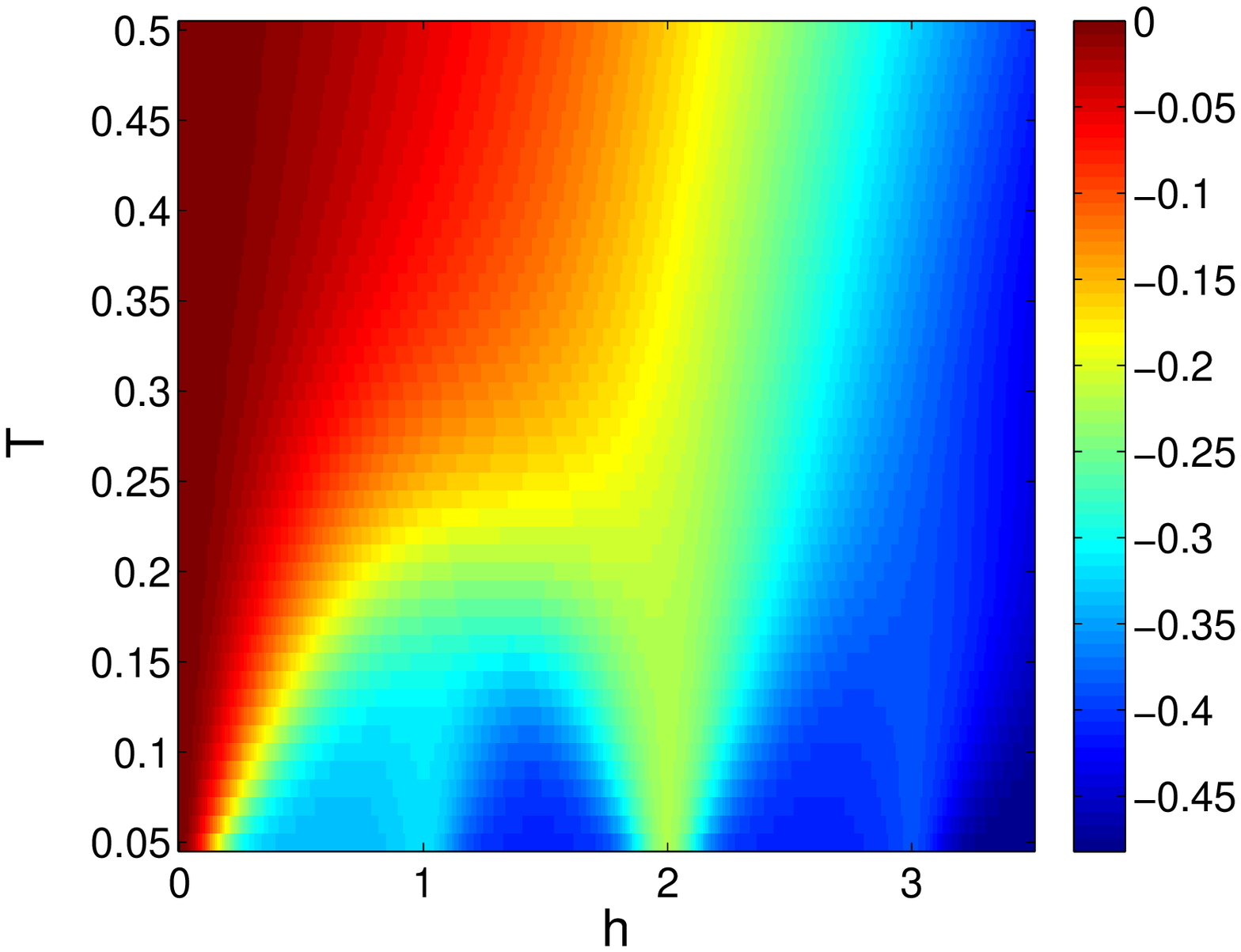}\label{fig:T0L5_ds}}
\subfigure[$c$]{\includegraphics[scale=0.43,clip]{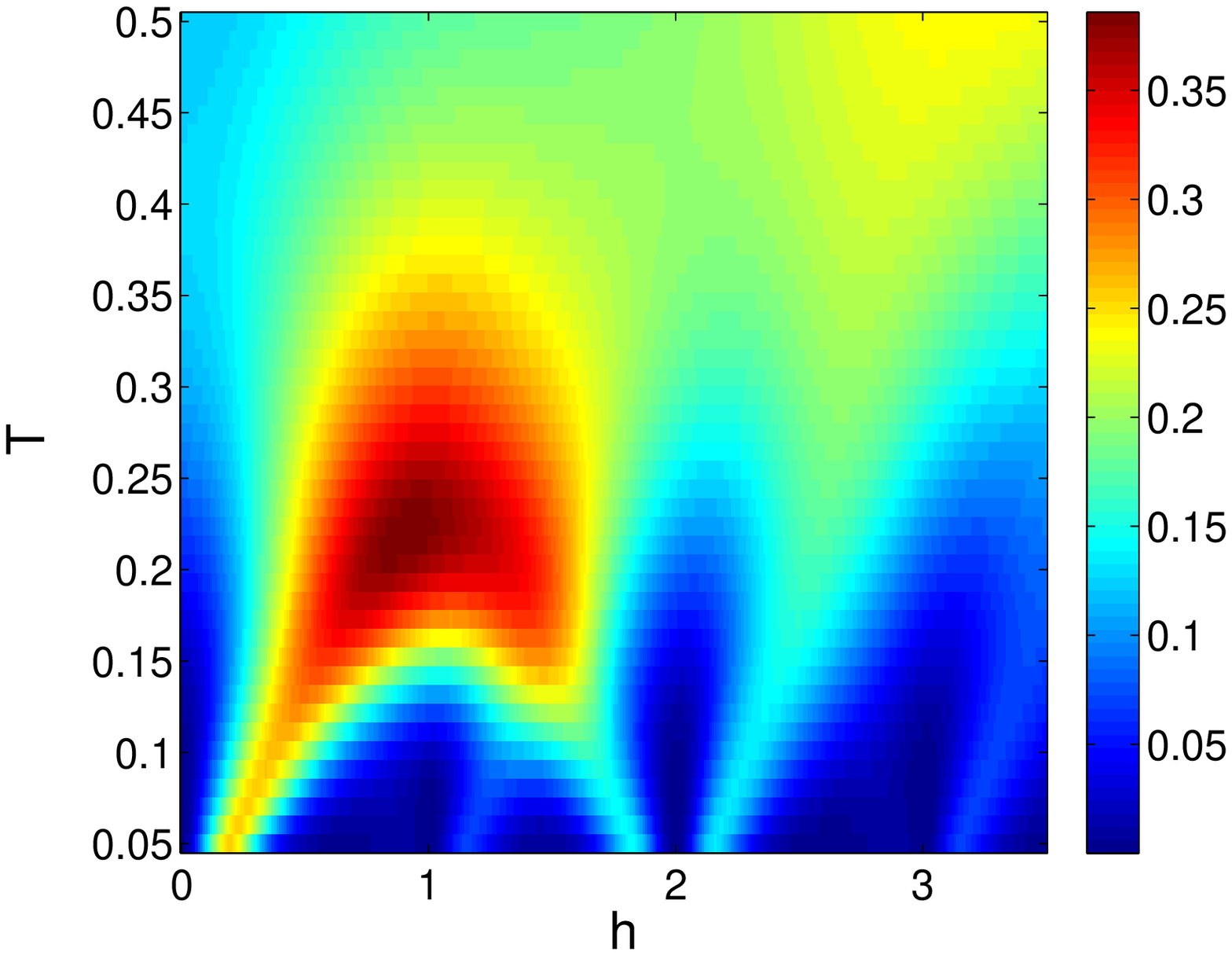}\label{fig:T0L5_c}}
\caption{The same quantities as in Fig.~\ref{fig:T0L2} for $L=5$.}\label{fig:T0L5}
\end{figure}
\begin{figure}[t]
\centering
\subfigure[$L=2$]{\includegraphics[scale=0.43,clip]{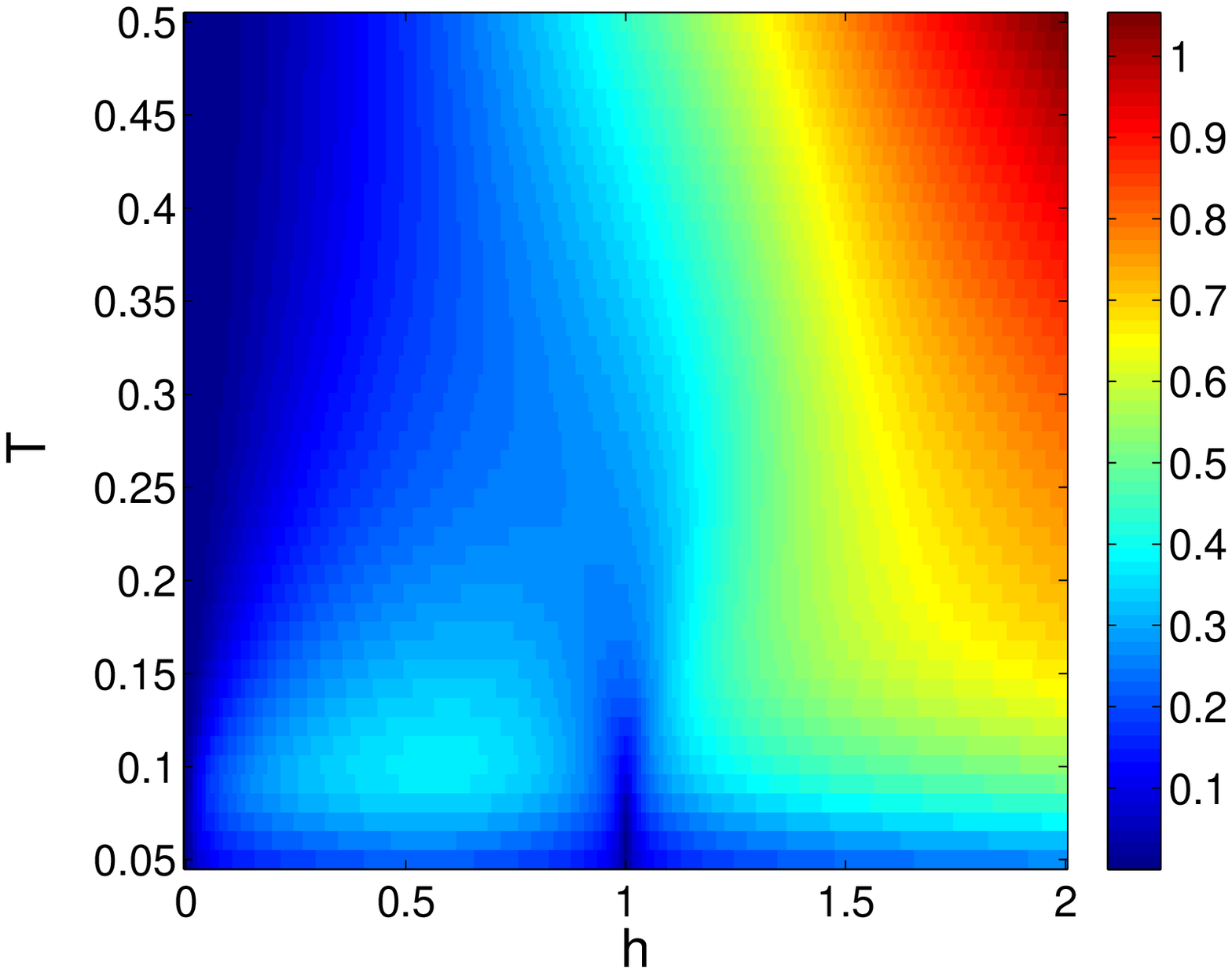}}
\subfigure[$L=3$]{\includegraphics[scale=0.43,clip]{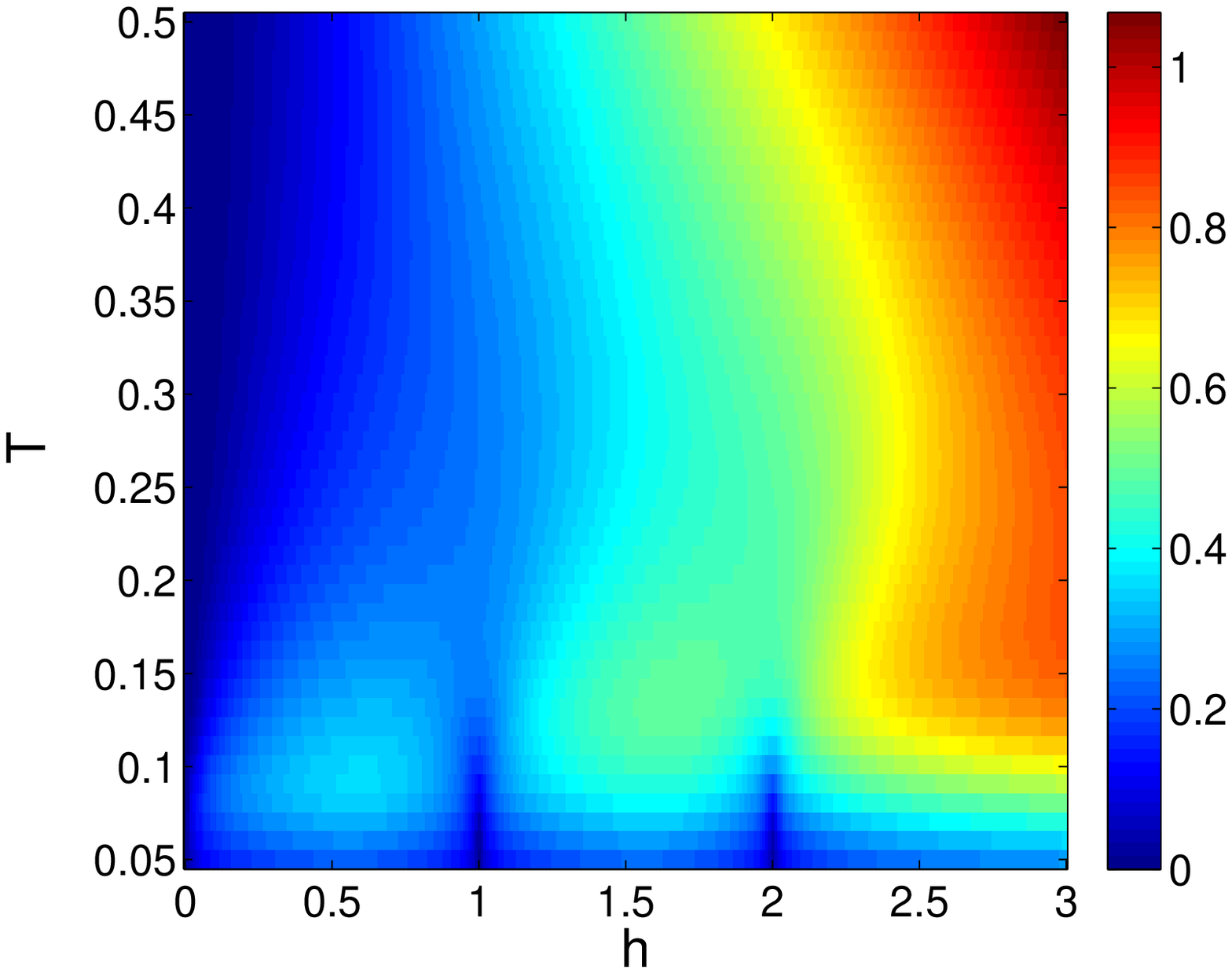}}
\subfigure[$L=4$]{\includegraphics[scale=0.43,clip]{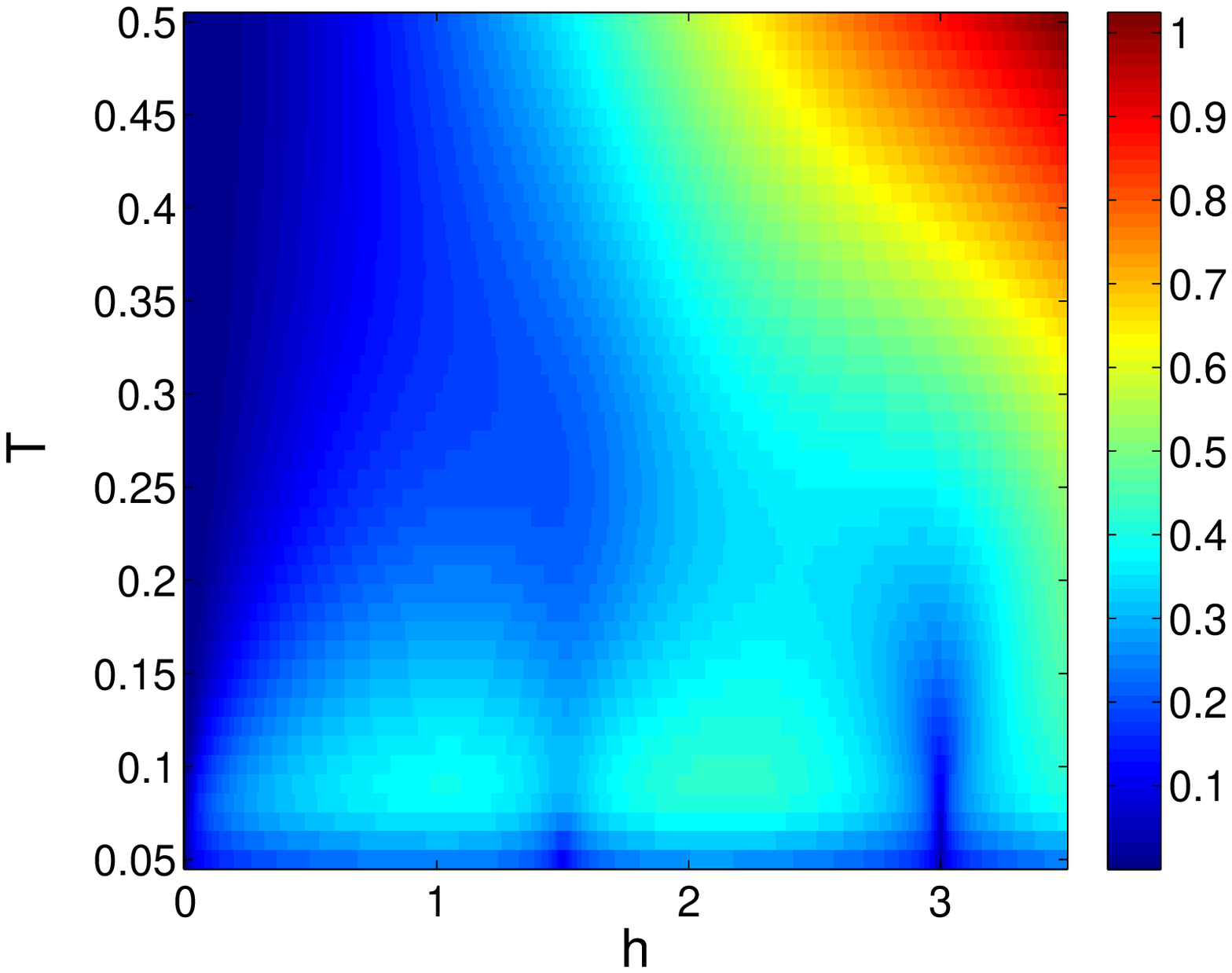}}
\subfigure[$L=5$]{\includegraphics[scale=0.43,clip]{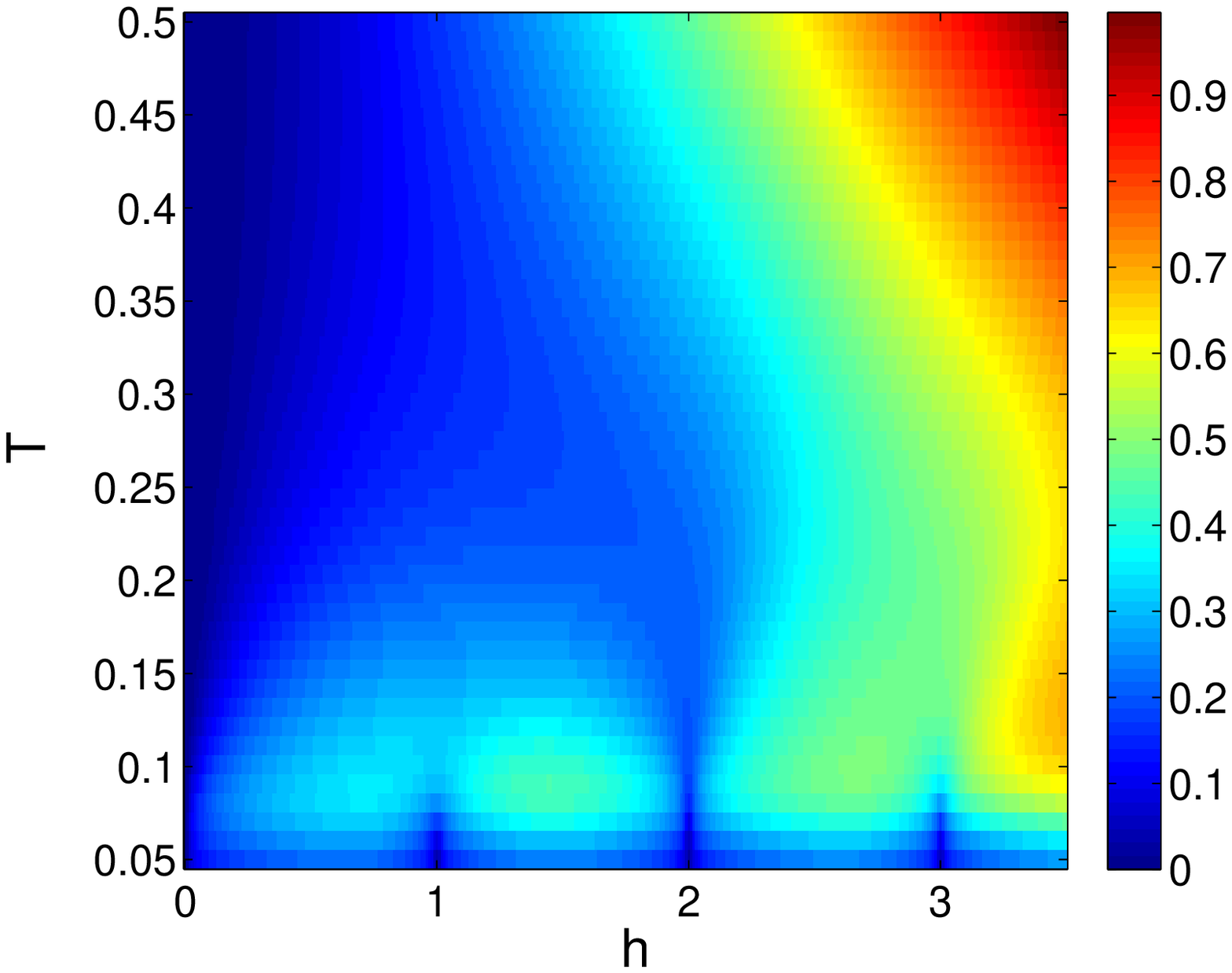}}
\caption{The adiabatic temperature changes $\Delta T_{ad}$ for the triangular T clusters of different sizes (a) $L=2$, (b) $L=3$, (c) $L=4$, and (d) $L=5$.}\label{fig:TdT}
\end{figure}
\begin{figure}[t]
\centering
\subfigure[$m$]{\includegraphics[scale=0.43,clip]{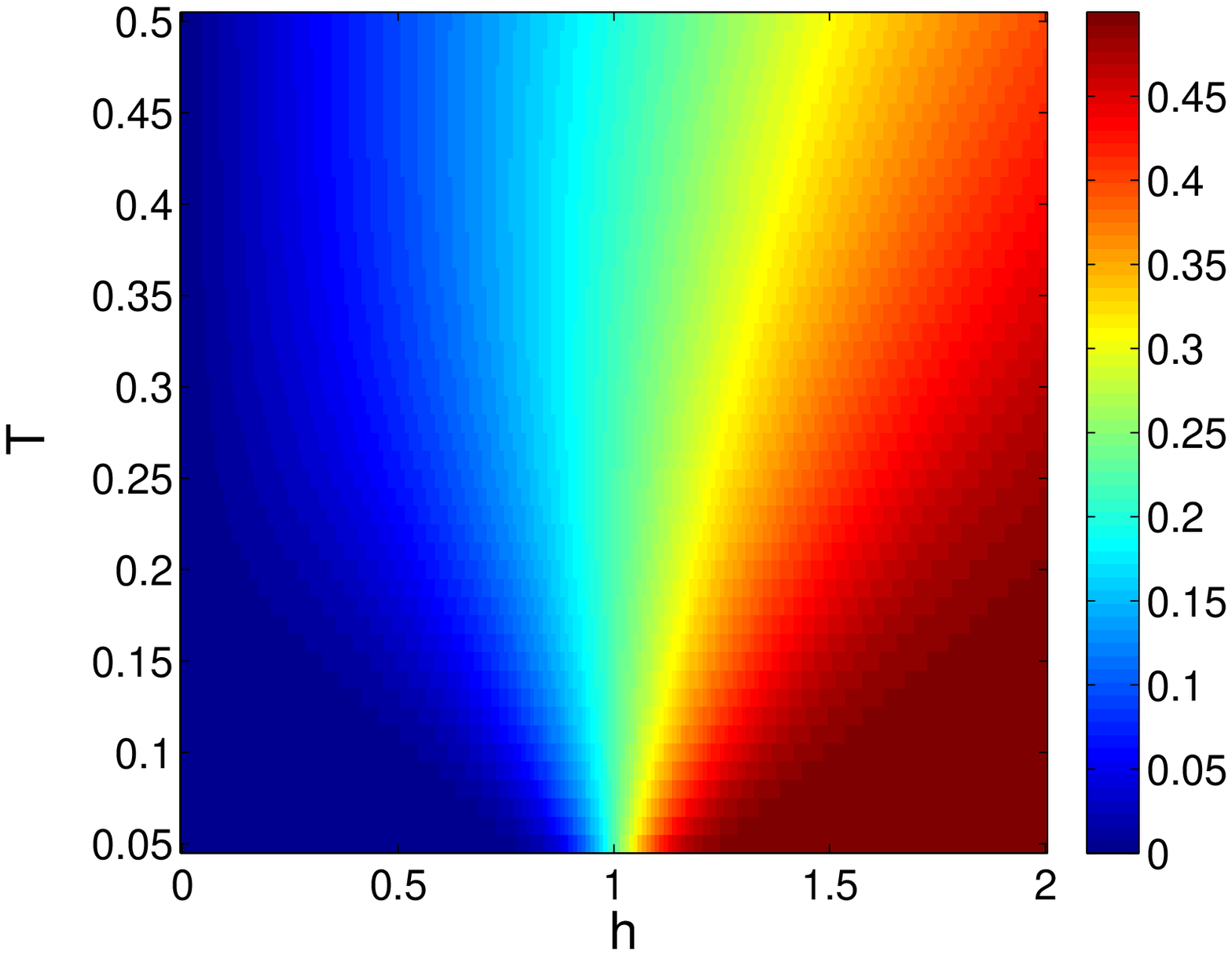}}
\subfigure[$(\partial m/\partial T)|_h$]{\includegraphics[scale=0.43,clip]{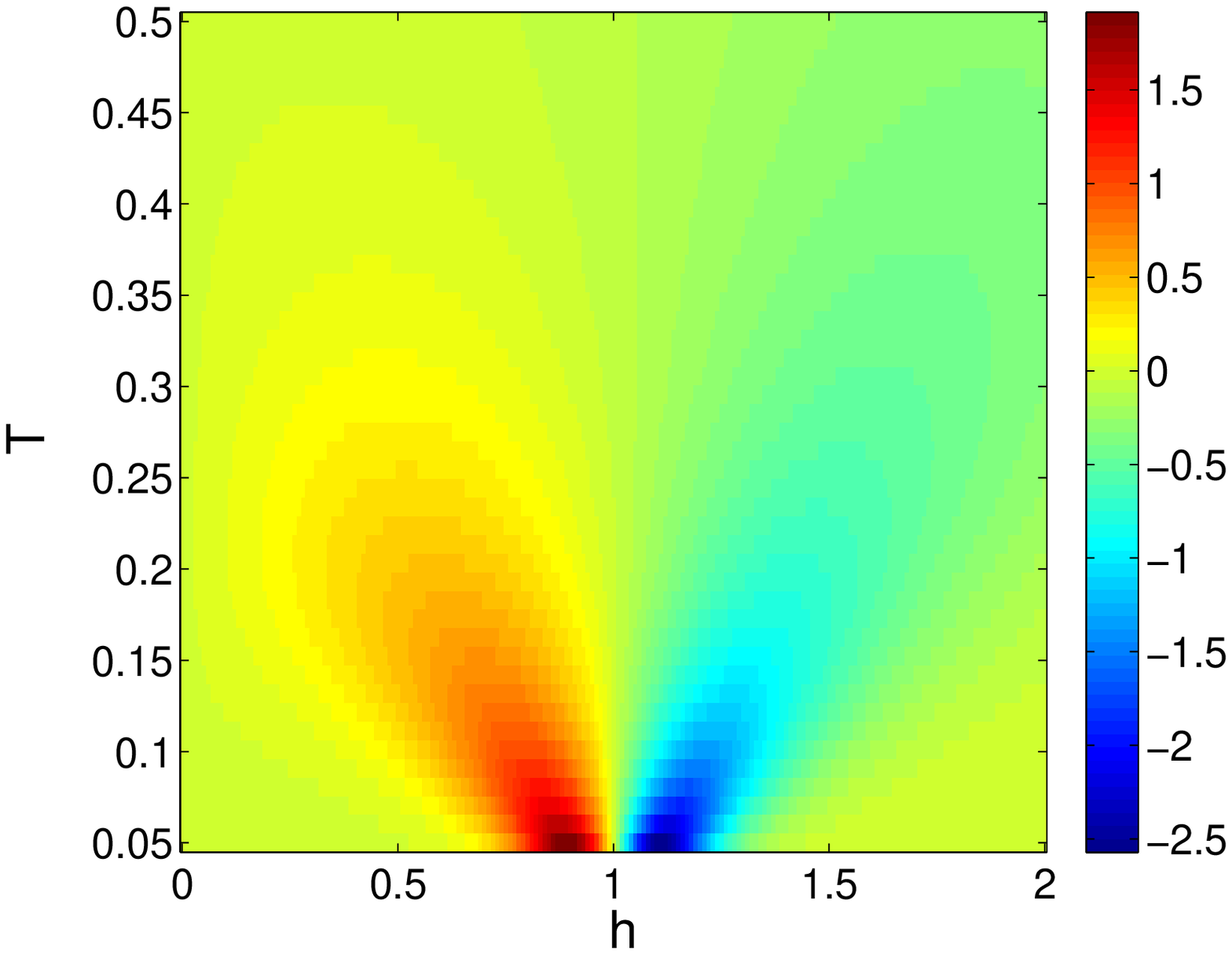}}
\subfigure[$\Delta S/N$]{\includegraphics[scale=0.43,clip]{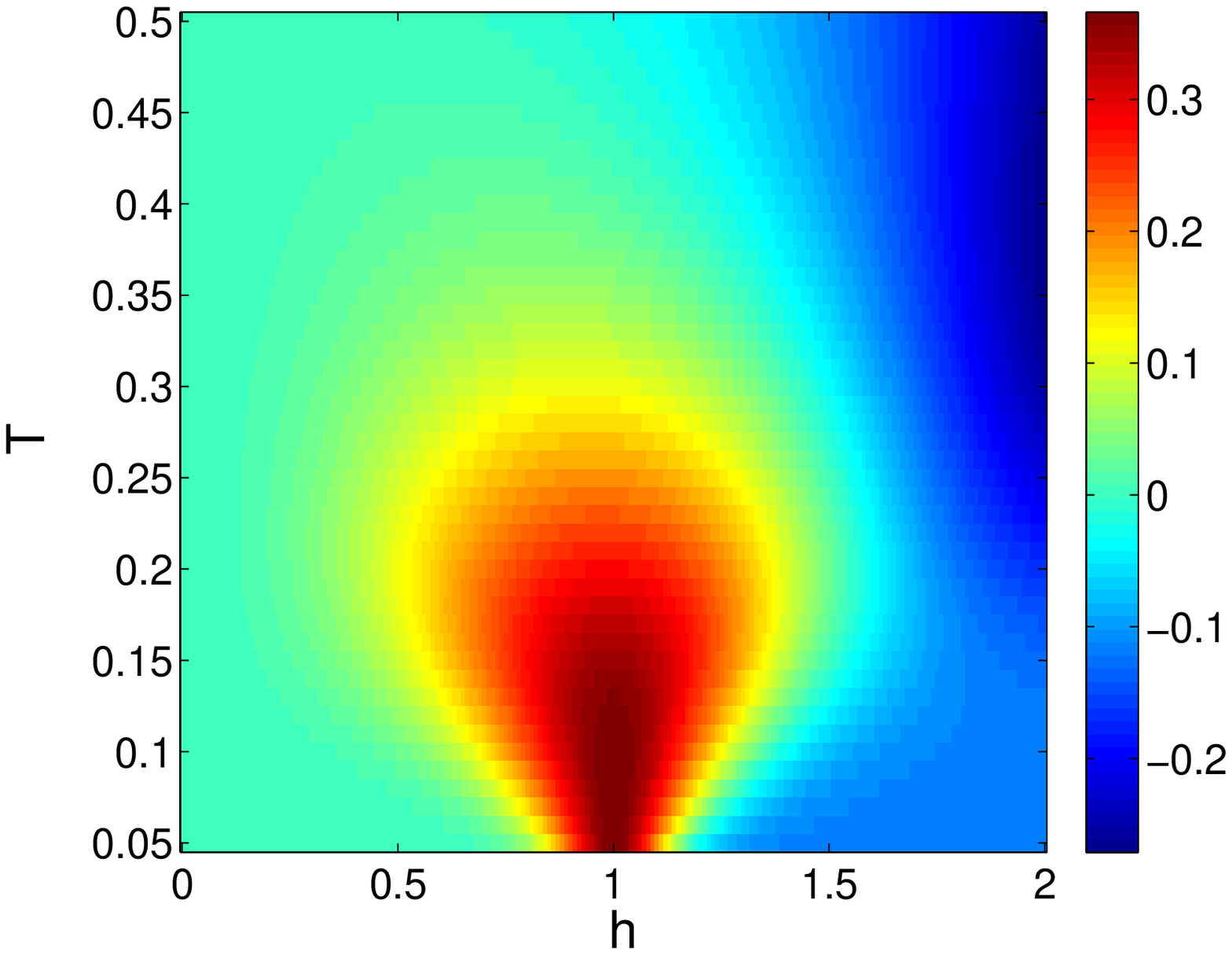}}
\subfigure[$c$]{\includegraphics[scale=0.43,clip]{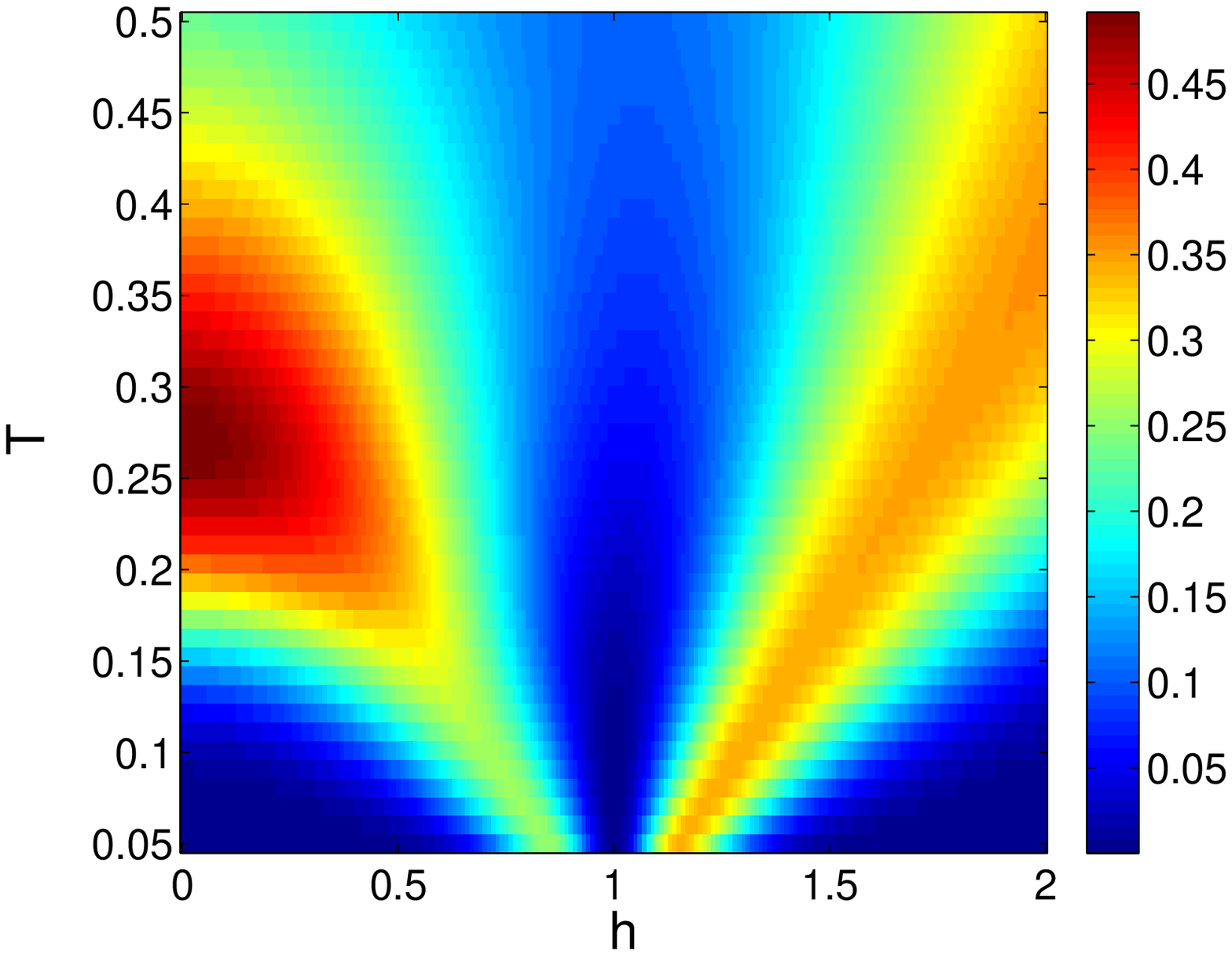}}
\caption{The same quantities as in Fig.~\ref{fig:T0L2} for H1 cluster with $L=3$.}\label{fig:H1L3}
\end{figure}
\begin{figure}[t]
\centering
\subfigure[T, $L=2$]{\includegraphics[scale=0.43,clip]{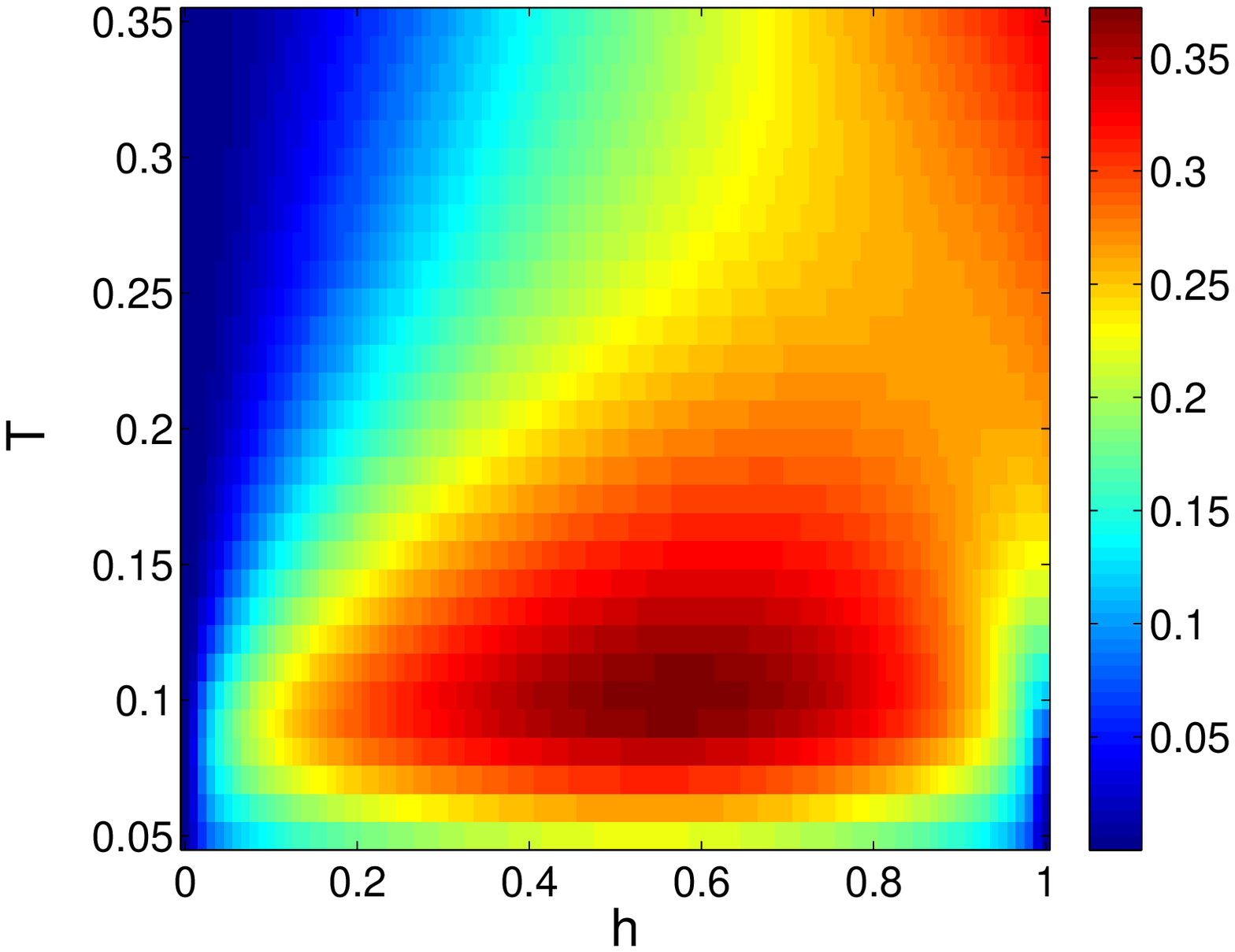}\label{fig:TdT_det}}
\subfigure[H1, $L=3$]{\includegraphics[scale=0.43,clip]{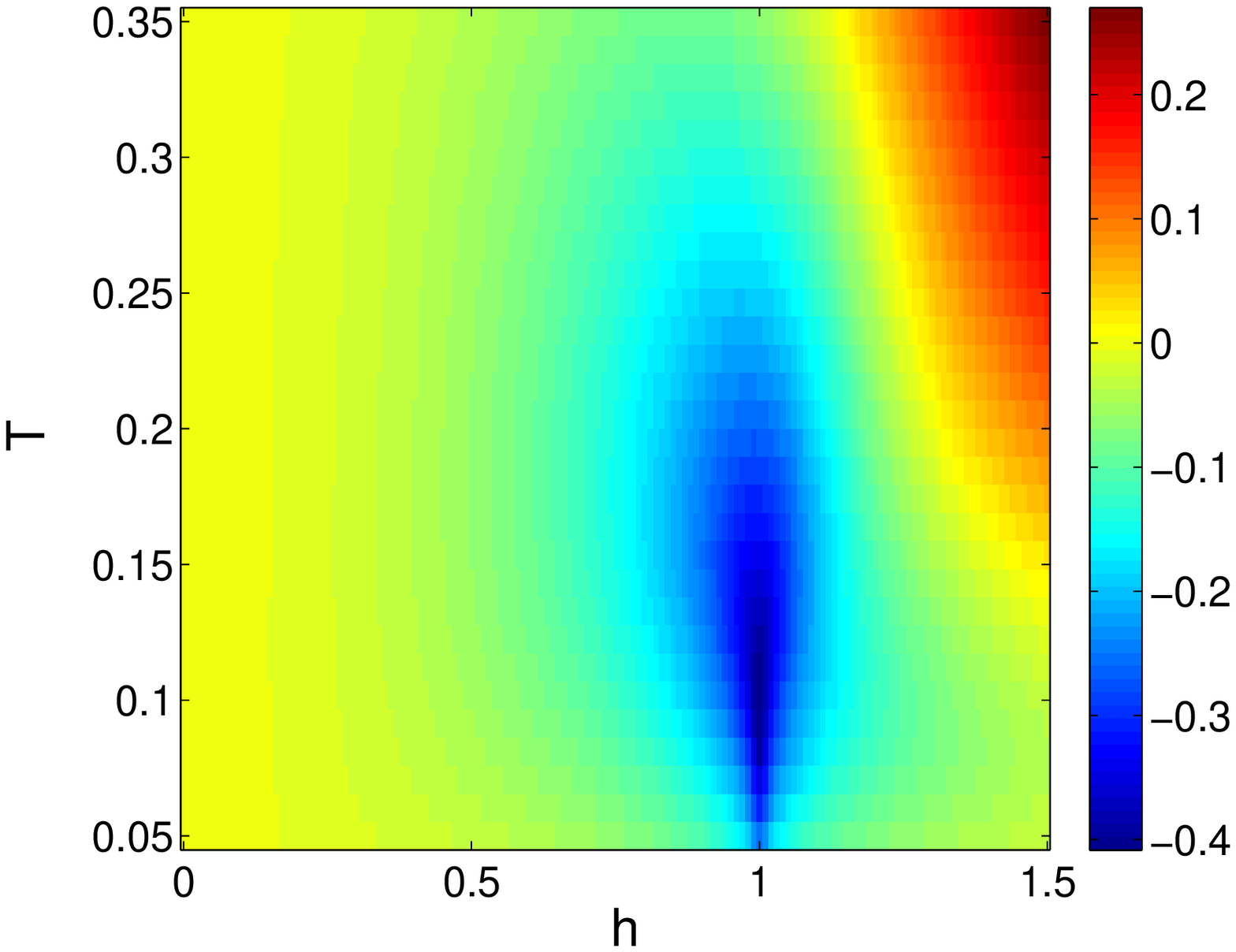}\label{fig:H1dT_det}}
\subfigure[TS, $L=2$]{\includegraphics[scale=0.43,clip]{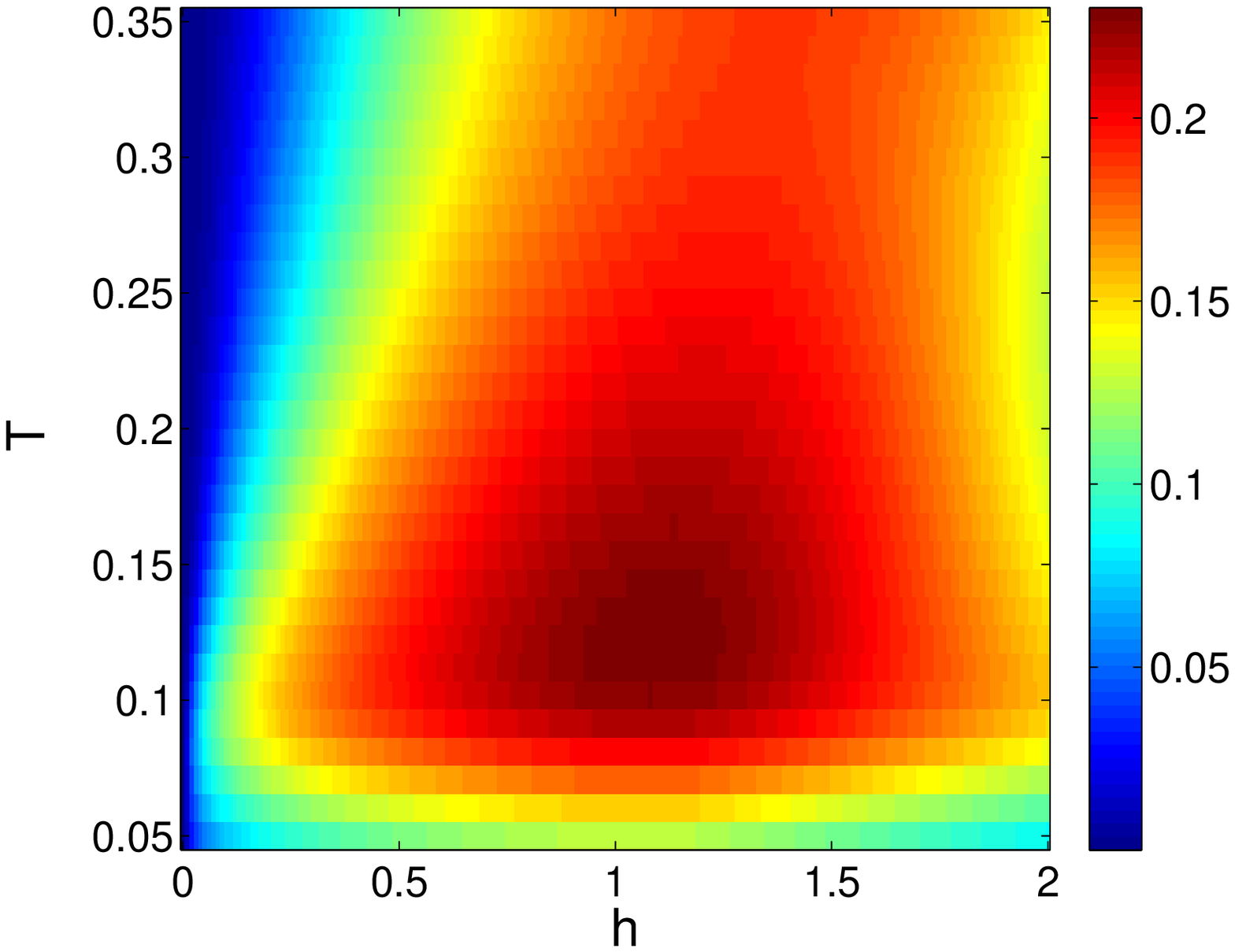}\label{fig:TdT_sel_det}}
\subfigure[T, $L \to \infty$]{\includegraphics[scale=0.43,clip]{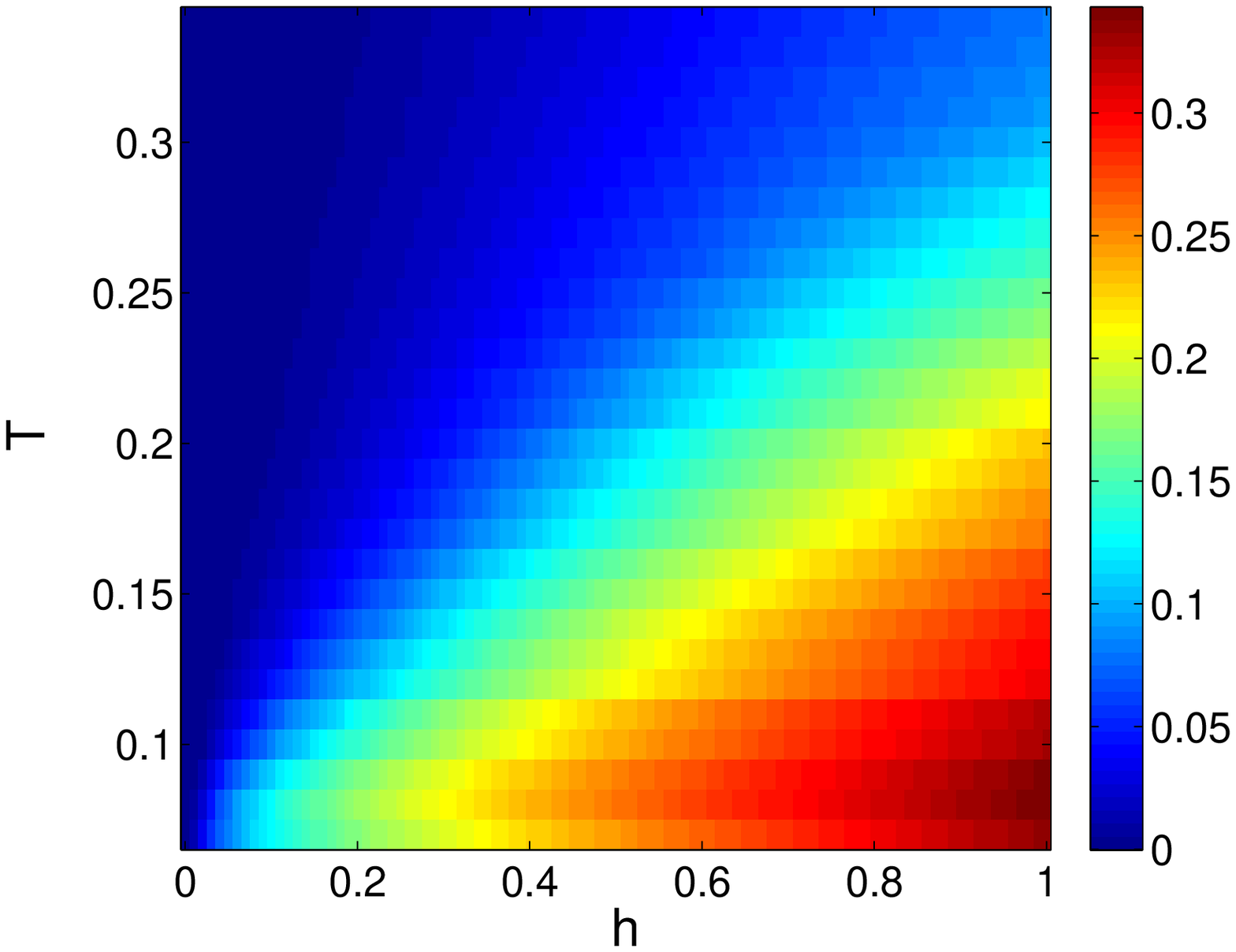}\label{fig:TLdT_det}}
\caption{The low-temperature and low-field adiabatic temperature changes $\Delta T_{ad}$ for (a) the triangular T cluster with $L=2$, (b) the hexagonal H1 cluster with $L=3$, (c) the triangular T cluster with $L=2$ and the field applied selectively only on the spins on one edge, and (d) the thermodynamic limit estimate from MC simulation.}\label{fig:dT_det}
\end{figure}

In the following, we will study the effect of thermal fluctuations on the magnetization and entropic processes in several selected domains, focusing on magnetocaloric properties. Our motivation are enhanced residual entropy density changes observed in ground state for certain domains. In particular, the largest residual entropies at zero field are observed in the triangular T clusters and they are drastically reduced by applying relatively moderate fields. On the other hand, for example, the nonfrustrated hexagonal H1 cluster with $L=3$, which is only trivially degenerate at zero field is relatively highly degenerate at the critical field intensity $h_1=1$. Thus, in the former case the applied field significantly reduces the entropy and one can anticipate an enhanced direct magnetocaloric effect (MCE) in an isothermal process at finite temperatures, while increasing the field from zero to $h_1=1$ in the latter case will result in an increase of the entropy, in which case one has an inverse MCE.\\
\hspace*{5mm} In Figs.~\ref{fig:T0L2}-\ref{fig:T0L5} we present in the temperature-field plane the magnetization $m$, the magnetization derivative with respect to temperature at a constant field $(\partial m/\partial T)|_h$, the isothermal entropy density change $\Delta S/N$ when the field is increased from zero to $h$, as well as the specific heat per spin $c$, for the triangular domain T with $L=2,3,4$ and $5$. As expected, owing to thermal fluctuations the sharp magnetization steps observed in ground state get gradually rounded and eventually dissolve at sufficiently high temperatures (Figs.~\ref{fig:T0L2_m}-\ref{fig:T0L5_m}). However, the effect of thermal fluctuations at a fixed field qualitatively changes from a magnetization decreasing in the first half of each step to a magnetization increasing at the second half of the step, as shown in Figs.~\ref{fig:T0L2_dmdt}-\ref{fig:T0L5_dmdt}. Similar behavior was also observed in the infinite lattice model~\cite{hwan,yao}. Figs.~\ref{fig:T0L2_ds}-\ref{fig:T0L5_ds} show that the relatively large entropy changes, observed in the field-increasing GS processes, persist within some range of low temperatures. For the triangular T shapes, the isothermal entropy changes from zero to finite fields are solely negative, i.e., the they all show the direct MCE. Nevertheless, from Eq.~(\ref{temp_change}) it follows that the adiabatic temperature change achieved by increasing the magnetic field from $0$ to $h$ will also depend on the total specific heat. In the present study, the latter is approximated by considering only the magnetic part $c$, obtained from relation~(\ref{spec_heat}), which is is presented in Figs.~\ref{fig:T0L2_c}-\ref{fig:T0L5_c}. Its behavior in the temperature-field plane is rather complicated and strongly cluster size-dependent. It is interesting to notice that the maxima associated with transitions between different magnetization plateuax are considerably shifted with respect to the magnetization jumps, which is expected in small clusters~\cite{ferd}, and the specific heat close to the fields corresponding to the magnetization jumps is negligible.\\ 
\hspace*{5mm} The resulting variations of the adiabatic temperature changes $\Delta T_{ad}$ for the triangular T clusters of different sizes $L$ are presented in Fig.~\ref{fig:TdT}. As expected in the direct MCE, the temperature changes are positive, i.e., the sample heats up when the external magnetic field is applied adiabatically. For practical applications it would be desirable to achieve significant temperature changes upon application of relatively small external fields. From this point of view, the simple triangle shape ($L=2$) displays fairly large $\Delta T_{ad} \approx 0.37$ at the field as small as $h=0.5$.\\
\hspace*{5mm} On the other hand, a rather different picture can be observed in clusters in which the frustration is relaxed or completely absent, such as in H1 cluster with $L=3$, shown in Fig.~\ref{fig:H1L3}. The low-temperature magnetization is switched from from zero to the saturation value in one step. Similar to the previous cases, below the saturation field thermal fluctuation increase the magnetization and above the saturation field decrease it. Nevertheless, the isotropic entropy changes can be either negative or positive. Namely, for the fields below the transition to the saturated phase it is positive but becomes negative for higher field values, which is a behavior typical for antiferromagnetic systems~\cite{rank}. The resulting adiabatic temperature changes for the clusters T with $L=2$ and H1 with $L=3$ are presented in Figs.~\ref{fig:TdT_det} and~\ref{fig:H1dT_det}, respectively. In the latter case the hexagonal H1 cluster is nonfrustrated  and displays an enhanced inverse MCE near the saturation field $h_s = 1$ and a direct MCE for larger fields and temperatures.

\section{Discussion}
\begin{figure}[t]
\centering
\subfigure[$m$]{\includegraphics[scale=0.43,clip]{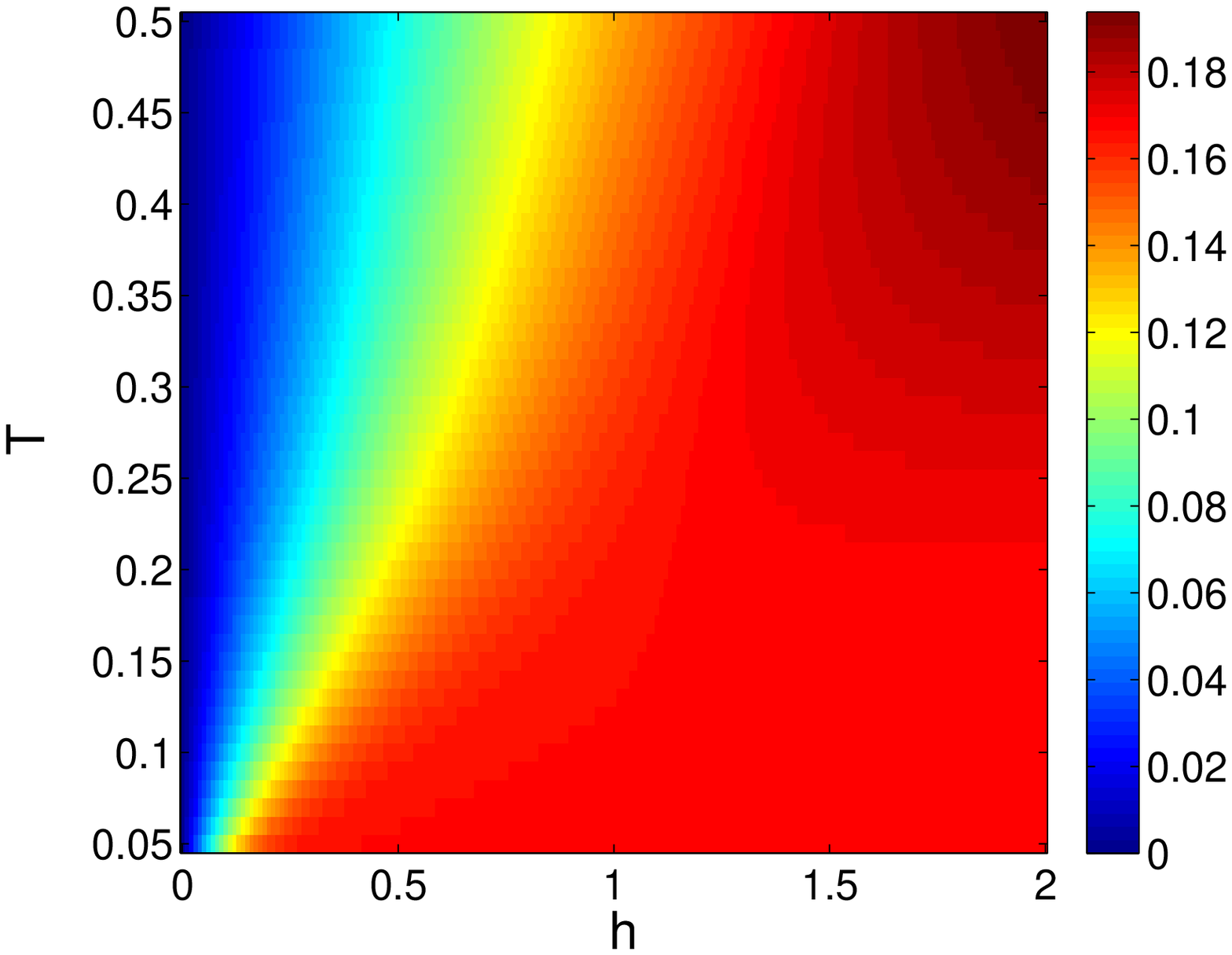}}
\subfigure[$(\partial m/\partial T)|_h$]{\includegraphics[scale=0.43,clip]{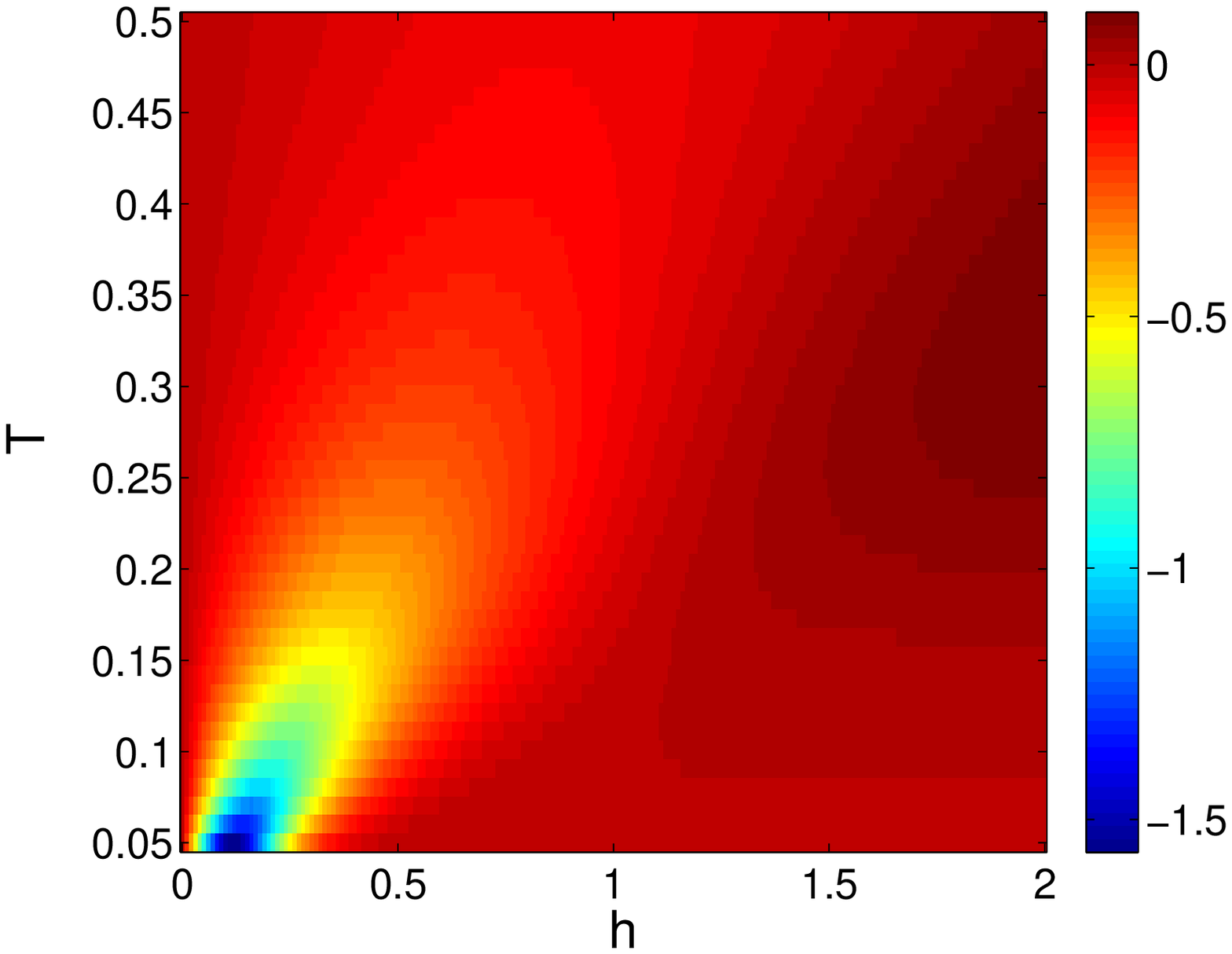}}
\subfigure[$\Delta S/N$]{\includegraphics[scale=0.43,clip]{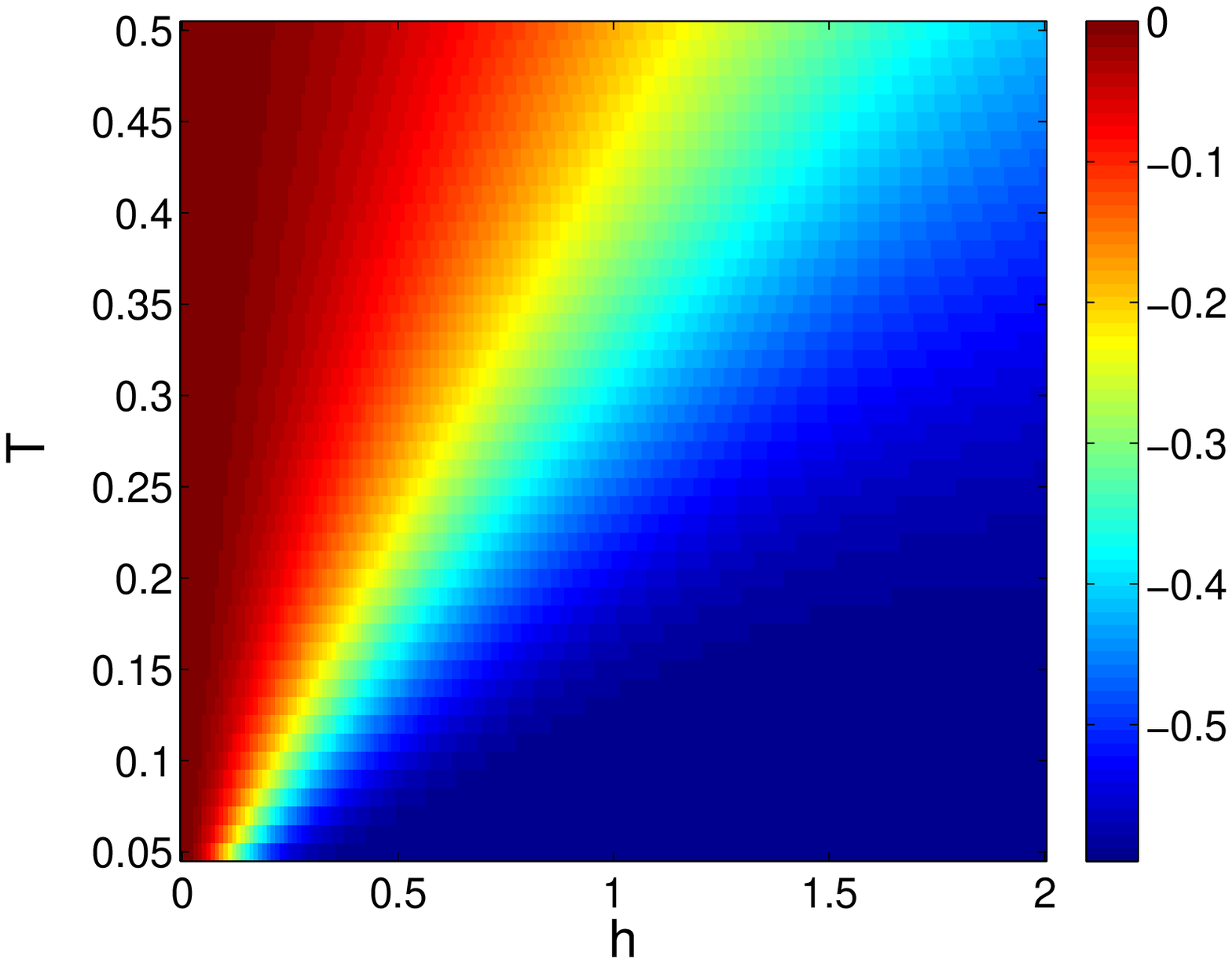}}
\subfigure[$c$]{\includegraphics[scale=0.43,clip]{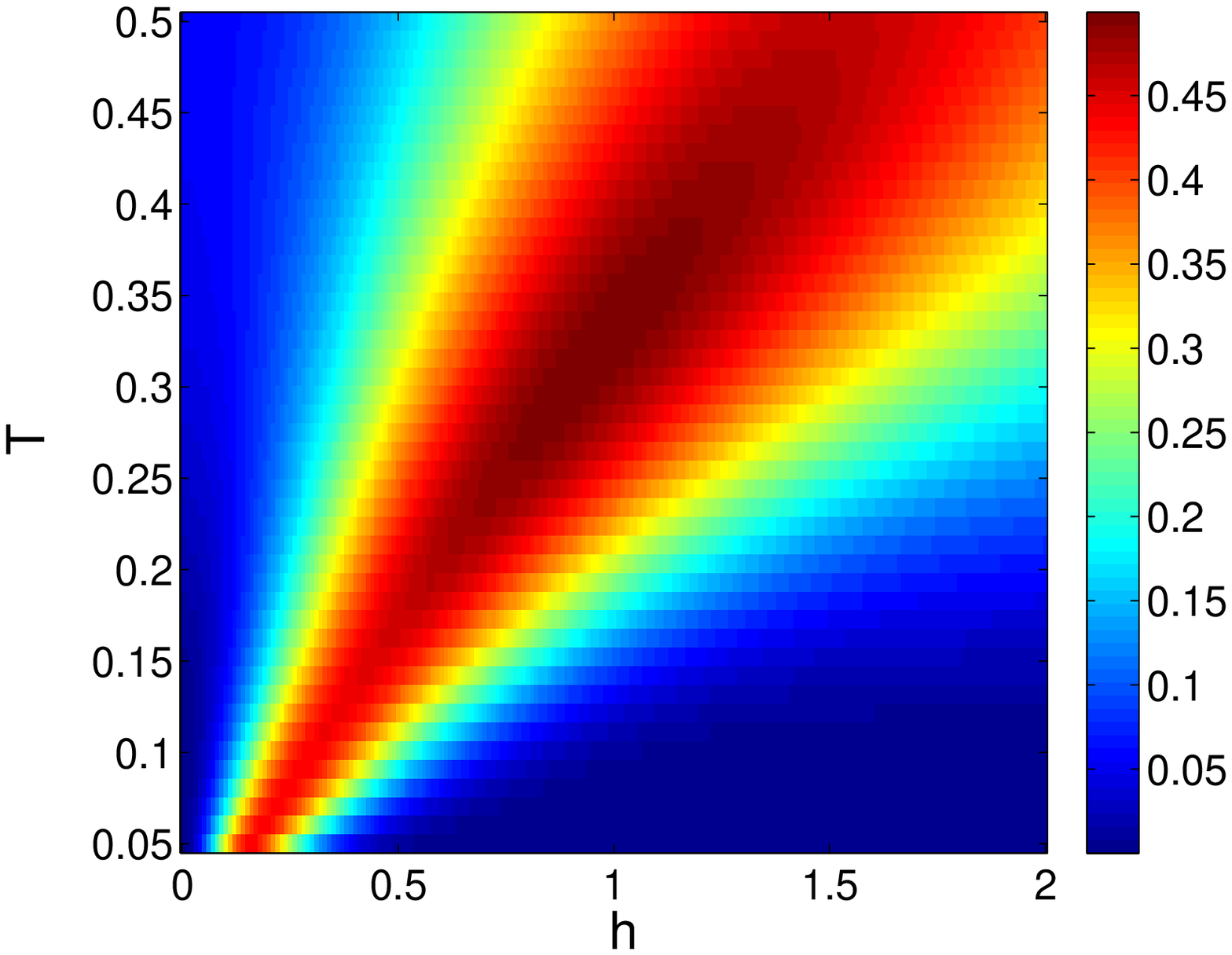}}
\caption{The same quantities as in Fig.~\ref{fig:T0L2} for T cluster with $L=2$ and selectively applied field only on two spins.}\label{fig:TL2_sel}
\end{figure}

\begin{figure}[t]
\centering
\subfigure[$m$]{\includegraphics[scale=0.43,clip]{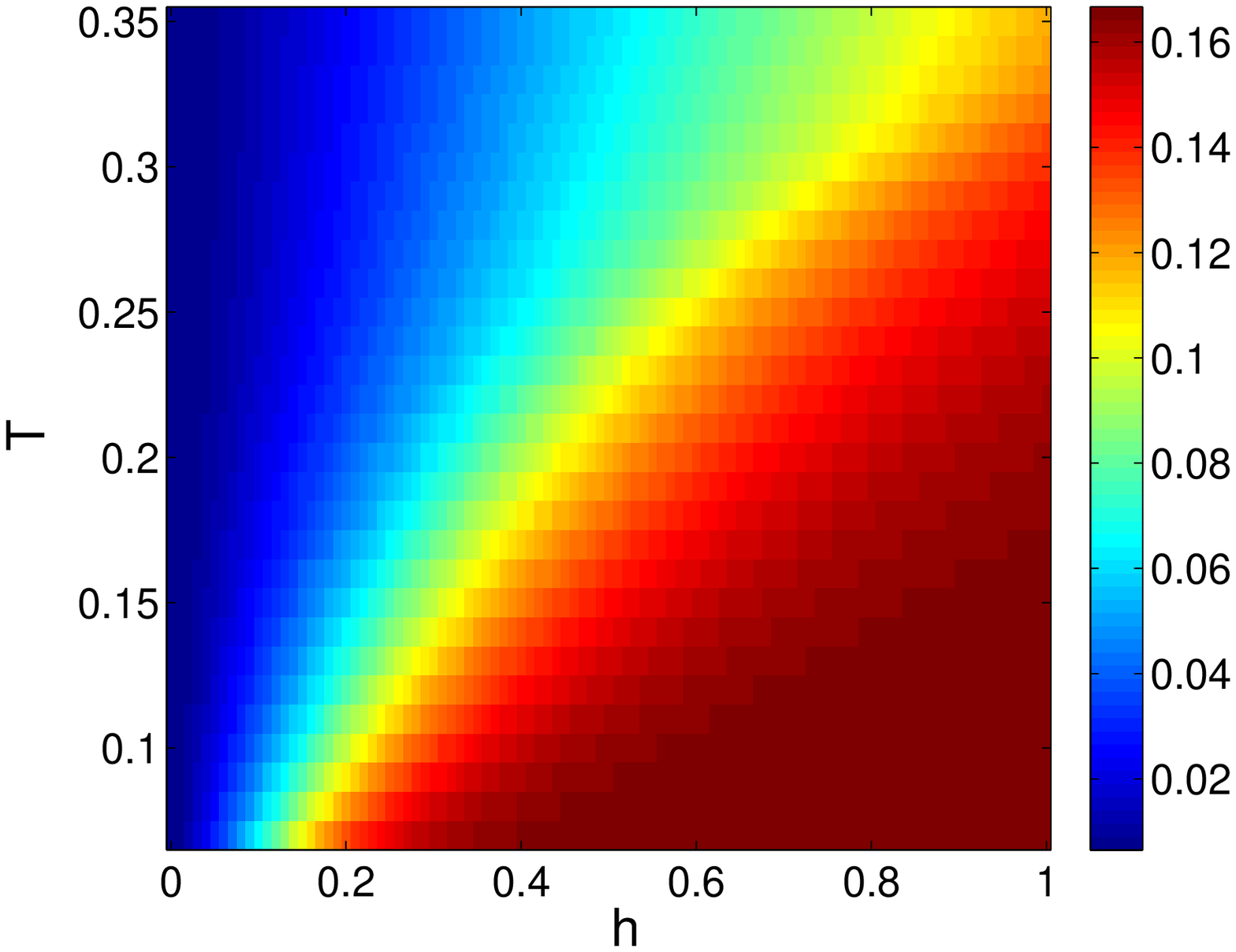}}
\subfigure[$(\partial m/\partial T)|_h$]{\includegraphics[scale=0.43,clip]{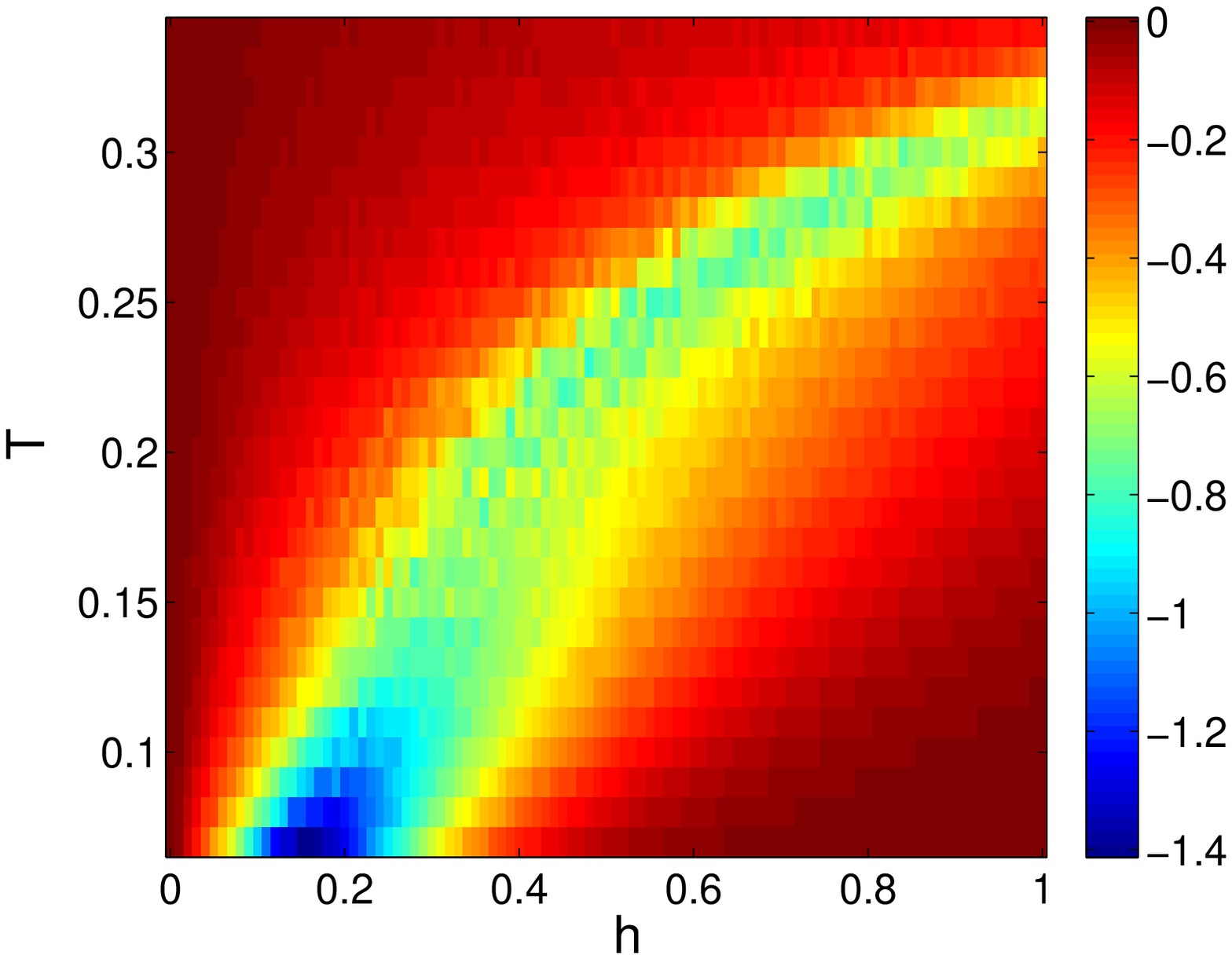}}
\subfigure[$\Delta S/N$]{\includegraphics[scale=0.43,clip]{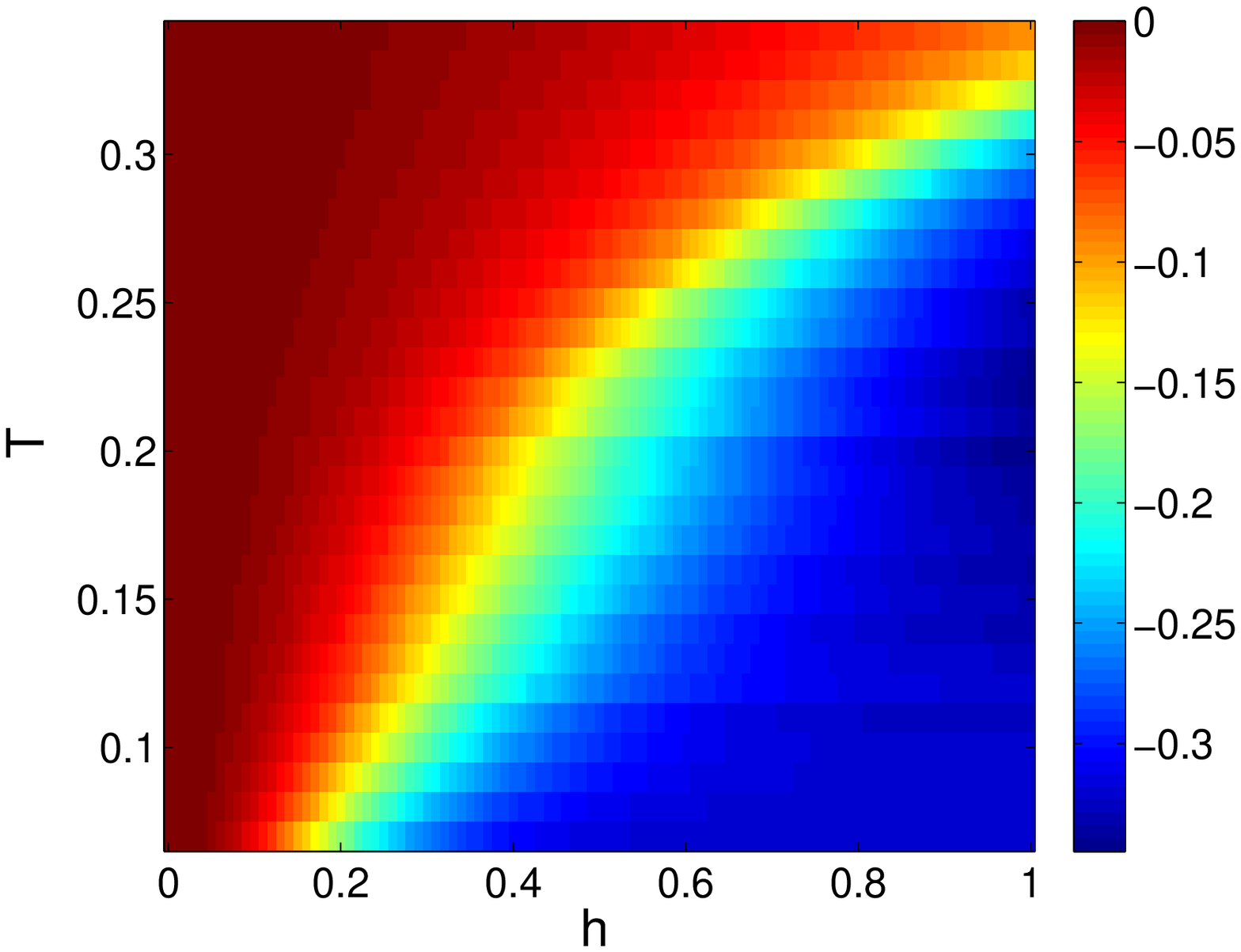}}
\subfigure[$c$]{\includegraphics[scale=0.43,clip]{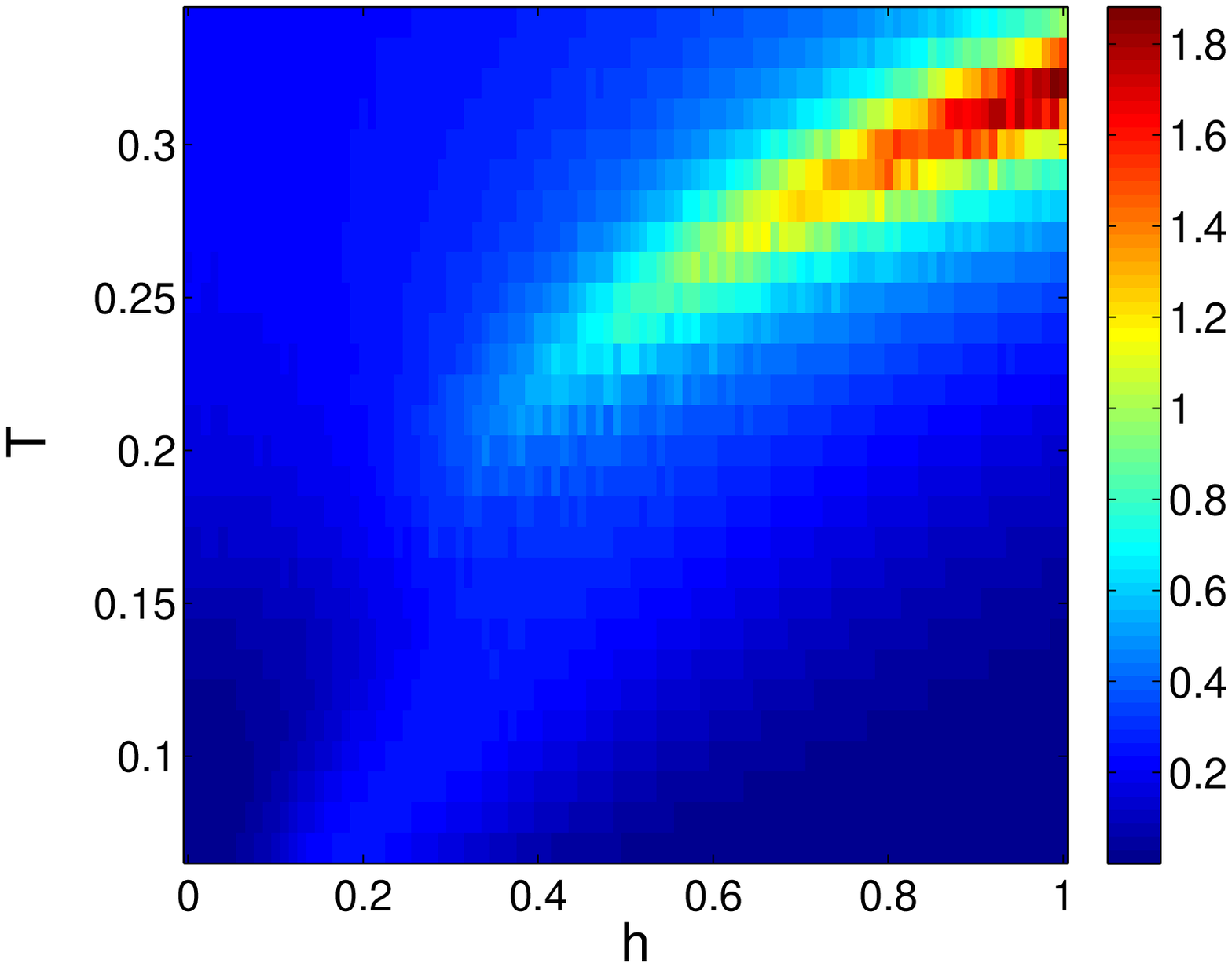}}
\caption{The same quantities as in Fig.~\ref{fig:T0L2} for the thermodynamic limit, obtained from MC simulations.}\label{fig:L24}
\end{figure}

Small triangular T clusters seem to show the largest MCE but it is possible to further enhance it? The ground state of such clusters is highly degenerate~\cite{milla1} but it is easy to see that this large degeneracy can be completely lifted by fixing the states of the spins along one edge. This can be achieved, for example, by applying sufficiently high external field only on these boundary spins. Then one can expect even larger abrupt entropy density change at smaller fields, compared with the case when the field is applied uniformly on all spins and the degeneracy is removed in several steps (see Fig.~\ref{fig:gs_S_tria0}). The behavior of the studied quantities for such a selective application of the external field is presented in Fig.~\ref{fig:TL2_sel}. As expected, a large negative entropy change as well as the magnetization change with respect to temperature are achieved at very small fields, however, the relatively large specific heat values diminish the integrand in Eq.~\ref{temp_change}. Thus, the resulting adiabatic temperature changes (Fig.~\ref{fig:TdT_sel_det}) are even lower that in the uniformly applied field case.\\
\hspace*{5mm} Finally, it is interesting to compare the magnetocaloric properties obtained for the finite frustrated T clusters with those for the infinite triangular lattice. The latter are obtained from Monte Carlo simulations and the corresponding quantities are presented in Fig.~\ref{fig:L24}, for low temperatures and moderate fields. Similar to the clusters behavior, the onset of the 1/3 magnetization plateau is accompanied by pronounced sensitivity to thermal fluctuations and significant isothermal entropy changes in a varying field. The specific heat maxima get larger values and, as expected~\cite{ferd}, occur closer the magnetization jump. Nevertheless, for the same fields, the entropy density changes are larger in the finite T clusters than the infinite lattice. The same applies for the adiabatic temperature changes that are juxtaposed in Fig.~\ref{fig:dT_det}, magnified in the low-temperature and low-field region.\\
\hspace*{5mm} In summary, we have studied thermodynamic and magnetocaloric properties of geometrically frustrated Ising spin clusters of selected shapes and sizes by exact enumeration. At zero temperature we focused on magnetization and entropic processes in an applied magnetic field. Depending on the domain shape and size, we obtained step-wise dependencies featuring a variety of states with with different numbers of magnetization plateaux of different heights and widths, as well as degeneracies. Perhaps the most interesting and unexpected observation was that in some instances the degeneracy gradually increased with the increasing field before being completely eliminated in the saturation phase. For some representative cases, we extend our calculations to finite temperatures by exact evaluation of densities of states in the energy-magnetization space. In zero field we focused on a peculiar behavior of some thermodynamic quantities. In particular, we showed that in the triangular T clusters the entropy attains its minimal (residual) value already well above $T=0$, while the rhombic R clusters minimize their entropies at much lower temperatures and for the cluster size $L \to \infty$ it only happens at $T=0$. In case of the magnetic susceptibility we demonstrated that, besides the previously observed bending in the infinite system and the simple triangular cluster, the inverse susceptibility can show more complicated dependence, including divergence on approach to zero temperature. In finite fields we studied various thermodynamic functions in the temperature-field parameter plane focusing on the cases showing an enhanced magnetocaloric effect. The exact results on the finite clusters were compared with the thermodynamic limit behavior obtained from Monte Carlo simulations and it was concluded that small triangular clusters display larger magnetocaloric effect at smaller fields than the infinite system.  

\section*{Acknowledgments}
This work was supported by the grant of the Slovak Research and Development Agency under the contract No. APVV-0132-11 and the Scientific Grant Agency of Ministry of Education of Slovak Republic (Grant No. 1/0234/12). The authors acknowledge the financial support by the ERDF EU (European Union European Regional Development Fund) grant provided under the contract No. ITMS26220120047 (activity 3.2.).

\end{document}